\documentclass[aps,amsmath,amssymb, prl,twocolumn,superscriptaddress]{revtex4-1}

\usepackage{graphicx}
\usepackage{dcolumn}
\usepackage{bm}
\usepackage[utf8]{inputenc}
\usepackage[ngerman,english]{babel}
\usepackage{float}
\usepackage{multirow}
\usepackage{longtable}
\usepackage{dcolumn}
\newcolumntype{d}{D{.}{.}{-1}}
\usepackage{setspace}
\usepackage{threeparttable}
\usepackage{tabularx}
\usepackage[symbol*]{footmisc}
\DefineFNsymbolsTM{myfnsymbols}{
  \textasteriskcentered *
  \textdagger    \dagger
  \textdaggerdbl \ddagger
  \textsection   \mathsection
  \textbardbl    \|%
  \textparagraph \mathparagraph
}%
\setfnsymbol{myfnsymbols}

\usepackage{siunitx}

\begin{document}


\title{Accessing the single-particle structure of the Pygmy Dipole Resonance in $^{208}$Pb}


\author{M. Spieker}
\email[]{mspieker@fsu.edu}
\affiliation{Department of Physics, Florida State University, Tallahassee, Florida, 32306, USA}

\author{A. Heusler}
\affiliation{Niebuhr-Str. 19c, D-10629 Berlin, Germany}


\author{B.\ A. Brown}
\affiliation{National Superconducting Cyclotron Laboratory, Michigan State University, East Lansing, Michigan 48824, USA}
\affiliation{Department of Physics and Astronomy, Michigan State University, East Lansing, Michigan 48824, USA}

\author{T. Faestermann}
\affiliation{Physik Department, Technische Universität München, D-85748 Garching, Germany}

\author{R. Hertenberger}
\affiliation{Fakultät für Physik, Ludwig-Maximilians-Universität München, D-85748 Garching, Germany}

\author{G. Potel}
\affiliation{Lawrence Livermore National Laboratory, Livermore, California, 94550, USA}

\author{M. Scheck}
\affiliation{School of Computing, Engineering, and Physical Sciences, University of the West of Scotland, Paisley, PA1 2BE,UK}
\affiliation{SUPA, Scottish Universities Physics Alliance, UK}

\author{N. Tsoneva}
\affiliation{Extreme Light Infrastructure (ELI-NP), Horia Hulubei National Institute of Physics and Nuclear Engineering (IFIN-HH), RO-077125, Bucharest-M\v{a}gurele, Romania}

\author{M. Weinert}
\affiliation{Institut f\"{u}r Kernphysik, Universit\"{a}t zu K\"{o}ln, Z\"{u}lpicher Straße 77, D-50937 K\"{o}ln, Germany}

\author{H.-F. Wirth}
\affiliation{Fakultät für Physik, Ludwig-Maximilians-Universität München, D-85748 Garching, Germany}


\author{A. Zilges}
\affiliation{Institut f\"{u}r Kernphysik, Universit\"{a}t zu K\"{o}ln, Z\"{u}lpicher Straße 77, D-50937 K\"{o}ln, Germany}


\date{\today}

\begin{abstract}

New experimental data on the neutron single-particle character of the Pygmy Dipole Resonance (PDR) in $^{208}$Pb are presented. They were obtained from $(d,p)$ and resonant proton scattering experiments performed at the Q3D spectrograph of the Maier-Leibnitz Laboratory in Garching, Germany. The new data are compared to the large suite of complementary, experimental data available for $^{208}$Pb and establish $(d,p)$ as an additional, valuable, experimental probe to study the PDR and its collectivity. Besides the single-particle character of the states, different features of the strength distributions are discussed and compared to Large-Scale-Shell-Model (LSSM) and energy-density functional (EDF) plus Quasiparticle-Phonon Model (QPM) theoretical approaches to elucidate the microscopic structure of the PDR in $^{208}$Pb.

\end{abstract}


\maketitle



Atomic nuclei with large proton-neutron asymmetry, like $^{208}$Pb, form a neutron skin\,\cite{Thi19a}. The neutron-skin thickness, $\Delta r_{np}$, is directly correlated to properties of neutron stars\,\cite{Thi19a,Hor01a,Hor01b,Fat12a,Fat13a}. This raised the interest of the science community in determining it experimentally\,\cite{Tam11a,Abr12a,Tam13a,Tar14a}. Following the first multi-messenger detection of a binary neutron star merger\,\cite{Abb17a} including gravitational waves\,\cite{Abb17b}, this interest has been recently reinforced\,\cite{Fat18a}. 




The electric dipole polarizability $\alpha_D$\,\cite{Tam11a,Tam13a,Ros13a,Has15a,Hag16a,Bir17a,Ton17a,Roc18a,Kau20a} is one key observable investigated to obtain constraints on $\Delta r_{np}$. For its precise determination, the low-lying $E1$ strength is extremely important. The term ``Pygmy Dipole Resonance'' (PDR) has been commonly used for the $E1$ strength around and below the neutron-separation energy, $S_n$\,\cite{Bar61a,Paar07a,Sav13a,Bra15a,Bra19a}. The PDR strength might also correlate more strongly with $\Delta r_{np}$\,\cite{Pie06a,Tso08a,Car10a,Pie11a,Bar13a,Roc18a} and, thus, provide tighter constraints. However, the possibly stronger correlation has been critically discussed\,\cite{Ina11a,Ina13a,Rei13a,Roc15a}. In any case, it would be necessary to distinguish the PDR from other $E1$ modes like the low-energy tail of the Giant Dipole Resonance (GDR) (see, {\it e.g.},\,\cite{End10a,Pol12a,Ton17a,Sav18a,Wie18a}). It has been shown that the PDR strength strongly impacts neutron-capture rates in the $s$ and $r$ process\,\cite{Gor98a,Lit09b,Tso15a,Ton17a,Lar19a}. A precise understanding of its microscopic structure is also essential to pin down how the PDR contributes to the $\gamma$-ray strength function ($\gamma$SF) often used to calculate $(n,\gamma)$ rates\,\cite{Lar19a}, i.e. whether there is a dependence of the $\gamma$SF's shape on excitation energy, spin-parity quantum number or even specific nuclear structure\,\cite{Ang12a,Bas16a,Gut16a,Mar17a,Cam18a,Isa19a,Sim20a,Sch20a}.


Depending on the mass region of the nuclear chart, the low-lying $E1$ response to isovector and isoscalar probes, or to probes testing surface rather than bulk properties, is different\,\cite{Sav06a,End09a,End10a,Der13a,Der14a,Cre14a,Pel14a,Lan14a,Cre15a,Krz16a,Nak17a,Cre18a,Sav18a}. While in lighter nuclei usually state-to-state differences were observed, some heavier nuclei featured the so-called isospin splitting of the low-lying $E1$ response (see the review articles\,\cite{Sav13a,Bra19a}). These different responses emphasized that different underlying structures would indeed need to be disentangled experimentally, if stringent comparisons to microscopic models wanted to be made.

Besides its isospin structure, the degree of collectivity of the PDR is still under debate\,\cite{Vre01a,Rye02a,Lit09a,Lan09a,Roc12a,Vre12a,Bia12a,Bar15a,Pap15a,Bra19a,Rie19a}. Often, collectivity is accessed in terms of the number of one-particle-one-hole (1p--1h) excitations acting coherently and, therefore, causing enhanced transition strength \cite{Lan09a,Roc12a,Vre12a}. A recent theoretical study of the PDR in $^{68}$Ni\,\cite{Bra19a}, which used a fully self-consistent nonrelativistic mean-field approach based on Skyrme Hartree-Fock plus random phase approximation (HF+RPA), reinforced that coherence between several 1p--1h configurations is rather observed in the isoscalar than in the isovector channel. Qualitatively comparable results had been obtained for $^{132}$Sn and $^{208}$Pb by employing similar theoretical approaches\,\cite{Roc12a,Vre12a}. These theoretical results question the usefulness of studying the PDR's collectivity based on the isovector $E1$ strengths alone. 


In this work, we present a detailed, high-resolution $(d,p)$ experimental study of the PDR in $^{208}$Pb and complement it with available experimental data to discuss the PDR's microscopic structure and its influence on experimental observables by comparing to state-of-the-art, theoretical models. The neutron 1p--1h configurations contributing to forming the PDR are accessed from $(d,p)$ data up to the proton-separation energy, $S_p$, and, for a limited number of states, from the results of resonant proton scattering via isobaric analog resonances ($(p,p')_{\mathrm{IAR}}$)\,\cite{Heu07a,Heu10a,Heu14a,Heu14b,Heu20a,Heu20b}, which probes components that could not be populated in the selective one-neutron transfer reaction. An unprecedented access to the theoretical wave functions was achieved. 
 
When discussing its collectivity within the HF+RPA approach, Roca-Maza {\it et al.} identified the PDR of $^{208}$Pb above 7\,MeV\,\cite{Roc12a}. Following a comparison of Nuclear Resonance Fluorescence data and Quasiparticle-Phonon Model (QPM) calculations, Ryezayeva {\it et al.} had argued that the PDR should indeed correspond to the strength observed around $S_n$\,\cite{Rye02a}. The lower-lying $1^-$ states should have a more pure single-particle character\,\cite{Rye02a}. Poltoratska {\it et al.}\,\cite{Pol12a} considered, however, all low-lying $E1$ strength up to $\sim$\,8.3\,MeV to belong to the PDR in agreement with a $(^{17}\mathrm{O},^{17}\mathrm{O}'\gamma)$ experiment\,\cite{Cre14a}, performed to study its isospin character. Based on QPM calculations for $^{206}$Pb, dominantly the neutron 1p--1h states below $S_n$ were identified to belong to the PDR of the $N = 124$ Pb isotope\,\cite{Ton17a}. The importance of including two-particle-two-hole (2p--2h) configurations to describe the isovector $B(E1)$ strength fragmentation was pointed out in \cite{Sch10a} using Large-Scale-Shell-Model (LSSM) calculations\,\cite{Bro00a}. Also the possibility of tetrahedral configurations in $^{208}$Pb was presented and some of the lower-lying states, including the $1^-_1$ state, were discussed to originate from this exotic type of excitation\,\cite{Heu17a}.

\begin{figure}[t]
\centering
\includegraphics[width=1\linewidth]{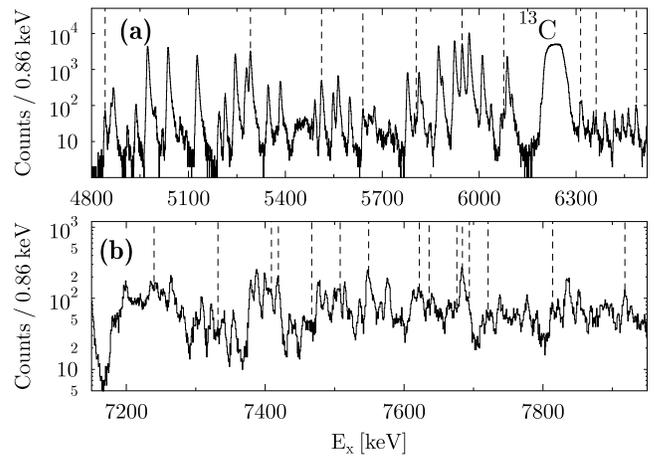}
\caption{\label{fig:spectra}{{\bf (a)}, {\bf (b)} $^{207}\mathrm{Pb}(d,p)^{208}\mathrm{Pb}$ spectra taken at $\theta = 25^{\circ}$ for two different magnetic settings. Only a part of the spectrum is shown in panel {\bf (b)}. Contamination from the $^{12}$C$(d,p)^{13}$C reaction is observed (labeled with $^{13}$C) due to the Carbon backing of the target. The kinematic correction with the Q3D multipole element was applied to the $^{207}$Pb$(d,p)$ reaction causing the peaks observed from $^{12}$C$(d,p)$ to be significantly broader (compare \cite{Eng79a}). Known $J^{\pi} = 1^-$ states of $^{208}$Pb\,\cite{Mar07a,ENSDF,Heu16a}, which could be resolved, are highlighted with vertical, dashed lines. All other states, seen in the spectra, correspond to excited states of $^{208}$Pb with a 3p$_{1/2}$ neutron-hole component in their wave function. Below $S_n$, many of them were experimentally observed before\,\cite{Vol73a,Sch97a,Val01a,Mar07a,ENSDF}.}} 
 \end{figure}

The new data, presented here, were obtained from a series of experiments performed to study excited states in $^{208}$Pb with the high-resolution Q3D spectrograph of the Maier-Leibnitz Laboratory (MLL) in Garching, Germany\,\cite{MLL,MLL2}. For the $(d,p)$ experiments, the deuterons were accelerated to 22\,MeV and impinged onto a 0.11-mg/cm$^2$ thick, highly-enriched $^{207}$Pb target (99\,$\%$ enrichment) on a Carbon backing. After the reaction, the residual particles were momentum-analyzed with the Q3D and detected in the focal-plane detection system\,\cite{Wir00,Wir01}. By adjusting the horizontal entrance slits, half ($\pm 1.6^{\circ}$) of the Q3D's maximum angular acceptance was used and an energy resolution of better than 6\,keV (FWHM) achieved. This facilitated the analysis of the dense excitation spectra seen in Fig.\,\ref{fig:spectra}.

\begin{figure}[t]
\centering
\includegraphics[width=1\linewidth]{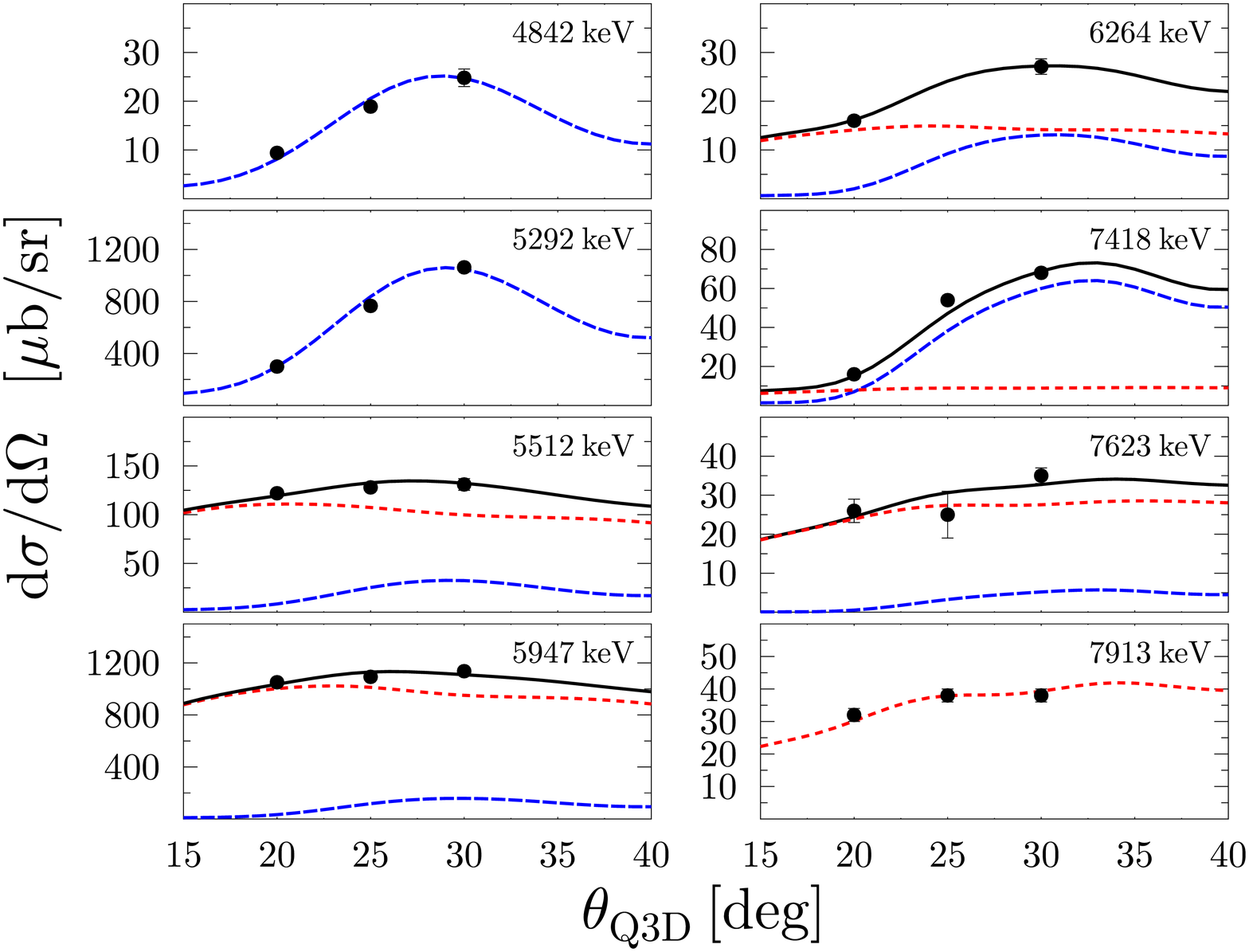}
\caption{\label{fig:angdists}{(color online) Measured $(d,p)$ angular distributions (differential cross sections $d\sigma / d\Omega$) for selected $J^{\pi}=1^-$ states (circles) in comparison to DWBA calculations (lines). Two different 1p--1h configurations, (3p$_{1/2})^{-1}$(4s$_{1/2}$)$^{+1}$ (blue, longer dashed lines) and (3p$_{1/2})^{-1}$(3d$_{3/2}$)$^{+1}$ (red, shorter dashed lines),  have been assumed to describe the experimental distributions. Black, solid lines correspond to superpositions of these two individual configurations. No multistep transfer was considered. For states above $S_n$, an effective neutron-separation energy of $S_n = 8.5$\,MeV had to be used. Otherwise, the shape of the angular distribution would have been heavily distorted. A similar approach had been chosen in \cite{Kov74a}. The unique features of the $l=0$ and $l=2$ transfers remain unchanged. For the $1^-$ state at 6264\,keV, a Carbon contaminant prevented a cross-section measurement at $\theta = 25^{\circ}$.}}
 \end{figure}

\begin{figure}[pt]
\centering
\includegraphics[width=1\linewidth]{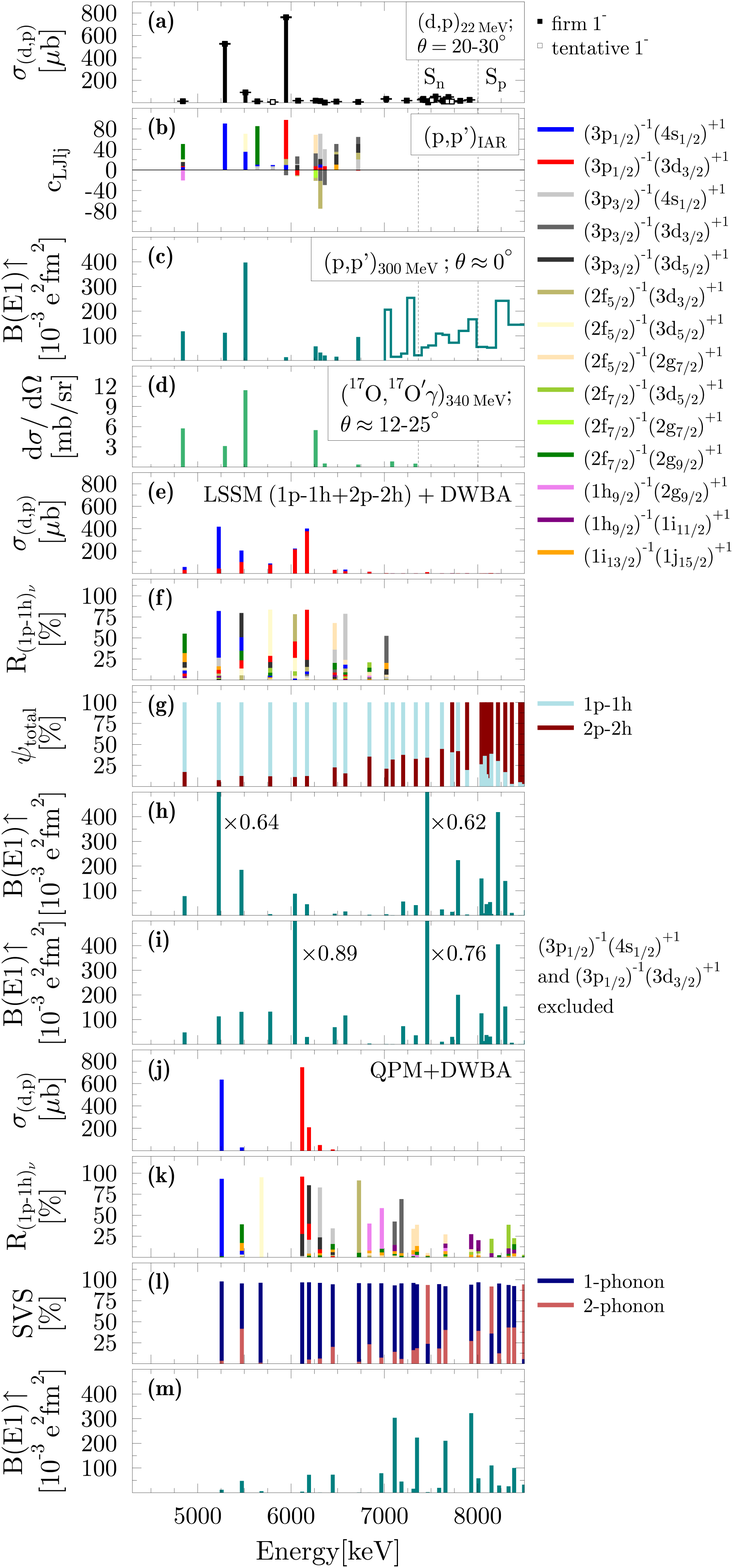}
\caption{\label{fig:exp_sm_comp}{(color online) {\bf (a)} Angle-integrated $(d,p)$ cross sections $\sigma_{(d,p)}$, {\bf (b)} $c_{LJlj}$ amplitudes from $(p,p')_{\mathrm{IAR}}$\,\cite{Heu07a,Heu10a,Heu20a,Heu20b}, {\bf (c)}} isovector $B(E1)$ strengths from $(p,p')$\,\cite{Pol12a}, and {\bf (d)} differential cross sections from $(^{17}\mathrm{O},^{17}\mathrm{O}'\gamma)$\,\cite{Cre14a}. The latter probe the isoscalar character of the $1^-$ states\,\cite{Cre14a}. {\bf (e)} $\sigma_{(d,p)}$ predicted by combining LSSM spectroscopic factors with DWBA calcultions. {\bf (f)} Decomposition of the LSSM wave functions into neutron 1p--1h components relative to the total wave function $\psi_{\mathrm{total}}$. {\bf (g)} 1p--1h and 2p--2h contributions to $\psi_{\mathrm{total}}$. LSSM isovector $B(E1)$ strength predicted when {\bf (h)} including all or {\bf (i)} exluding the specified contributions. {\bf (j)}-{\bf (m)} same as {\bf(e)}-{\bf(h)} but for EDF+QPM. SVS stands for ``state-vector structure''\,\cite{Tso16a,suppl}.}
 \end{figure}

The $(d,p)$ data were analyzed at three scattering angles; 20$^\circ$, 25$^\circ$, and 30$^\circ$. This allows to distinguish the two different transfer configurations through which the known $1^-$ states of $^{208}$Pb\,\cite{Mar07a,ENSDF} can be populated from the $J^{\pi} = 1/2^-$ ground state of $^{207}$Pb; namely (3p$_{1/2})^{-1}$(4s$_{1/2})^{+1}$ ($l=0$) and (3p$_{1/2})^{-1}$(3d$_{3/2})^{+1}$ ($l=2$). The angular distributions are shown in Fig.\,\ref{fig:angdists} alongside Distorted-Wave-Born-Approximation (DWBA) calculations performed with the coupled-channels program \textsc{chuck3} \cite{chuck}. The global optical-model parameters (OMP) of \cite{bec69a} were used for the protons and of \cite{Dae80a} for the deuterons with adjustments to the real potential of the volume Woods-Saxon part from \cite{Deb70a}. With the exception of using an effective neutron-separation energy for states above $S_n$, the same OMP were used for all excited states. As shown in Fig.\,\ref{fig:angdists}, the measured and DWBA angular distributions are in excellent agreement. The dominant contributions of the most strongly excited $1^-$ states at 5292\,keV, 5512\,keV, and 5947\,keV were previously identified\,\cite{Vol73a,Sch97a,Val98a,Val01a,Heu07a,Heu10a} and confirmed here. Only small additional (3p$_{1/2})^{-1}$(4s$_{1/2})^{+1}$ contributions were needed to explain the experimental angular distributions for the 5512-keV and 5947-keV states. In return, this highlights the sensitivity of the present experiment to such small contributions. The $(p,p')_{\mathrm{IAR}}$ data on the $3d_{3/2}$ and $4s_{1/2}$ IARs in $^{209}$Bi confirm the dominant structure assignments for the 5292-keV and 5947-keV state [compare Fig.\,\ref{fig:exp_sm_comp}\,{\bf (b)}], respectively. Also at higher excitation energies, superpositions of the two configurations were often needed to explain the experimental $(d,p)$ data as shown for three examples in Fig.\,\ref{fig:angdists}. As indicated by the $(p,p')_{\mathrm{IAR}}$ data, other 1p--1h configurations are important as well and might dominate the structure of the states [compare Fig.\,\ref{fig:exp_sm_comp}\,{\bf (b)}]. In total, 11 out of the 15 amplitudes were studied experimentally\,\cite{Heu07a,Heu10a,Heu14a,Heu14b,Heu20a,Heu20b}. More details on the determination of the relative $c_{LJlj}$ amplitudes for the different neutron 1p--1h configurations from $(p,p')_{\mathrm{IAR}}$ are presented in\,\cite{Heu07a,Heu10a,Heu14a,Heu14b,Heu16a,Heu20a}.

The model-indepedent, angle-integrated $(d,p)$ cross sections and $c_{LJlj}$ amplitudes from $(p,p')_{\mathrm{IAR}}$ are shown in Fig.\,\ref{fig:exp_sm_comp} in comparison to a selection of other experimental data on the PDR in $^{208}$Pb\,\cite{Pol12a,Cre14a}. The $(d,p)$ strength pattern [Fig.\,\ref{fig:exp_sm_comp}\,{\bf (a)}] is dominated by the two strongly populated $1^-$ states at 5292\,keV and 5947\,keV, corresponding to the major fragments of the (3p$_{1/2})^{-1}$(4s$_{1/2})^{+1}$ [$S = 0.77(4)$] and (3p$_{1/2})^{-1}$(3d$_{3/2})^{+1}$ [$S =  0.66(4)$] neutron 1p--1h strength (compare Fig.\,\ref{fig:angdists}), respectively. The stated spectroscopic factors, $S$, are model-dependent but were determined consistently, i.e. using the same OMP. This is different from the approach chosen in \cite{Val98a,Val01a}, where OMP were varied depending on the $l$ transfer introducing a stronger model dependency. While the 5292-keV state shows appreciable $B(E1)$ strength\,\cite{Rye02a,Pol12a} and is also comparably strongly populated in $(^{17}\mathrm{O},^{17}\mathrm{O}'\gamma)$\,\cite{Cre14a}, the 5947-keV state is, strikingly, barely excited with the electromagnetic probe and not at all with the hadronic probe [compare Figs.\,\ref{fig:exp_sm_comp}\,{\bf (a)}, {\bf (c)}, {\bf (d)}]. Remarkably, the group of states with excitation energies of 6264\,keV, 6314\,keV, 6362\,keV, and 6486\,keV features both gradually decreasing $(d,p)$ cross sections and isovector $B(E1)$ strengths, while only the 6264-keV state is strongly excited in $(^{17}\mathrm{O},^{17}\mathrm{O}'\gamma)$. For the 6264-keV and 6314-keV states, mixtures of $l=0$ and $l=2$ transfers were needed to describe the experimental angular distributions (see, exemplary, the 6264-keV state in Fig.\,\ref{fig:angdists}). One configuration was sufficient for the 6362-keV ($l=0$) and 6486-keV ($l=2$) states. Interestingly, the 6264-keV state is the only one of the four, which has (2f$_{5/2})^{-1}$(2g$_{7/2})^{+1}$ and (2f$_{7/2})^{-1}$(2g$_{7/2})^{+1}$ components in its wave function [compare Fig.\,\ref{fig:exp_sm_comp}\,{\bf (b)}].


Figs.\,\ref{fig:exp_sm_comp}\,{\bf (e)}-{\bf (m)} presents the results of LSSM\,\cite{Bro00a} and energy-density functional (EDF)+QPM \cite{Tso16a} calculations. To calculate the differential cross sections $d\sigma/d\Omega$, predicted spectroscopic factors, i.e. the overlap of the $^{207}$Pb ground state with excited $1^-$ states in $^{208}$Pb when adding a neutron, were combined with the DWBA calculations\,\cite{chuck}, which described the experimental data. The angle-integrated $\sigma_{(d,p)}$ cross sections were also determined between $\theta = 20^{\circ} - 30^{\circ}$.

The LSSM calculations [Fig.\,\ref{fig:exp_sm_comp}\,{\bf (e)}-{\bf (i)}] were introduced in \cite{Bro00a,Sch10a}. In addition to the $\sigma_{(d,p)}$ values (see supplement for truncation at 1p--1h level\,\cite{suppl}), we provide the decomposition of the wave functions into the different neutron 1p--1h components ($> 1$\,$\%$) relative to the total wave function $\psi_{\mathrm{total}}$ [Fig.\,\ref{fig:exp_sm_comp}\,{\bf (f)}], and the contributions of 1p--1h and 2p--2h components to $\psi_{\mathrm{total}}$ [compare Fig.\,\ref{fig:exp_sm_comp}\,{\bf (g)}]. The predicted excitation energy, 5226\,keV, of the major (3p$_{1/2})^{-1}$(4s$_{1/2})^{+1}$ [$S_{LSSM} = 0.56$] fragment is very close to the experimental 5292-keV state [$S_{exp}=0.77(4)$]. Considering the other strong fragment with $S_{LSSM}=0.16$ at 5469\,keV provides a centroid energy of 5280\,keV with $S_{LSSM} = 0.72$ in almost perfect agreement with the experimental data. Below 6.25\,MeV, the (3p$_{1/2})^{-1}$(3d$_{3/2})^{+1}$ strength is much more fragmented than experimentally observed. The strongest fragment is predicted at 6171\,keV with $S_{LSSM} = 0.34$. The LSSM centroid of the $l=2$ strength is found at 5912\,keV with $S_{LSSM} = 0.78$ ($E_x < 6.25$\,MeV), which again compares well to the experimental centroid at 5904\,keV [$S_{exp} = 0.73(4)$] when considering the 5512-keV and 5947-keV states. The summed, angle-integrated cross sections below $S_n$ are $\sum \sigma_{(d,p)_{exp}}= 1524(17)$\,$\mu$b and $\sum \sigma_{(d,p)_{LSSM}}= 1470$\,$\mu$b. However, the fragmentation of the LSSM spectroscopic strength between $S_n$ and $S_p$ is not as observed in experiment.  For firm $1^-$ states above $S_n$, $\sum \sigma_{(d,p)_{exp}}$ is 254(9)\,$\mu$b while the LSSM predicts only 22\,$\mu$b. 13\,$\%$ of the $d_{3/2}$ and 9\,$\%$ of the $s_{1/2}$ strength are pushed to energies higher than 8.6\,MeV in the LSSM. The data suggest that this strength is located below $S_p$.

Many neutron 1p--1h excitations contribute to $\psi_{\mathrm{total}}$ with the strongest component never exceeding 56\,$\%$. Given the experimental limitation of only being able to determine three to four amplitudes when studying one IAR, this seems largely consistent with the $(p,p')_{\mathrm{IAR}}$ data. As seen from the comparison of Figs.\,\ref{fig:exp_sm_comp}\,{\bf (b)} and \ref{fig:exp_sm_comp}\,{\bf (f)}, most $1^-$ states cannot be considered as simple neutron 1p--1h states. The $(d,p)$ data prove that almost all $1^-$ states have at least small (3p$_{1/2})^{-1}$(4s$_{1/2})^{+1}$ and (3p$_{1/2})^{-1}$(3d$_{3/2})^{+1}$ components, many of which were below the sensitivity limit in $(p,p')_{\mathrm{IAR}}$. With only a few exceptions, the neutron 1p--1h contribution makes up around 80\,$\%$ of $\psi_{\mathrm{total}}$ in the LSSM at lower energies (compare $R_{(1p-1h)_{\nu}}$ in Fig.\,\ref{fig:exp_sm_comp}\,{\bf (f)} for the first ten $1^-$ states). The strongest neutron 1p--1h component in the wave function of the $1^-_1$ is identified as (2f$_{7/2})^{-1}$(2g$_{9/2})^{+1}$ in both $(p,p')_{\mathrm{IAR}}$ and the LSSM. The experimental data for the $1^-_1$ support that less than 60\,$\%$ of $\psi_{\mathrm{total}}$ are due to neutron 1p--1h components\,\cite{Heu14b,Heu17a}. For almost all lower-lying $1^-$ states, the 2p--2h contribution already exceeds 10\,$\%$ [compare Fig.\,\ref{fig:exp_sm_comp}\,{\bf (g)}]. A clear structural change is observed above 7.5\,MeV, where 2p--2h configurations begin to dominate the wave functions. Note that Poltoratska {\it et al.}\,\cite{Pol12a} experimentally observed a structure change at $\sim 8.2$\,MeV, where GDR-type wave functions began to dominate. At $\sim 8.4$\,MeV, the 1p--1h contribution to $\psi_{\mathrm{total}}$ drops well below 10\,$\%$ in the LSSM.

The LSSM $B(E1)$ strength distribution\,\cite{Sch10a} is shown in Fig.\,\ref{fig:exp_sm_comp}\,{\bf (h)}. Problematically, the most enhanced $B(E1)$ value is observed for the major (3p$_{1/2})^{-1}$(4s$_{1/2})^{+1}$ fragment, i.e. the $1^-_2$ state in conflict with experiment [compare Fig.\,\ref{fig:exp_sm_comp}\,{\bf (c)}]. The major (3p$_{1/2})^{-1}$(3d$_{3/2})^{+1}$ fragment has an experimental $B(E1) = 13(1) \times 10^{-3}$\,$\mathrm{e^2fm^2}$. The LSSM calculations predict $B(E1) = 45 \times 10^{-3}$\,$\mathrm{e^2fm^2}$. Also, the large $B(E1)$ value from the LSSM state at 7.5\,MeV is not observed for a specific, experimental state. We note that, in general, there is a large amount of cancellation between the shell-model components of the $E1$ matrix elements in the PDR region in contrast to states of the GDR (compare the supplement \cite{suppl} for more details), which introduces a pronounced sensitivity to the Hamiltonian and large uncertainties for the theoretical $B(E1)$ values. To further highlight this problem, we have excluded the (3p$_{1/2})^{-1}$(4s$_{1/2})^{+1}$ and (3p$_{1/2})^{-1}$(3d$_{3/2})^{+1}$ contributions to the $B(E1)$ strengths in Fig.\,\ref{fig:exp_sm_comp}\,{\bf (i)}. The strength fragmentation below 7\,MeV changes drastically. Missing microscopic configurations or incorrect individual contributions, thus, influence the shape of the $\gamma$SF.

\begin{figure}[t]
\centering
\includegraphics[width=1\linewidth]{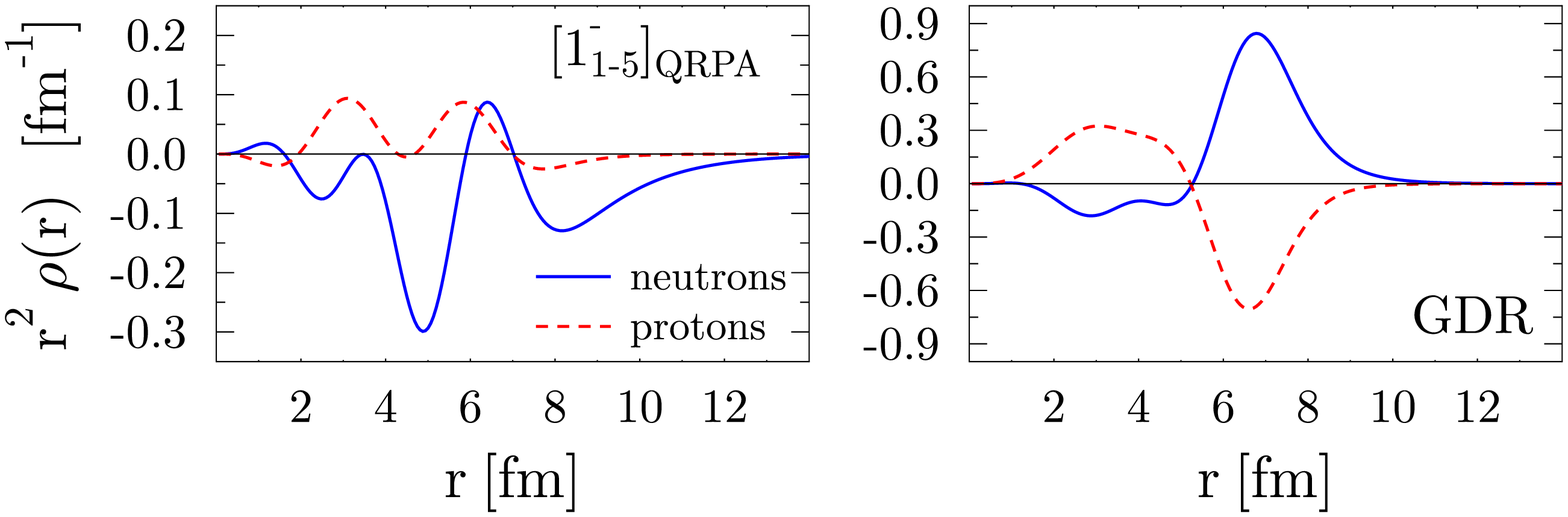}
\caption{\label{fig:trd}{(color online) Summed transition densities for the first five $1^-_{\mathrm{QRPA}}$ states, which contain the (3p$_{1/2})^{-1}$(4s$_{1/2})^{+1}$ and (3p$_{1/2})^{-1}$(3d$_{3/2})^{+1}$ components. All five states are dominant neutron 1p--1h states (compare \cite{suppl}). For comparison, the summed transition densities for the GDR are shown.}} 
 \end{figure}

To study the neutron-skin structure of the $1^-$ states with (3p$_{1/2})^{-1}$(4s$_{1/2})^{+1}$, (3p$_{1/2})^{-1}$(3d$_{3/2})^{+1}$ and other neutron 1p--1h components, EDF+QRPA and EDF+QPM calculations were performed (see \cite{Tso16a} for a review). In contrast to \cite{Rye02a,Tam11a,Pol12a}, single-particle energies were neither determined from nor adjusted to data. Instead, they were directly obtained at the mean-field level from the EDF\,\cite{Tso16a}. Fig.\,\ref{fig:exp_sm_comp}\,{\bf (j)} presents the QPM+DWBA predictions for $\sigma_{(d,p)}$. Results obtained at the QRPA level and further details are given in the supplement \cite{suppl}. The QPM also predicts a dominant (3p$_{1/2})^{-1}$(4s$_{1/2})^{+1}$ [$S_{5.32\,\mathrm{MeV}}=0.92$] fragment but expects, different from the LSSM and in agreement with experiment, the (3p$_{1/2})^{-1}$(3d$_{3/2})^{+1}$ strength to be mainly concentrated in one state [$S_{6.12\,\mathrm{MeV}}=0.68$]. The QPM predicts $\sum \sigma_{(d,p)_{QPM}} = 1676$\,$\mu$b below $S_n$. However, also the QPM does not fragment the $l=0$ and $l=2$ strength sufficiently to describe the strength above $S_n$. While the QPM reproduces the experimental $B(E1)$ strength distribution around and above $S_n$, i.e. where the states' structure becomes more complex [Figs.\,\ref{fig:exp_sm_comp}\,{\bf (k)}, {\bf (l)}], it does not generate sufficiently enhanced strength at lower energies [Fig.\,\ref{fig:exp_sm_comp}\,{\bf (m)}]. Due to the doubly magic nature of $^{208}$Pb, the 1p--1h structure of the QRPA phonons dominates the configuration mixing and polarization contributions (compare \cite{suppl}). In order to improve the comparison with experiment, dynamic effects beyond the static mean field would need to be implemented. As 2p--2h contributions in the LSSM, multiphonon contributions are small below 8\,MeV. Interestingly, the $1^-_{2,QPM}$ state seems to correspond to the LSSM and experimental $1^-_1$ state. It has a significant 2-phonon admixture [compare Fig.\,\ref{fig:exp_sm_comp}\,{\bf (l)}]. Fig.\,\ref{fig:trd} presents the summed transition densities for the first five QRPA $1^-$ phonons, which contain the (3p$_{1/2})^{-1}$(4s$_{1/2})^{+1}$ and (3p$_{1/2})^{-1}$(3d$_{3/2})^{+1}$ spectroscopic strengths. The summed transition densities show features which are compatible with the oscillation of the neutron skin\,\cite{Tso08a} and clearly different from the GDR.

In summary, we performed the first extensive study of the single-particle structure of the PDR in $^{208}$Pb based on experimental data. The LSSM and EDF+QPM calculations were able to account for the main features of the $(d,p)$ data. However, both models do not generate enough spectroscopic strengths above $S_n$. Such shortcomings could have significant influence on $(n,\gamma)$ rates when determined via surrogate methods using theoretical nuclear-structure input\,\cite{Lar19a,Rat19a,Esch12a}. The extended comparison, including the $(p,p')_{\mathrm{IAR}}$ data, suggests that the LSSM wave functions might be slightly too complex. At lower energies, the QRPA $1^-$ phonons might not be sufficiently admixed to several QPM $1^-$ states. Most $1^-$ states can, however, not be considered as simple neutron 1p--1h states as many neutron 1p--1h excitations contribute to their respective wave function. We pointed out the big cancellation effects between individual $E1$ matrix elements, which are observed for the PDR in contrast to the GDR. Yet, if the PDR states' structure only contains one or a few 1p--1h components, individual matrix elements could also not add coherently and generate enhanced $B(E1)$ strengths. Enhanced strength is observed below $S_n$. In contrast to previous claims, the transition densities for the low-lying $1^-$ states with dominant neutron 1p--1h character clearly resemble features of a dipole-type neutron-skin oscillation. The present work proves the value of complementary, experimental data on and the theoretical analysis of the PDR's 1p--1h structure to access the microscopic wave functions. Similar studies will help to further understand the microscopic origin of the low-lying isovector and isoscalar $B(E1)$ strengths. High-resolution, one-nucleon transfer experiments on stable nuclides in different mass regions, where the change of the underlying single-particle structure can be tracked as both proton and neutron number change, are planned. Further developments at next-generation exotic beam facilities might allow access to the PDR with one-nucleon transfer in inverse kinematics using, {\it e.g.}, solenoidal spectrometers\,\cite{hel20a,sol20a,sol18a,iss20a,Tan20a}.

\begin{acknowledgments}
B.A.B. acknowledges support by the National Science Foundation under Grant No. PHY-1811855. M.Sch. acknowledges financial support by the UK-STFC. N.T. was supported by Extreme Light Infrastructure Nuclear Physics (ELI-NP) Phase II, a project co-financed by the Romanian Government and
the European Union through the European Regional Development Fund ``the
Competitiveness Operational Programme'' (1/07.07.2016, COP, ID 1334). A.Z. acknowledges support by the Deutsche Forschungsgemeinschaft under grant ZI 510/9-1. M.Sp. wants to thank K.\,Kemper and J.\,Piekarewicz for helpful and inspiring discussions.
\end{acknowledgments}

\bibliography{208Pb}

\providecommand{\noopsort}[1]{}\providecommand{\singleletter}[1]{#1}%
\begin{thebibliography}{105}%
\makeatletter
\providecommand \@ifxundefined [1]{%
 \@ifx{#1\undefined}
}%
\providecommand \@ifnum [1]{%
 \ifnum #1\expandafter \@firstoftwo
 \else \expandafter \@secondoftwo
 \fi
}%
\providecommand \@ifx [1]{%
 \ifx #1\expandafter \@firstoftwo
 \else \expandafter \@secondoftwo
 \fi
}%
\providecommand \natexlab [1]{#1}%
\providecommand \enquote  [1]{``#1''}%
\providecommand \bibnamefont  [1]{#1}%
\providecommand \bibfnamefont [1]{#1}%
\providecommand \citenamefont [1]{#1}%
\providecommand \href@noop [0]{\@secondoftwo}%
\providecommand \href [0]{\begingroup \@sanitize@url \@href}%
\providecommand \@href[1]{\@@startlink{#1}\@@href}%
\providecommand \@@href[1]{\endgroup#1\@@endlink}%
\providecommand \@sanitize@url [0]{\catcode `\\12\catcode `\$12\catcode
  `\&12\catcode `\#12\catcode `\^12\catcode `\_12\catcode `\%12\relax}%
\providecommand \@@startlink[1]{}%
\providecommand \@@endlink[0]{}%
\providecommand \url  [0]{\begingroup\@sanitize@url \@url }%
\providecommand \@url [1]{\endgroup\@href {#1}{\urlprefix }}%
\providecommand \urlprefix  [0]{URL }%
\providecommand \Eprint [0]{\href }%
\providecommand \doibase [0]{http://dx.doi.org/}%
\providecommand \selectlanguage [0]{\@gobble}%
\providecommand \bibinfo  [0]{\@secondoftwo}%
\providecommand \bibfield  [0]{\@secondoftwo}%
\providecommand \translation [1]{[#1]}%
\providecommand \BibitemOpen [0]{}%
\providecommand \bibitemStop [0]{}%
\providecommand \bibitemNoStop [0]{.\EOS\space}%
\providecommand \EOS [0]{\spacefactor3000\relax}%
\providecommand \BibitemShut  [1]{\csname bibitem#1\endcsname}%
\let\auto@bib@innerbib\@empty
\bibitem [{\citenamefont {Thiel}\ \emph {et~al.}(2019)\citenamefont {Thiel},
  \citenamefont {Sfienti}, \citenamefont {Piekarewicz}, \citenamefont
  {Horowitz},\ and\ \citenamefont {Vanderhaeghen}}]{Thi19a}%
  \BibitemOpen
  \bibfield  {author} {\bibinfo {author} {\bibfnamefont {M.}~\bibnamefont
  {Thiel}}, \bibinfo {author} {\bibfnamefont {C.}~\bibnamefont {Sfienti}},
  \bibinfo {author} {\bibfnamefont {J.}~\bibnamefont {Piekarewicz}}, \bibinfo
  {author} {\bibfnamefont {C.~J.}\ \bibnamefont {Horowitz}}, \ and\ \bibinfo
  {author} {\bibfnamefont {M.}~\bibnamefont {Vanderhaeghen}},\ }\href {\doibase
  10.1088/1361-6471/ab2c6d} {\bibfield  {journal} {\bibinfo  {journal} {Journal
  of Physics G: Nuclear and Particle Physics}\ }\textbf {\bibinfo {volume}
  {46}},\ \bibinfo {pages} {093003} (\bibinfo {year} {2019})}\BibitemShut
  {NoStop}%
\bibitem [{\citenamefont {Horowitz}\ and\ \citenamefont
  {Piekarewicz}(2001{\natexlab{a}})}]{Hor01a}%
  \BibitemOpen
  \bibfield  {author} {\bibinfo {author} {\bibfnamefont {C.~J.}\ \bibnamefont
  {Horowitz}}\ and\ \bibinfo {author} {\bibfnamefont {J.}~\bibnamefont
  {Piekarewicz}},\ }\href {\doibase 10.1103/PhysRevLett.86.5647} {\bibfield
  {journal} {\bibinfo  {journal} {Phys. Rev. Lett.}\ }\textbf {\bibinfo
  {volume} {86}},\ \bibinfo {pages} {5647} (\bibinfo {year}
  {2001}{\natexlab{a}})}\BibitemShut {NoStop}%
\bibitem [{\citenamefont {Horowitz}\ and\ \citenamefont
  {Piekarewicz}(2001{\natexlab{b}})}]{Hor01b}%
  \BibitemOpen
  \bibfield  {author} {\bibinfo {author} {\bibfnamefont {C.~J.}\ \bibnamefont
  {Horowitz}}\ and\ \bibinfo {author} {\bibfnamefont {J.}~\bibnamefont
  {Piekarewicz}},\ }\href {\doibase 10.1103/PhysRevC.64.062802} {\bibfield
  {journal} {\bibinfo  {journal} {Phys. Rev. C}\ }\textbf {\bibinfo {volume}
  {64}},\ \bibinfo {pages} {062802} (\bibinfo {year}
  {2001}{\natexlab{b}})}\BibitemShut {NoStop}%
\bibitem [{\citenamefont {Fattoyev}\ and\ \citenamefont
  {Piekarewicz}(2012)}]{Fat12a}%
  \BibitemOpen
  \bibfield  {author} {\bibinfo {author} {\bibfnamefont {F.~J.}\ \bibnamefont
  {Fattoyev}}\ and\ \bibinfo {author} {\bibfnamefont {J.}~\bibnamefont
  {Piekarewicz}},\ }\href {\doibase 10.1103/PhysRevC.86.015802} {\bibfield
  {journal} {\bibinfo  {journal} {Phys. Rev. C}\ }\textbf {\bibinfo {volume}
  {86}},\ \bibinfo {pages} {015802} (\bibinfo {year} {2012})}\BibitemShut
  {NoStop}%
\bibitem [{\citenamefont {Fattoyev}\ and\ \citenamefont
  {Piekarewicz}(2013)}]{Fat13a}%
  \BibitemOpen
  \bibfield  {author} {\bibinfo {author} {\bibfnamefont {F.~J.}\ \bibnamefont
  {Fattoyev}}\ and\ \bibinfo {author} {\bibfnamefont {J.}~\bibnamefont
  {Piekarewicz}},\ }\href {\doibase 10.1103/PhysRevLett.111.162501} {\bibfield
  {journal} {\bibinfo  {journal} {Phys. Rev. Lett.}\ }\textbf {\bibinfo
  {volume} {111}},\ \bibinfo {pages} {162501} (\bibinfo {year}
  {2013})}\BibitemShut {NoStop}%
\bibitem [{\citenamefont {Tamii}\ \emph {et~al.}(2011)\citenamefont {Tamii},
  \citenamefont {Poltoratska}, \citenamefont {von Neumann-Cosel}, \citenamefont
  {Fujita}, \citenamefont {Adachi}, \citenamefont {Bertulani}, \citenamefont
  {Carter}, \citenamefont {Dozono}, \citenamefont {Fujita}, \citenamefont
  {Fujita}, \citenamefont {Hatanaka}, \citenamefont {Ishikawa}, \citenamefont
  {Itoh}, \citenamefont {Kawabata}, \citenamefont {Kalmykov}, \citenamefont
  {Krumbholz}, \citenamefont {Litvinova}, \citenamefont {Matsubara},
  \citenamefont {Nakanishi}, \citenamefont {Neveling}, \citenamefont {Okamura},
  \citenamefont {Ong}, \citenamefont {\"Ozel-Tashenov}, \citenamefont
  {Ponomarev}, \citenamefont {Richter}, \citenamefont {Rubio}, \citenamefont
  {Sakaguchi}, \citenamefont {Sakemi}, \citenamefont {Sasamoto}, \citenamefont
  {Shimbara}, \citenamefont {Shimizu}, \citenamefont {Smit}, \citenamefont
  {Suzuki}, \citenamefont {Tameshige}, \citenamefont {Wambach}, \citenamefont
  {Yamada}, \citenamefont {Yosoi},\ and\ \citenamefont {Zenihiro}}]{Tam11a}%
  \BibitemOpen
  \bibfield  {author} {\bibinfo {author} {\bibfnamefont {A.}~\bibnamefont
  {Tamii}}, \bibinfo {author} {\bibfnamefont {I.}~\bibnamefont {Poltoratska}},
  \bibinfo {author} {\bibfnamefont {P.}~\bibnamefont {von Neumann-Cosel}},
  \bibinfo {author} {\bibfnamefont {Y.}~\bibnamefont {Fujita}}, \bibinfo
  {author} {\bibfnamefont {T.}~\bibnamefont {Adachi}}, \bibinfo {author}
  {\bibfnamefont {C.~A.}\ \bibnamefont {Bertulani}}, \bibinfo {author}
  {\bibfnamefont {J.}~\bibnamefont {Carter}}, \bibinfo {author} {\bibfnamefont
  {M.}~\bibnamefont {Dozono}}, \bibinfo {author} {\bibfnamefont
  {H.}~\bibnamefont {Fujita}}, \bibinfo {author} {\bibfnamefont
  {K.}~\bibnamefont {Fujita}}, \bibinfo {author} {\bibfnamefont
  {K.}~\bibnamefont {Hatanaka}}, \bibinfo {author} {\bibfnamefont
  {D.}~\bibnamefont {Ishikawa}}, \bibinfo {author} {\bibfnamefont
  {M.}~\bibnamefont {Itoh}}, \bibinfo {author} {\bibfnamefont {T.}~\bibnamefont
  {Kawabata}}, \bibinfo {author} {\bibfnamefont {Y.}~\bibnamefont {Kalmykov}},
  \bibinfo {author} {\bibfnamefont {A.~M.}\ \bibnamefont {Krumbholz}}, \bibinfo
  {author} {\bibfnamefont {E.}~\bibnamefont {Litvinova}}, \bibinfo {author}
  {\bibfnamefont {H.}~\bibnamefont {Matsubara}}, \bibinfo {author}
  {\bibfnamefont {K.}~\bibnamefont {Nakanishi}}, \bibinfo {author}
  {\bibfnamefont {R.}~\bibnamefont {Neveling}}, \bibinfo {author}
  {\bibfnamefont {H.}~\bibnamefont {Okamura}}, \bibinfo {author} {\bibfnamefont
  {H.~J.}\ \bibnamefont {Ong}}, \bibinfo {author} {\bibfnamefont
  {B.}~\bibnamefont {\"Ozel-Tashenov}}, \bibinfo {author} {\bibfnamefont
  {V.~Y.}\ \bibnamefont {Ponomarev}}, \bibinfo {author} {\bibfnamefont
  {A.}~\bibnamefont {Richter}}, \bibinfo {author} {\bibfnamefont
  {B.}~\bibnamefont {Rubio}}, \bibinfo {author} {\bibfnamefont
  {H.}~\bibnamefont {Sakaguchi}}, \bibinfo {author} {\bibfnamefont
  {Y.}~\bibnamefont {Sakemi}}, \bibinfo {author} {\bibfnamefont
  {Y.}~\bibnamefont {Sasamoto}}, \bibinfo {author} {\bibfnamefont
  {Y.}~\bibnamefont {Shimbara}}, \bibinfo {author} {\bibfnamefont
  {Y.}~\bibnamefont {Shimizu}}, \bibinfo {author} {\bibfnamefont {F.~D.}\
  \bibnamefont {Smit}}, \bibinfo {author} {\bibfnamefont {T.}~\bibnamefont
  {Suzuki}}, \bibinfo {author} {\bibfnamefont {Y.}~\bibnamefont {Tameshige}},
  \bibinfo {author} {\bibfnamefont {J.}~\bibnamefont {Wambach}}, \bibinfo
  {author} {\bibfnamefont {R.}~\bibnamefont {Yamada}}, \bibinfo {author}
  {\bibfnamefont {M.}~\bibnamefont {Yosoi}}, \ and\ \bibinfo {author}
  {\bibfnamefont {J.}~\bibnamefont {Zenihiro}},\ }\href {\doibase
  10.1103/PhysRevLett.107.062502} {\bibfield  {journal} {\bibinfo  {journal}
  {Phys. Rev. Lett.}\ }\textbf {\bibinfo {volume} {107}},\ \bibinfo {pages}
  {062502} (\bibinfo {year} {2011})}\BibitemShut {NoStop}%
\bibitem [{\citenamefont {Abrahamyan}\ \emph {et~al.}(2012)\citenamefont
  {Abrahamyan}, \citenamefont {Ahmed}, \citenamefont {Albataineh},
  \citenamefont {Aniol}, \citenamefont {Armstrong}, \citenamefont {Armstrong},
  \citenamefont {Averett}, \citenamefont {Babineau}, \citenamefont {Barbieri},
  \citenamefont {Bellini}, \citenamefont {Beminiwattha}, \citenamefont
  {Benesch}, \citenamefont {Benmokhtar}, \citenamefont {Bielarski},
  \citenamefont {Boeglin}, \citenamefont {Camsonne}, \citenamefont {Canan},
  \citenamefont {Carter}, \citenamefont {Cates}, \citenamefont {Chen},
  \citenamefont {Chen}, \citenamefont {Hen}, \citenamefont {Cusanno},
  \citenamefont {Dalton}, \citenamefont {De~Leo}, \citenamefont {de~Jager},
  \citenamefont {Deconinck}, \citenamefont {Decowski}, \citenamefont {Deng},
  \citenamefont {Deur}, \citenamefont {Dutta}, \citenamefont {Etile},
  \citenamefont {Flay}, \citenamefont {Franklin}, \citenamefont {Friend},
  \citenamefont {Frullani}, \citenamefont {Fuchey}, \citenamefont {Garibaldi},
  \citenamefont {Gasser}, \citenamefont {Gilman}, \citenamefont {Giusa},
  \citenamefont {Glamazdin}, \citenamefont {Gomez}, \citenamefont {Grames},
  \citenamefont {Gu}, \citenamefont {Hansen}, \citenamefont {Hansknecht},
  \citenamefont {Higinbotham}, \citenamefont {Holmes}, \citenamefont
  {Holmstrom}, \citenamefont {Horowitz}, \citenamefont {Hoskins}, \citenamefont
  {Huang}, \citenamefont {Hyde}, \citenamefont {Itard}, \citenamefont {Jen},
  \citenamefont {Jensen}, \citenamefont {Jin}, \citenamefont {Johnston},
  \citenamefont {Kelleher}, \citenamefont {Kliakhandler}, \citenamefont {King},
  \citenamefont {Kowalski}, \citenamefont {Kumar}, \citenamefont {Leacock},
  \citenamefont {Leckey}, \citenamefont {Lee}, \citenamefont {LeRose},
  \citenamefont {Lindgren}, \citenamefont {Liyanage}, \citenamefont {Lubinsky},
  \citenamefont {Mammei}, \citenamefont {Mammoliti}, \citenamefont
  {Margaziotis}, \citenamefont {Markowitz}, \citenamefont {McCreary},
  \citenamefont {McNulty}, \citenamefont {Mercado}, \citenamefont {Meziani},
  \citenamefont {Michaels}, \citenamefont {Mihovilovic}, \citenamefont
  {Muangma}, \citenamefont {Mu\~noz Camacho}, \citenamefont {Nanda},
  \citenamefont {Nelyubin}, \citenamefont {Nuruzzaman}, \citenamefont {Oh},
  \citenamefont {Palmer}, \citenamefont {Parno}, \citenamefont {Paschke},
  \citenamefont {Phillips}, \citenamefont {Poelker}, \citenamefont
  {Pomatsalyuk}, \citenamefont {Posik}, \citenamefont {Puckett}, \citenamefont
  {Quinn}, \citenamefont {Rakhman}, \citenamefont {Reimer}, \citenamefont
  {Riordan}, \citenamefont {Rogan}, \citenamefont {Ron}, \citenamefont {Russo},
  \citenamefont {Saenboonruang}, \citenamefont {Saha}, \citenamefont
  {Sawatzky}, \citenamefont {Shahinyan}, \citenamefont {Silwal}, \citenamefont
  {Sirca}, \citenamefont {Slifer}, \citenamefont {Solvignon}, \citenamefont
  {Souder}, \citenamefont {Sperduto}, \citenamefont {Subedi}, \citenamefont
  {Suleiman}, \citenamefont {Sulkosky}, \citenamefont {Sutera}, \citenamefont
  {Tobias}, \citenamefont {Troth}, \citenamefont {Urciuoli}, \citenamefont
  {Waidyawansa}, \citenamefont {Wang}, \citenamefont {Wexler}, \citenamefont
  {Wilson}, \citenamefont {Wojtsekhowski}, \citenamefont {Yan}, \citenamefont
  {Yao}, \citenamefont {Ye}, \citenamefont {Ye}, \citenamefont {Yim},
  \citenamefont {Zana}, \citenamefont {Zhan}, \citenamefont {Zhang},
  \citenamefont {Zhang}, \citenamefont {Zheng},\ and\ \citenamefont
  {Zhu}}]{Abr12a}%
  \BibitemOpen
  \bibfield  {author} {\bibinfo {author} {\bibfnamefont {S.}~\bibnamefont
  {Abrahamyan}}, \bibinfo {author} {\bibfnamefont {Z.}~\bibnamefont {Ahmed}},
  \bibinfo {author} {\bibfnamefont {H.}~\bibnamefont {Albataineh}}, \bibinfo
  {author} {\bibfnamefont {K.}~\bibnamefont {Aniol}}, \bibinfo {author}
  {\bibfnamefont {D.~S.}\ \bibnamefont {Armstrong}}, \bibinfo {author}
  {\bibfnamefont {W.}~\bibnamefont {Armstrong}}, \bibinfo {author}
  {\bibfnamefont {T.}~\bibnamefont {Averett}}, \bibinfo {author} {\bibfnamefont
  {B.}~\bibnamefont {Babineau}}, \bibinfo {author} {\bibfnamefont
  {A.}~\bibnamefont {Barbieri}}, \bibinfo {author} {\bibfnamefont
  {V.}~\bibnamefont {Bellini}}, \bibinfo {author} {\bibfnamefont
  {R.}~\bibnamefont {Beminiwattha}}, \bibinfo {author} {\bibfnamefont
  {J.}~\bibnamefont {Benesch}}, \bibinfo {author} {\bibfnamefont
  {F.}~\bibnamefont {Benmokhtar}}, \bibinfo {author} {\bibfnamefont
  {T.}~\bibnamefont {Bielarski}}, \bibinfo {author} {\bibfnamefont
  {W.}~\bibnamefont {Boeglin}}, \bibinfo {author} {\bibfnamefont
  {A.}~\bibnamefont {Camsonne}}, \bibinfo {author} {\bibfnamefont
  {M.}~\bibnamefont {Canan}}, \bibinfo {author} {\bibfnamefont
  {P.}~\bibnamefont {Carter}}, \bibinfo {author} {\bibfnamefont {G.~D.}\
  \bibnamefont {Cates}}, \bibinfo {author} {\bibfnamefont {C.}~\bibnamefont
  {Chen}}, \bibinfo {author} {\bibfnamefont {J.-P.}\ \bibnamefont {Chen}},
  \bibinfo {author} {\bibfnamefont {O.}~\bibnamefont {Hen}}, \bibinfo {author}
  {\bibfnamefont {F.}~\bibnamefont {Cusanno}}, \bibinfo {author} {\bibfnamefont
  {M.~M.}\ \bibnamefont {Dalton}}, \bibinfo {author} {\bibfnamefont
  {R.}~\bibnamefont {De~Leo}}, \bibinfo {author} {\bibfnamefont
  {K.}~\bibnamefont {de~Jager}}, \bibinfo {author} {\bibfnamefont
  {W.}~\bibnamefont {Deconinck}}, \bibinfo {author} {\bibfnamefont
  {P.}~\bibnamefont {Decowski}}, \bibinfo {author} {\bibfnamefont
  {X.}~\bibnamefont {Deng}}, \bibinfo {author} {\bibfnamefont {A.}~\bibnamefont
  {Deur}}, \bibinfo {author} {\bibfnamefont {D.}~\bibnamefont {Dutta}},
  \bibinfo {author} {\bibfnamefont {A.}~\bibnamefont {Etile}}, \bibinfo
  {author} {\bibfnamefont {D.}~\bibnamefont {Flay}}, \bibinfo {author}
  {\bibfnamefont {G.~B.}\ \bibnamefont {Franklin}}, \bibinfo {author}
  {\bibfnamefont {M.}~\bibnamefont {Friend}}, \bibinfo {author} {\bibfnamefont
  {S.}~\bibnamefont {Frullani}}, \bibinfo {author} {\bibfnamefont
  {E.}~\bibnamefont {Fuchey}}, \bibinfo {author} {\bibfnamefont
  {F.}~\bibnamefont {Garibaldi}}, \bibinfo {author} {\bibfnamefont
  {E.}~\bibnamefont {Gasser}}, \bibinfo {author} {\bibfnamefont
  {R.}~\bibnamefont {Gilman}}, \bibinfo {author} {\bibfnamefont
  {A.}~\bibnamefont {Giusa}}, \bibinfo {author} {\bibfnamefont
  {A.}~\bibnamefont {Glamazdin}}, \bibinfo {author} {\bibfnamefont
  {J.}~\bibnamefont {Gomez}}, \bibinfo {author} {\bibfnamefont
  {J.}~\bibnamefont {Grames}}, \bibinfo {author} {\bibfnamefont
  {C.}~\bibnamefont {Gu}}, \bibinfo {author} {\bibfnamefont {O.}~\bibnamefont
  {Hansen}}, \bibinfo {author} {\bibfnamefont {J.}~\bibnamefont {Hansknecht}},
  \bibinfo {author} {\bibfnamefont {D.~W.}\ \bibnamefont {Higinbotham}},
  \bibinfo {author} {\bibfnamefont {R.~S.}\ \bibnamefont {Holmes}}, \bibinfo
  {author} {\bibfnamefont {T.}~\bibnamefont {Holmstrom}}, \bibinfo {author}
  {\bibfnamefont {C.~J.}\ \bibnamefont {Horowitz}}, \bibinfo {author}
  {\bibfnamefont {J.}~\bibnamefont {Hoskins}}, \bibinfo {author} {\bibfnamefont
  {J.}~\bibnamefont {Huang}}, \bibinfo {author} {\bibfnamefont {C.~E.}\
  \bibnamefont {Hyde}}, \bibinfo {author} {\bibfnamefont {F.}~\bibnamefont
  {Itard}}, \bibinfo {author} {\bibfnamefont {C.-M.}\ \bibnamefont {Jen}},
  \bibinfo {author} {\bibfnamefont {E.}~\bibnamefont {Jensen}}, \bibinfo
  {author} {\bibfnamefont {G.}~\bibnamefont {Jin}}, \bibinfo {author}
  {\bibfnamefont {S.}~\bibnamefont {Johnston}}, \bibinfo {author}
  {\bibfnamefont {A.}~\bibnamefont {Kelleher}}, \bibinfo {author}
  {\bibfnamefont {K.}~\bibnamefont {Kliakhandler}}, \bibinfo {author}
  {\bibfnamefont {P.~M.}\ \bibnamefont {King}}, \bibinfo {author}
  {\bibfnamefont {S.}~\bibnamefont {Kowalski}}, \bibinfo {author}
  {\bibfnamefont {K.~S.}\ \bibnamefont {Kumar}}, \bibinfo {author}
  {\bibfnamefont {J.}~\bibnamefont {Leacock}}, \bibinfo {author} {\bibfnamefont
  {J.}~\bibnamefont {Leckey}}, \bibinfo {author} {\bibfnamefont {J.~H.}\
  \bibnamefont {Lee}}, \bibinfo {author} {\bibfnamefont {J.~J.}\ \bibnamefont
  {LeRose}}, \bibinfo {author} {\bibfnamefont {R.}~\bibnamefont {Lindgren}},
  \bibinfo {author} {\bibfnamefont {N.}~\bibnamefont {Liyanage}}, \bibinfo
  {author} {\bibfnamefont {N.}~\bibnamefont {Lubinsky}}, \bibinfo {author}
  {\bibfnamefont {J.}~\bibnamefont {Mammei}}, \bibinfo {author} {\bibfnamefont
  {F.}~\bibnamefont {Mammoliti}}, \bibinfo {author} {\bibfnamefont {D.~J.}\
  \bibnamefont {Margaziotis}}, \bibinfo {author} {\bibfnamefont
  {P.}~\bibnamefont {Markowitz}}, \bibinfo {author} {\bibfnamefont
  {A.}~\bibnamefont {McCreary}}, \bibinfo {author} {\bibfnamefont
  {D.}~\bibnamefont {McNulty}}, \bibinfo {author} {\bibfnamefont
  {L.}~\bibnamefont {Mercado}}, \bibinfo {author} {\bibfnamefont {Z.-E.}\
  \bibnamefont {Meziani}}, \bibinfo {author} {\bibfnamefont {R.~W.}\
  \bibnamefont {Michaels}}, \bibinfo {author} {\bibfnamefont {M.}~\bibnamefont
  {Mihovilovic}}, \bibinfo {author} {\bibfnamefont {N.}~\bibnamefont
  {Muangma}}, \bibinfo {author} {\bibfnamefont {C.}~\bibnamefont {Mu\~noz
  Camacho}}, \bibinfo {author} {\bibfnamefont {S.}~\bibnamefont {Nanda}},
  \bibinfo {author} {\bibfnamefont {V.}~\bibnamefont {Nelyubin}}, \bibinfo
  {author} {\bibfnamefont {N.}~\bibnamefont {Nuruzzaman}}, \bibinfo {author}
  {\bibfnamefont {Y.}~\bibnamefont {Oh}}, \bibinfo {author} {\bibfnamefont
  {A.}~\bibnamefont {Palmer}}, \bibinfo {author} {\bibfnamefont
  {D.}~\bibnamefont {Parno}}, \bibinfo {author} {\bibfnamefont {K.~D.}\
  \bibnamefont {Paschke}}, \bibinfo {author} {\bibfnamefont {S.~K.}\
  \bibnamefont {Phillips}}, \bibinfo {author} {\bibfnamefont {B.}~\bibnamefont
  {Poelker}}, \bibinfo {author} {\bibfnamefont {R.}~\bibnamefont
  {Pomatsalyuk}}, \bibinfo {author} {\bibfnamefont {M.}~\bibnamefont {Posik}},
  \bibinfo {author} {\bibfnamefont {A.~J.~R.}\ \bibnamefont {Puckett}},
  \bibinfo {author} {\bibfnamefont {B.}~\bibnamefont {Quinn}}, \bibinfo
  {author} {\bibfnamefont {A.}~\bibnamefont {Rakhman}}, \bibinfo {author}
  {\bibfnamefont {P.~E.}\ \bibnamefont {Reimer}}, \bibinfo {author}
  {\bibfnamefont {S.}~\bibnamefont {Riordan}}, \bibinfo {author} {\bibfnamefont
  {P.}~\bibnamefont {Rogan}}, \bibinfo {author} {\bibfnamefont
  {G.}~\bibnamefont {Ron}}, \bibinfo {author} {\bibfnamefont {G.}~\bibnamefont
  {Russo}}, \bibinfo {author} {\bibfnamefont {K.}~\bibnamefont
  {Saenboonruang}}, \bibinfo {author} {\bibfnamefont {A.}~\bibnamefont {Saha}},
  \bibinfo {author} {\bibfnamefont {B.}~\bibnamefont {Sawatzky}}, \bibinfo
  {author} {\bibfnamefont {A.}~\bibnamefont {Shahinyan}}, \bibinfo {author}
  {\bibfnamefont {R.}~\bibnamefont {Silwal}}, \bibinfo {author} {\bibfnamefont
  {S.}~\bibnamefont {Sirca}}, \bibinfo {author} {\bibfnamefont
  {K.}~\bibnamefont {Slifer}}, \bibinfo {author} {\bibfnamefont
  {P.}~\bibnamefont {Solvignon}}, \bibinfo {author} {\bibfnamefont {P.~A.}\
  \bibnamefont {Souder}}, \bibinfo {author} {\bibfnamefont {M.~L.}\
  \bibnamefont {Sperduto}}, \bibinfo {author} {\bibfnamefont {R.}~\bibnamefont
  {Subedi}}, \bibinfo {author} {\bibfnamefont {R.}~\bibnamefont {Suleiman}},
  \bibinfo {author} {\bibfnamefont {V.}~\bibnamefont {Sulkosky}}, \bibinfo
  {author} {\bibfnamefont {C.~M.}\ \bibnamefont {Sutera}}, \bibinfo {author}
  {\bibfnamefont {W.~A.}\ \bibnamefont {Tobias}}, \bibinfo {author}
  {\bibfnamefont {W.}~\bibnamefont {Troth}}, \bibinfo {author} {\bibfnamefont
  {G.~M.}\ \bibnamefont {Urciuoli}}, \bibinfo {author} {\bibfnamefont
  {B.}~\bibnamefont {Waidyawansa}}, \bibinfo {author} {\bibfnamefont
  {D.}~\bibnamefont {Wang}}, \bibinfo {author} {\bibfnamefont {J.}~\bibnamefont
  {Wexler}}, \bibinfo {author} {\bibfnamefont {R.}~\bibnamefont {Wilson}},
  \bibinfo {author} {\bibfnamefont {B.}~\bibnamefont {Wojtsekhowski}}, \bibinfo
  {author} {\bibfnamefont {X.}~\bibnamefont {Yan}}, \bibinfo {author}
  {\bibfnamefont {H.}~\bibnamefont {Yao}}, \bibinfo {author} {\bibfnamefont
  {Y.}~\bibnamefont {Ye}}, \bibinfo {author} {\bibfnamefont {Z.}~\bibnamefont
  {Ye}}, \bibinfo {author} {\bibfnamefont {V.}~\bibnamefont {Yim}}, \bibinfo
  {author} {\bibfnamefont {L.}~\bibnamefont {Zana}}, \bibinfo {author}
  {\bibfnamefont {X.}~\bibnamefont {Zhan}}, \bibinfo {author} {\bibfnamefont
  {J.}~\bibnamefont {Zhang}}, \bibinfo {author} {\bibfnamefont
  {Y.}~\bibnamefont {Zhang}}, \bibinfo {author} {\bibfnamefont
  {X.}~\bibnamefont {Zheng}}, \ and\ \bibinfo {author} {\bibfnamefont
  {P.}~\bibnamefont {Zhu}} (\bibinfo {collaboration} {PREX Collaboration}),\
  }\href {\doibase 10.1103/PhysRevLett.108.112502} {\bibfield  {journal}
  {\bibinfo  {journal} {Phys. Rev. Lett.}\ }\textbf {\bibinfo {volume} {108}},\
  \bibinfo {pages} {112502} (\bibinfo {year} {2012})}\BibitemShut {NoStop}%
\bibitem [{\citenamefont {Tamii}\ \emph {et~al.}(2013)\citenamefont {Tamii},
  \citenamefont {von Neumann-Cosel},\ and\ \citenamefont
  {Poltoratska}}]{Tam13a}%
  \BibitemOpen
  \bibfield  {author} {\bibinfo {author} {\bibfnamefont {A.}~\bibnamefont
  {Tamii}}, \bibinfo {author} {\bibfnamefont {P.}~\bibnamefont {von
  Neumann-Cosel}}, \ and\ \bibinfo {author} {\bibfnamefont {I.}~\bibnamefont
  {Poltoratska}},\ }\href {\doibase 10.1140/epja/i2014-14028-7} {\bibfield
  {journal} {\bibinfo  {journal} {Eur. Phys. J. A}\ }\textbf {\bibinfo {volume}
  {50}},\ \bibinfo {pages} {28} (\bibinfo {year} {2013})}\BibitemShut {NoStop}%
\bibitem [{\citenamefont {Tarbert}\ \emph {et~al.}(2014)\citenamefont
  {Tarbert}, \citenamefont {Watts}, \citenamefont {Glazier}, \citenamefont
  {Aguar}, \citenamefont {Ahrens}, \citenamefont {Annand}, \citenamefont
  {Arends}, \citenamefont {Beck}, \citenamefont {Bekrenev}, \citenamefont
  {Boillat}, \citenamefont {Braghieri}, \citenamefont {Branford}, \citenamefont
  {Briscoe}, \citenamefont {Brudvik}, \citenamefont {Cherepnya}, \citenamefont
  {Codling}, \citenamefont {Downie}, \citenamefont {Foehl}, \citenamefont
  {Grabmayr}, \citenamefont {Gregor}, \citenamefont {Heid}, \citenamefont
  {Hornidge}, \citenamefont {Jahn}, \citenamefont {Kashevarov}, \citenamefont
  {Knezevic}, \citenamefont {Kondratiev}, \citenamefont {Korolija},
  \citenamefont {Kotulla}, \citenamefont {Krambrich}, \citenamefont {Krusche},
  \citenamefont {Lang}, \citenamefont {Lisin}, \citenamefont {Livingston},
  \citenamefont {Lugert}, \citenamefont {MacGregor}, \citenamefont {Manley},
  \citenamefont {Martinez}, \citenamefont {McGeorge}, \citenamefont
  {Mekterovic}, \citenamefont {Metag}, \citenamefont {Nefkens}, \citenamefont
  {Nikolaev}, \citenamefont {Novotny}, \citenamefont {Owens}, \citenamefont
  {Pedroni}, \citenamefont {Polonski}, \citenamefont {Prakhov}, \citenamefont
  {Price}, \citenamefont {Rosner}, \citenamefont {Rost}, \citenamefont
  {Rostomyan}, \citenamefont {Schadmand}, \citenamefont {Schumann},
  \citenamefont {Sober}, \citenamefont {Starostin}, \citenamefont {Supek},
  \citenamefont {Thomas}, \citenamefont {Unverzagt}, \citenamefont {Walcher},
  \citenamefont {Zana},\ and\ \citenamefont {Zehr}}]{Tar14a}%
  \BibitemOpen
  \bibfield  {author} {\bibinfo {author} {\bibfnamefont {C.~M.}\ \bibnamefont
  {Tarbert}}, \bibinfo {author} {\bibfnamefont {D.~P.}\ \bibnamefont {Watts}},
  \bibinfo {author} {\bibfnamefont {D.~I.}\ \bibnamefont {Glazier}}, \bibinfo
  {author} {\bibfnamefont {P.}~\bibnamefont {Aguar}}, \bibinfo {author}
  {\bibfnamefont {J.}~\bibnamefont {Ahrens}}, \bibinfo {author} {\bibfnamefont
  {J.~R.~M.}\ \bibnamefont {Annand}}, \bibinfo {author} {\bibfnamefont {H.~J.}\
  \bibnamefont {Arends}}, \bibinfo {author} {\bibfnamefont {R.}~\bibnamefont
  {Beck}}, \bibinfo {author} {\bibfnamefont {V.}~\bibnamefont {Bekrenev}},
  \bibinfo {author} {\bibfnamefont {B.}~\bibnamefont {Boillat}}, \bibinfo
  {author} {\bibfnamefont {A.}~\bibnamefont {Braghieri}}, \bibinfo {author}
  {\bibfnamefont {D.}~\bibnamefont {Branford}}, \bibinfo {author}
  {\bibfnamefont {W.~J.}\ \bibnamefont {Briscoe}}, \bibinfo {author}
  {\bibfnamefont {J.}~\bibnamefont {Brudvik}}, \bibinfo {author} {\bibfnamefont
  {S.}~\bibnamefont {Cherepnya}}, \bibinfo {author} {\bibfnamefont
  {R.}~\bibnamefont {Codling}}, \bibinfo {author} {\bibfnamefont {E.~J.}\
  \bibnamefont {Downie}}, \bibinfo {author} {\bibfnamefont {K.}~\bibnamefont
  {Foehl}}, \bibinfo {author} {\bibfnamefont {P.}~\bibnamefont {Grabmayr}},
  \bibinfo {author} {\bibfnamefont {R.}~\bibnamefont {Gregor}}, \bibinfo
  {author} {\bibfnamefont {E.}~\bibnamefont {Heid}}, \bibinfo {author}
  {\bibfnamefont {D.}~\bibnamefont {Hornidge}}, \bibinfo {author}
  {\bibfnamefont {O.}~\bibnamefont {Jahn}}, \bibinfo {author} {\bibfnamefont
  {V.~L.}\ \bibnamefont {Kashevarov}}, \bibinfo {author} {\bibfnamefont
  {A.}~\bibnamefont {Knezevic}}, \bibinfo {author} {\bibfnamefont
  {R.}~\bibnamefont {Kondratiev}}, \bibinfo {author} {\bibfnamefont
  {M.}~\bibnamefont {Korolija}}, \bibinfo {author} {\bibfnamefont
  {M.}~\bibnamefont {Kotulla}}, \bibinfo {author} {\bibfnamefont
  {D.}~\bibnamefont {Krambrich}}, \bibinfo {author} {\bibfnamefont
  {B.}~\bibnamefont {Krusche}}, \bibinfo {author} {\bibfnamefont
  {M.}~\bibnamefont {Lang}}, \bibinfo {author} {\bibfnamefont {V.}~\bibnamefont
  {Lisin}}, \bibinfo {author} {\bibfnamefont {K.}~\bibnamefont {Livingston}},
  \bibinfo {author} {\bibfnamefont {S.}~\bibnamefont {Lugert}}, \bibinfo
  {author} {\bibfnamefont {I.~J.~D.}\ \bibnamefont {MacGregor}}, \bibinfo
  {author} {\bibfnamefont {D.~M.}\ \bibnamefont {Manley}}, \bibinfo {author}
  {\bibfnamefont {M.}~\bibnamefont {Martinez}}, \bibinfo {author}
  {\bibfnamefont {J.~C.}\ \bibnamefont {McGeorge}}, \bibinfo {author}
  {\bibfnamefont {D.}~\bibnamefont {Mekterovic}}, \bibinfo {author}
  {\bibfnamefont {V.}~\bibnamefont {Metag}}, \bibinfo {author} {\bibfnamefont
  {B.~M.~K.}\ \bibnamefont {Nefkens}}, \bibinfo {author} {\bibfnamefont
  {A.}~\bibnamefont {Nikolaev}}, \bibinfo {author} {\bibfnamefont
  {R.}~\bibnamefont {Novotny}}, \bibinfo {author} {\bibfnamefont {R.~O.}\
  \bibnamefont {Owens}}, \bibinfo {author} {\bibfnamefont {P.}~\bibnamefont
  {Pedroni}}, \bibinfo {author} {\bibfnamefont {A.}~\bibnamefont {Polonski}},
  \bibinfo {author} {\bibfnamefont {S.~N.}\ \bibnamefont {Prakhov}}, \bibinfo
  {author} {\bibfnamefont {J.~W.}\ \bibnamefont {Price}}, \bibinfo {author}
  {\bibfnamefont {G.}~\bibnamefont {Rosner}}, \bibinfo {author} {\bibfnamefont
  {M.}~\bibnamefont {Rost}}, \bibinfo {author} {\bibfnamefont {T.}~\bibnamefont
  {Rostomyan}}, \bibinfo {author} {\bibfnamefont {S.}~\bibnamefont
  {Schadmand}}, \bibinfo {author} {\bibfnamefont {S.}~\bibnamefont {Schumann}},
  \bibinfo {author} {\bibfnamefont {D.}~\bibnamefont {Sober}}, \bibinfo
  {author} {\bibfnamefont {A.}~\bibnamefont {Starostin}}, \bibinfo {author}
  {\bibfnamefont {I.}~\bibnamefont {Supek}}, \bibinfo {author} {\bibfnamefont
  {A.}~\bibnamefont {Thomas}}, \bibinfo {author} {\bibfnamefont
  {M.}~\bibnamefont {Unverzagt}}, \bibinfo {author} {\bibfnamefont
  {T.}~\bibnamefont {Walcher}}, \bibinfo {author} {\bibfnamefont
  {L.}~\bibnamefont {Zana}}, \ and\ \bibinfo {author} {\bibfnamefont
  {F.}~\bibnamefont {Zehr}} (\bibinfo {collaboration} {Crystal Ball at MAMI and
  A2 Collaboration}),\ }\href {\doibase 10.1103/PhysRevLett.112.242502}
  {\bibfield  {journal} {\bibinfo  {journal} {Phys. Rev. Lett.}\ }\textbf
  {\bibinfo {volume} {112}},\ \bibinfo {pages} {242502} (\bibinfo {year}
  {2014})}\BibitemShut {NoStop}%
\bibitem [{\citenamefont {Abbott}\ \emph
  {et~al.}(2017{\natexlab{a}})\citenamefont {Abbott}, \citenamefont {Abbott},
  \citenamefont {Abbott}, \citenamefont {Acernese}, \citenamefont {Ackley}
  \emph {et~al.}}]{Abb17a}%
  \BibitemOpen
  \bibfield  {author} {\bibinfo {author} {\bibfnamefont {B.~P.}\ \bibnamefont
  {Abbott}}, \bibinfo {author} {\bibfnamefont {R.}~\bibnamefont {Abbott}},
  \bibinfo {author} {\bibfnamefont {T.~D.}\ \bibnamefont {Abbott}}, \bibinfo
  {author} {\bibfnamefont {F.}~\bibnamefont {Acernese}}, \bibinfo {author}
  {\bibfnamefont {K.}~\bibnamefont {Ackley}},  \emph {et~al.},\ }\href
  {\doibase 10.3847/2041-8213/aa91c9} {\bibfield  {journal} {\bibinfo
  {journal} {The Astrophysical Journal}\ }\textbf {\bibinfo {volume} {848}},\
  \bibinfo {pages} {L12} (\bibinfo {year} {2017}{\natexlab{a}})}\BibitemShut
  {NoStop}%
\bibitem [{\citenamefont {Abbott}\ \emph
  {et~al.}(2017{\natexlab{b}})\citenamefont {Abbott} \emph {et~al.}}]{Abb17b}%
  \BibitemOpen
  \bibfield  {author} {\bibinfo {author} {\bibfnamefont {B.~P.}\ \bibnamefont
  {Abbott}} \emph {et~al.} (\bibinfo {collaboration} {LIGO Scientific
  Collaboration and Virgo Collaboration}),\ }\href {\doibase
  10.1103/PhysRevLett.119.161101} {\bibfield  {journal} {\bibinfo  {journal}
  {Phys. Rev. Lett.}\ }\textbf {\bibinfo {volume} {119}},\ \bibinfo {pages}
  {161101} (\bibinfo {year} {2017}{\natexlab{b}})}\BibitemShut {NoStop}%
\bibitem [{\citenamefont {Fattoyev}\ \emph {et~al.}(2018)\citenamefont
  {Fattoyev}, \citenamefont {Piekarewicz},\ and\ \citenamefont
  {Horowitz}}]{Fat18a}%
  \BibitemOpen
  \bibfield  {author} {\bibinfo {author} {\bibfnamefont {F.~J.}\ \bibnamefont
  {Fattoyev}}, \bibinfo {author} {\bibfnamefont {J.}~\bibnamefont
  {Piekarewicz}}, \ and\ \bibinfo {author} {\bibfnamefont {C.~J.}\ \bibnamefont
  {Horowitz}},\ }\href {\doibase 10.1103/PhysRevLett.120.172702} {\bibfield
  {journal} {\bibinfo  {journal} {Phys. Rev. Lett.}\ }\textbf {\bibinfo
  {volume} {120}},\ \bibinfo {pages} {172702} (\bibinfo {year}
  {2018})}\BibitemShut {NoStop}%
\bibitem [{\citenamefont {Rossi}\ \emph {et~al.}(2013)\citenamefont {Rossi},
  \citenamefont {Adrich}, \citenamefont {Aksouh}, \citenamefont {Alvarez-Pol},
  \citenamefont {Aumann}, \citenamefont {Benlliure}, \citenamefont {B\"ohmer},
  \citenamefont {Boretzky}, \citenamefont {Casarejos}, \citenamefont
  {Chartier}, \citenamefont {Chatillon}, \citenamefont {Cortina-Gil},
  \citenamefont {Datta~Pramanik}, \citenamefont {Emling}, \citenamefont
  {Ershova}, \citenamefont {Fernandez-Dominguez}, \citenamefont {Geissel},
  \citenamefont {Gorska}, \citenamefont {Heil}, \citenamefont {Johansson},
  \citenamefont {Junghans}, \citenamefont {Kelic-Heil}, \citenamefont
  {Kiselev}, \citenamefont {Klimkiewicz}, \citenamefont {Kratz}, \citenamefont
  {Kr\"ucken}, \citenamefont {Kurz}, \citenamefont {Labiche}, \citenamefont
  {Le~Bleis}, \citenamefont {Lemmon}, \citenamefont {Litvinov}, \citenamefont
  {Mahata}, \citenamefont {Maierbeck}, \citenamefont {Movsesyan}, \citenamefont
  {Nilsson}, \citenamefont {Nociforo}, \citenamefont {Palit}, \citenamefont
  {Paschalis}, \citenamefont {Plag}, \citenamefont {Reifarth}, \citenamefont
  {Savran}, \citenamefont {Scheit}, \citenamefont {Simon}, \citenamefont
  {S\"ummerer}, \citenamefont {Wagner}, \citenamefont
  {Walu\ifmmode~\acute{s}\else \'{s}\fi{}}, \citenamefont {Weick},\ and\
  \citenamefont {Winkler}}]{Ros13a}%
  \BibitemOpen
  \bibfield  {author} {\bibinfo {author} {\bibfnamefont {D.~M.}\ \bibnamefont
  {Rossi}}, \bibinfo {author} {\bibfnamefont {P.}~\bibnamefont {Adrich}},
  \bibinfo {author} {\bibfnamefont {F.}~\bibnamefont {Aksouh}}, \bibinfo
  {author} {\bibfnamefont {H.}~\bibnamefont {Alvarez-Pol}}, \bibinfo {author}
  {\bibfnamefont {T.}~\bibnamefont {Aumann}}, \bibinfo {author} {\bibfnamefont
  {J.}~\bibnamefont {Benlliure}}, \bibinfo {author} {\bibfnamefont
  {M.}~\bibnamefont {B\"ohmer}}, \bibinfo {author} {\bibfnamefont
  {K.}~\bibnamefont {Boretzky}}, \bibinfo {author} {\bibfnamefont
  {E.}~\bibnamefont {Casarejos}}, \bibinfo {author} {\bibfnamefont
  {M.}~\bibnamefont {Chartier}}, \bibinfo {author} {\bibfnamefont
  {A.}~\bibnamefont {Chatillon}}, \bibinfo {author} {\bibfnamefont
  {D.}~\bibnamefont {Cortina-Gil}}, \bibinfo {author} {\bibfnamefont
  {U.}~\bibnamefont {Datta~Pramanik}}, \bibinfo {author} {\bibfnamefont
  {H.}~\bibnamefont {Emling}}, \bibinfo {author} {\bibfnamefont
  {O.}~\bibnamefont {Ershova}}, \bibinfo {author} {\bibfnamefont
  {B.}~\bibnamefont {Fernandez-Dominguez}}, \bibinfo {author} {\bibfnamefont
  {H.}~\bibnamefont {Geissel}}, \bibinfo {author} {\bibfnamefont
  {M.}~\bibnamefont {Gorska}}, \bibinfo {author} {\bibfnamefont
  {M.}~\bibnamefont {Heil}}, \bibinfo {author} {\bibfnamefont {H.~T.}\
  \bibnamefont {Johansson}}, \bibinfo {author} {\bibfnamefont {A.}~\bibnamefont
  {Junghans}}, \bibinfo {author} {\bibfnamefont {A.}~\bibnamefont
  {Kelic-Heil}}, \bibinfo {author} {\bibfnamefont {O.}~\bibnamefont {Kiselev}},
  \bibinfo {author} {\bibfnamefont {A.}~\bibnamefont {Klimkiewicz}}, \bibinfo
  {author} {\bibfnamefont {J.~V.}\ \bibnamefont {Kratz}}, \bibinfo {author}
  {\bibfnamefont {R.}~\bibnamefont {Kr\"ucken}}, \bibinfo {author}
  {\bibfnamefont {N.}~\bibnamefont {Kurz}}, \bibinfo {author} {\bibfnamefont
  {M.}~\bibnamefont {Labiche}}, \bibinfo {author} {\bibfnamefont
  {T.}~\bibnamefont {Le~Bleis}}, \bibinfo {author} {\bibfnamefont
  {R.}~\bibnamefont {Lemmon}}, \bibinfo {author} {\bibfnamefont {Y.~A.}\
  \bibnamefont {Litvinov}}, \bibinfo {author} {\bibfnamefont {K.}~\bibnamefont
  {Mahata}}, \bibinfo {author} {\bibfnamefont {P.}~\bibnamefont {Maierbeck}},
  \bibinfo {author} {\bibfnamefont {A.}~\bibnamefont {Movsesyan}}, \bibinfo
  {author} {\bibfnamefont {T.}~\bibnamefont {Nilsson}}, \bibinfo {author}
  {\bibfnamefont {C.}~\bibnamefont {Nociforo}}, \bibinfo {author}
  {\bibfnamefont {R.}~\bibnamefont {Palit}}, \bibinfo {author} {\bibfnamefont
  {S.}~\bibnamefont {Paschalis}}, \bibinfo {author} {\bibfnamefont
  {R.}~\bibnamefont {Plag}}, \bibinfo {author} {\bibfnamefont {R.}~\bibnamefont
  {Reifarth}}, \bibinfo {author} {\bibfnamefont {D.}~\bibnamefont {Savran}},
  \bibinfo {author} {\bibfnamefont {H.}~\bibnamefont {Scheit}}, \bibinfo
  {author} {\bibfnamefont {H.}~\bibnamefont {Simon}}, \bibinfo {author}
  {\bibfnamefont {K.}~\bibnamefont {S\"ummerer}}, \bibinfo {author}
  {\bibfnamefont {A.}~\bibnamefont {Wagner}}, \bibinfo {author} {\bibfnamefont
  {W.}~\bibnamefont {Walu\ifmmode~\acute{s}\else \'{s}\fi{}}}, \bibinfo
  {author} {\bibfnamefont {H.}~\bibnamefont {Weick}}, \ and\ \bibinfo {author}
  {\bibfnamefont {M.}~\bibnamefont {Winkler}},\ }\href {\doibase
  10.1103/PhysRevLett.111.242503} {\bibfield  {journal} {\bibinfo  {journal}
  {Phys. Rev. Lett.}\ }\textbf {\bibinfo {volume} {111}},\ \bibinfo {pages}
  {242503} (\bibinfo {year} {2013})}\BibitemShut {NoStop}%
\bibitem [{\citenamefont {Hashimoto}\ \emph {et~al.}(2015)\citenamefont
  {Hashimoto}, \citenamefont {Krumbholz}, \citenamefont {Reinhard},
  \citenamefont {Tamii}, \citenamefont {von Neumann-Cosel}, \citenamefont
  {Adachi}, \citenamefont {Aoi}, \citenamefont {Bertulani}, \citenamefont
  {Fujita}, \citenamefont {Fujita}, \citenamefont {Ganio\ifmmode~\check{g}\else
  \v{g}\fi{}lu}, \citenamefont {Hatanaka}, \citenamefont {Ideguchi},
  \citenamefont {Iwamoto}, \citenamefont {Kawabata}, \citenamefont {Khai},
  \citenamefont {Krugmann}, \citenamefont {Martin}, \citenamefont {Matsubara},
  \citenamefont {Miki}, \citenamefont {Neveling}, \citenamefont {Okamura},
  \citenamefont {Ong}, \citenamefont {Poltoratska}, \citenamefont {Ponomarev},
  \citenamefont {Richter}, \citenamefont {Sakaguchi}, \citenamefont {Shimbara},
  \citenamefont {Shimizu}, \citenamefont {Simonis}, \citenamefont {Smit},
  \citenamefont {S\"usoy}, \citenamefont {Suzuki}, \citenamefont {Thies},
  \citenamefont {Yosoi},\ and\ \citenamefont {Zenihiro}}]{Has15a}%
  \BibitemOpen
  \bibfield  {author} {\bibinfo {author} {\bibfnamefont {T.}~\bibnamefont
  {Hashimoto}}, \bibinfo {author} {\bibfnamefont {A.~M.}\ \bibnamefont
  {Krumbholz}}, \bibinfo {author} {\bibfnamefont {P.-G.}\ \bibnamefont
  {Reinhard}}, \bibinfo {author} {\bibfnamefont {A.}~\bibnamefont {Tamii}},
  \bibinfo {author} {\bibfnamefont {P.}~\bibnamefont {von Neumann-Cosel}},
  \bibinfo {author} {\bibfnamefont {T.}~\bibnamefont {Adachi}}, \bibinfo
  {author} {\bibfnamefont {N.}~\bibnamefont {Aoi}}, \bibinfo {author}
  {\bibfnamefont {C.~A.}\ \bibnamefont {Bertulani}}, \bibinfo {author}
  {\bibfnamefont {H.}~\bibnamefont {Fujita}}, \bibinfo {author} {\bibfnamefont
  {Y.}~\bibnamefont {Fujita}}, \bibinfo {author} {\bibfnamefont
  {E.}~\bibnamefont {Ganio\ifmmode~\check{g}\else \v{g}\fi{}lu}}, \bibinfo
  {author} {\bibfnamefont {K.}~\bibnamefont {Hatanaka}}, \bibinfo {author}
  {\bibfnamefont {E.}~\bibnamefont {Ideguchi}}, \bibinfo {author}
  {\bibfnamefont {C.}~\bibnamefont {Iwamoto}}, \bibinfo {author} {\bibfnamefont
  {T.}~\bibnamefont {Kawabata}}, \bibinfo {author} {\bibfnamefont {N.~T.}\
  \bibnamefont {Khai}}, \bibinfo {author} {\bibfnamefont {A.}~\bibnamefont
  {Krugmann}}, \bibinfo {author} {\bibfnamefont {D.}~\bibnamefont {Martin}},
  \bibinfo {author} {\bibfnamefont {H.}~\bibnamefont {Matsubara}}, \bibinfo
  {author} {\bibfnamefont {K.}~\bibnamefont {Miki}}, \bibinfo {author}
  {\bibfnamefont {R.}~\bibnamefont {Neveling}}, \bibinfo {author}
  {\bibfnamefont {H.}~\bibnamefont {Okamura}}, \bibinfo {author} {\bibfnamefont
  {H.~J.}\ \bibnamefont {Ong}}, \bibinfo {author} {\bibfnamefont
  {I.}~\bibnamefont {Poltoratska}}, \bibinfo {author} {\bibfnamefont {V.~Y.}\
  \bibnamefont {Ponomarev}}, \bibinfo {author} {\bibfnamefont {A.}~\bibnamefont
  {Richter}}, \bibinfo {author} {\bibfnamefont {H.}~\bibnamefont {Sakaguchi}},
  \bibinfo {author} {\bibfnamefont {Y.}~\bibnamefont {Shimbara}}, \bibinfo
  {author} {\bibfnamefont {Y.}~\bibnamefont {Shimizu}}, \bibinfo {author}
  {\bibfnamefont {J.}~\bibnamefont {Simonis}}, \bibinfo {author} {\bibfnamefont
  {F.~D.}\ \bibnamefont {Smit}}, \bibinfo {author} {\bibfnamefont
  {G.}~\bibnamefont {S\"usoy}}, \bibinfo {author} {\bibfnamefont
  {T.}~\bibnamefont {Suzuki}}, \bibinfo {author} {\bibfnamefont {J.~H.}\
  \bibnamefont {Thies}}, \bibinfo {author} {\bibfnamefont {M.}~\bibnamefont
  {Yosoi}}, \ and\ \bibinfo {author} {\bibfnamefont {J.}~\bibnamefont
  {Zenihiro}},\ }\href {\doibase 10.1103/PhysRevC.92.031305} {\bibfield
  {journal} {\bibinfo  {journal} {Phys. Rev. C}\ }\textbf {\bibinfo {volume}
  {92}},\ \bibinfo {pages} {031305} (\bibinfo {year} {2015})}\BibitemShut
  {NoStop}%
\bibitem [{\citenamefont {Hagen}\ \emph {et~al.}(2016)\citenamefont {Hagen},
  \citenamefont {Ekstr{\"o}m}, \citenamefont {Fors{\'e}n} \emph
  {et~al.}}]{Hag16a}%
  \BibitemOpen
  \bibfield  {author} {\bibinfo {author} {\bibfnamefont {G.}~\bibnamefont
  {Hagen}}, \bibinfo {author} {\bibfnamefont {A.}~\bibnamefont {Ekstr{\"o}m}},
  \bibinfo {author} {\bibfnamefont {C.}~\bibnamefont {Fors{\'e}n}},  \emph
  {et~al.},\ }\href {\doibase https://doi.org/10.1038/nphys3529} {\bibfield
  {journal} {\bibinfo  {journal} {Nature Phys.}\ }\textbf {\bibinfo {volume}
  {12}},\ \bibinfo {pages} {186} (\bibinfo {year} {2016})}\BibitemShut
  {NoStop}%
\bibitem [{\citenamefont {Birkhan}\ \emph {et~al.}(2017)\citenamefont
  {Birkhan}, \citenamefont {Miorelli}, \citenamefont {Bacca}, \citenamefont
  {Bassauer}, \citenamefont {Bertulani}, \citenamefont {Hagen}, \citenamefont
  {Matsubara}, \citenamefont {von Neumann-Cosel}, \citenamefont {Papenbrock},
  \citenamefont {Pietralla}, \citenamefont {Ponomarev}, \citenamefont
  {Richter}, \citenamefont {Schwenk},\ and\ \citenamefont {Tamii}}]{Bir17a}%
  \BibitemOpen
  \bibfield  {author} {\bibinfo {author} {\bibfnamefont {J.}~\bibnamefont
  {Birkhan}}, \bibinfo {author} {\bibfnamefont {M.}~\bibnamefont {Miorelli}},
  \bibinfo {author} {\bibfnamefont {S.}~\bibnamefont {Bacca}}, \bibinfo
  {author} {\bibfnamefont {S.}~\bibnamefont {Bassauer}}, \bibinfo {author}
  {\bibfnamefont {C.~A.}\ \bibnamefont {Bertulani}}, \bibinfo {author}
  {\bibfnamefont {G.}~\bibnamefont {Hagen}}, \bibinfo {author} {\bibfnamefont
  {H.}~\bibnamefont {Matsubara}}, \bibinfo {author} {\bibfnamefont
  {P.}~\bibnamefont {von Neumann-Cosel}}, \bibinfo {author} {\bibfnamefont
  {T.}~\bibnamefont {Papenbrock}}, \bibinfo {author} {\bibfnamefont
  {N.}~\bibnamefont {Pietralla}}, \bibinfo {author} {\bibfnamefont {V.~Y.}\
  \bibnamefont {Ponomarev}}, \bibinfo {author} {\bibfnamefont {A.}~\bibnamefont
  {Richter}}, \bibinfo {author} {\bibfnamefont {A.}~\bibnamefont {Schwenk}}, \
  and\ \bibinfo {author} {\bibfnamefont {A.}~\bibnamefont {Tamii}},\ }\href
  {\doibase 10.1103/PhysRevLett.118.252501} {\bibfield  {journal} {\bibinfo
  {journal} {Phys. Rev. Lett.}\ }\textbf {\bibinfo {volume} {118}},\ \bibinfo
  {pages} {252501} (\bibinfo {year} {2017})}\BibitemShut {NoStop}%
\bibitem [{\citenamefont {Tonchev}\ \emph {et~al.}(2017)\citenamefont
  {Tonchev}, \citenamefont {Tsoneva}, \citenamefont {Bhatia}, \citenamefont
  {Arnold}, \citenamefont {Goriely}, \citenamefont {Hammond}, \citenamefont
  {Kelley}, \citenamefont {Kwan}, \citenamefont {Lenske}, \citenamefont
  {Piekarewicz}, \citenamefont {Raut}, \citenamefont {Rusev}, \citenamefont
  {Shizuma},\ and\ \citenamefont {Tornow}}]{Ton17a}%
  \BibitemOpen
  \bibfield  {author} {\bibinfo {author} {\bibfnamefont {A.}~\bibnamefont
  {Tonchev}}, \bibinfo {author} {\bibfnamefont {N.}~\bibnamefont {Tsoneva}},
  \bibinfo {author} {\bibfnamefont {C.}~\bibnamefont {Bhatia}}, \bibinfo
  {author} {\bibfnamefont {C.}~\bibnamefont {Arnold}}, \bibinfo {author}
  {\bibfnamefont {S.}~\bibnamefont {Goriely}}, \bibinfo {author} {\bibfnamefont
  {S.}~\bibnamefont {Hammond}}, \bibinfo {author} {\bibfnamefont
  {J.}~\bibnamefont {Kelley}}, \bibinfo {author} {\bibfnamefont
  {E.}~\bibnamefont {Kwan}}, \bibinfo {author} {\bibfnamefont {H.}~\bibnamefont
  {Lenske}}, \bibinfo {author} {\bibfnamefont {J.}~\bibnamefont {Piekarewicz}},
  \bibinfo {author} {\bibfnamefont {R.}~\bibnamefont {Raut}}, \bibinfo {author}
  {\bibfnamefont {G.}~\bibnamefont {Rusev}}, \bibinfo {author} {\bibfnamefont
  {T.}~\bibnamefont {Shizuma}}, \ and\ \bibinfo {author} {\bibfnamefont
  {W.}~\bibnamefont {Tornow}},\ }\href {\doibase
  https://doi.org/10.1016/j.physletb.2017.07.062} {\bibfield  {journal}
  {\bibinfo  {journal} {Physics Letters B}\ }\textbf {\bibinfo {volume}
  {773}},\ \bibinfo {pages} {20 } (\bibinfo {year} {2017})}\BibitemShut
  {NoStop}%
\bibitem [{\citenamefont {Roca-Maza}\ and\ \citenamefont
  {Paar}(2018)}]{Roc18a}%
  \BibitemOpen
  \bibfield  {author} {\bibinfo {author} {\bibfnamefont {X.}~\bibnamefont
  {Roca-Maza}}\ and\ \bibinfo {author} {\bibfnamefont {N.}~\bibnamefont
  {Paar}},\ }\href {\doibase https://doi.org/10.1016/j.ppnp.2018.04.001}
  {\bibfield  {journal} {\bibinfo  {journal} {Progress in Particle and Nuclear
  Physics}\ }\textbf {\bibinfo {volume} {101}},\ \bibinfo {pages} {96 }
  (\bibinfo {year} {2018})}\BibitemShut {NoStop}%
\bibitem [{\citenamefont {Kaufmann}\ \emph {et~al.}(2020)\citenamefont
  {Kaufmann}, \citenamefont {Simonis}, \citenamefont {Bacca}, \citenamefont
  {Billowes}, \citenamefont {Bissell}, \citenamefont {Blaum}, \citenamefont
  {Cheal}, \citenamefont {Ruiz}, \citenamefont {Gins}, \citenamefont {Gorges},
  \citenamefont {Hagen}, \citenamefont {Heylen}, \citenamefont
  {Kanellakopoulos}, \citenamefont {Malbrunot-Ettenauer}, \citenamefont
  {Miorelli}, \citenamefont {Neugart}, \citenamefont {Neyens}, \citenamefont
  {N\"ortersh\"auser}, \citenamefont {S\'anchez}, \citenamefont {Sailer},
  \citenamefont {Schwenk}, \citenamefont {Ratajczyk}, \citenamefont
  {Rodr\'{\i}guez}, \citenamefont {Wehner}, \citenamefont {Wraith},
  \citenamefont {Xie}, \citenamefont {Xu}, \citenamefont {Yang},\ and\
  \citenamefont {Yordanov}}]{Kau20a}%
  \BibitemOpen
  \bibfield  {author} {\bibinfo {author} {\bibfnamefont {S.}~\bibnamefont
  {Kaufmann}}, \bibinfo {author} {\bibfnamefont {J.}~\bibnamefont {Simonis}},
  \bibinfo {author} {\bibfnamefont {S.}~\bibnamefont {Bacca}}, \bibinfo
  {author} {\bibfnamefont {J.}~\bibnamefont {Billowes}}, \bibinfo {author}
  {\bibfnamefont {M.~L.}\ \bibnamefont {Bissell}}, \bibinfo {author}
  {\bibfnamefont {K.}~\bibnamefont {Blaum}}, \bibinfo {author} {\bibfnamefont
  {B.}~\bibnamefont {Cheal}}, \bibinfo {author} {\bibfnamefont {R.~F.~G.}\
  \bibnamefont {Ruiz}}, \bibinfo {author} {\bibfnamefont {W.}~\bibnamefont
  {Gins}}, \bibinfo {author} {\bibfnamefont {C.}~\bibnamefont {Gorges}},
  \bibinfo {author} {\bibfnamefont {G.}~\bibnamefont {Hagen}}, \bibinfo
  {author} {\bibfnamefont {H.}~\bibnamefont {Heylen}}, \bibinfo {author}
  {\bibfnamefont {A.}~\bibnamefont {Kanellakopoulos}}, \bibinfo {author}
  {\bibfnamefont {S.}~\bibnamefont {Malbrunot-Ettenauer}}, \bibinfo {author}
  {\bibfnamefont {M.}~\bibnamefont {Miorelli}}, \bibinfo {author}
  {\bibfnamefont {R.}~\bibnamefont {Neugart}}, \bibinfo {author} {\bibfnamefont
  {G.}~\bibnamefont {Neyens}}, \bibinfo {author} {\bibfnamefont
  {W.}~\bibnamefont {N\"ortersh\"auser}}, \bibinfo {author} {\bibfnamefont
  {R.}~\bibnamefont {S\'anchez}}, \bibinfo {author} {\bibfnamefont
  {S.}~\bibnamefont {Sailer}}, \bibinfo {author} {\bibfnamefont
  {A.}~\bibnamefont {Schwenk}}, \bibinfo {author} {\bibfnamefont
  {T.}~\bibnamefont {Ratajczyk}}, \bibinfo {author} {\bibfnamefont {L.~V.}\
  \bibnamefont {Rodr\'{\i}guez}}, \bibinfo {author} {\bibfnamefont
  {L.}~\bibnamefont {Wehner}}, \bibinfo {author} {\bibfnamefont
  {C.}~\bibnamefont {Wraith}}, \bibinfo {author} {\bibfnamefont
  {L.}~\bibnamefont {Xie}}, \bibinfo {author} {\bibfnamefont {Z.~Y.}\
  \bibnamefont {Xu}}, \bibinfo {author} {\bibfnamefont {X.~F.}\ \bibnamefont
  {Yang}}, \ and\ \bibinfo {author} {\bibfnamefont {D.~T.}\ \bibnamefont
  {Yordanov}},\ }\href {\doibase 10.1103/PhysRevLett.124.132502} {\bibfield
  {journal} {\bibinfo  {journal} {Phys. Rev. Lett.}\ }\textbf {\bibinfo
  {volume} {124}},\ \bibinfo {pages} {132502} (\bibinfo {year}
  {2020})}\BibitemShut {NoStop}%
\bibitem [{\citenamefont {Bartholomew}(1961)}]{Bar61a}%
  \BibitemOpen
  \bibfield  {author} {\bibinfo {author} {\bibfnamefont {G.}~\bibnamefont
  {Bartholomew}},\ }\href@noop {} {\bibfield  {journal} {\bibinfo  {journal}
  {Annu. Rev. Nucl. Sci.}\ }\textbf {\bibinfo {volume} {11}},\ \bibinfo {pages}
  {259} (\bibinfo {year} {1961})}\BibitemShut {NoStop}%
\bibitem [{\citenamefont {Paar}\ \emph {et~al.}(2007)\citenamefont {Paar},
  \citenamefont {Vretenar}, \citenamefont {Khan},\ and\ \citenamefont
  {Col{\`{o}}}}]{Paar07a}%
  \BibitemOpen
  \bibfield  {author} {\bibinfo {author} {\bibfnamefont {N.}~\bibnamefont
  {Paar}}, \bibinfo {author} {\bibfnamefont {D.}~\bibnamefont {Vretenar}},
  \bibinfo {author} {\bibfnamefont {E.}~\bibnamefont {Khan}}, \ and\ \bibinfo
  {author} {\bibfnamefont {G.}~\bibnamefont {Col{\`{o}}}},\ }\href {\doibase
  10.1088/0034-4885/70/5/r02} {\bibfield  {journal} {\bibinfo  {journal}
  {Reports on Progress in Physics}\ }\textbf {\bibinfo {volume} {70}},\
  \bibinfo {pages} {691} (\bibinfo {year} {2007})}\BibitemShut {NoStop}%
\bibitem [{\citenamefont {Savran}\ \emph {et~al.}(2013)\citenamefont {Savran},
  \citenamefont {Aumann},\ and\ \citenamefont {Zilges}}]{Sav13a}%
  \BibitemOpen
  \bibfield  {author} {\bibinfo {author} {\bibfnamefont {D.}~\bibnamefont
  {Savran}}, \bibinfo {author} {\bibfnamefont {T.}~\bibnamefont {Aumann}}, \
  and\ \bibinfo {author} {\bibfnamefont {A.}~\bibnamefont {Zilges}},\ }\href
  {\doibase https://doi.org/10.1016/j.ppnp.2013.02.003} {\bibfield  {journal}
  {\bibinfo  {journal} {Progress in Particle and Nuclear Physics}\ }\textbf
  {\bibinfo {volume} {70}},\ \bibinfo {pages} {210 } (\bibinfo {year}
  {2013})}\BibitemShut {NoStop}%
\bibitem [{\citenamefont {{Bracco, A.}}\ \emph {et~al.}(2015)\citenamefont
  {{Bracco, A.}}, \citenamefont {{Crespi, F. C. L.}},\ and\ \citenamefont
  {{Lanza, E. G.}}}]{Bra15a}%
  \BibitemOpen
  \bibfield  {author} {\bibinfo {author} {\bibnamefont {{Bracco, A.}}},
  \bibinfo {author} {\bibnamefont {{Crespi, F. C. L.}}}, \ and\ \bibinfo
  {author} {\bibnamefont {{Lanza, E. G.}}},\ }\href {\doibase
  10.1140/epja/i2015-15099-6} {\bibfield  {journal} {\bibinfo  {journal} {Eur.
  Phys. J. A}\ }\textbf {\bibinfo {volume} {51}},\ \bibinfo {pages} {99}
  (\bibinfo {year} {2015})}\BibitemShut {NoStop}%
\bibitem [{\citenamefont {Bracco}\ \emph {et~al.}(2019)\citenamefont {Bracco},
  \citenamefont {Lanza},\ and\ \citenamefont {Tamii}}]{Bra19a}%
  \BibitemOpen
  \bibfield  {author} {\bibinfo {author} {\bibfnamefont {A.}~\bibnamefont
  {Bracco}}, \bibinfo {author} {\bibfnamefont {E.}~\bibnamefont {Lanza}}, \
  and\ \bibinfo {author} {\bibfnamefont {A.}~\bibnamefont {Tamii}},\ }\href
  {\doibase https://doi.org/10.1016/j.ppnp.2019.02.001} {\bibfield  {journal}
  {\bibinfo  {journal} {Progress in Particle and Nuclear Physics}\ }\textbf
  {\bibinfo {volume} {106}},\ \bibinfo {pages} {360 } (\bibinfo {year}
  {2019})}\BibitemShut {NoStop}%
\bibitem [{\citenamefont {Piekarewicz}(2006)}]{Pie06a}%
  \BibitemOpen
  \bibfield  {author} {\bibinfo {author} {\bibfnamefont {J.}~\bibnamefont
  {Piekarewicz}},\ }\href {\doibase 10.1103/PhysRevC.73.044325} {\bibfield
  {journal} {\bibinfo  {journal} {Phys. Rev. C}\ }\textbf {\bibinfo {volume}
  {73}},\ \bibinfo {pages} {044325} (\bibinfo {year} {2006})}\BibitemShut
  {NoStop}%
\bibitem [{\citenamefont {Tsoneva}\ and\ \citenamefont
  {Lenske}(2008)}]{Tso08a}%
  \BibitemOpen
  \bibfield  {author} {\bibinfo {author} {\bibfnamefont {N.}~\bibnamefont
  {Tsoneva}}\ and\ \bibinfo {author} {\bibfnamefont {H.}~\bibnamefont
  {Lenske}},\ }\href {\doibase 10.1103/PhysRevC.77.024321} {\bibfield
  {journal} {\bibinfo  {journal} {Phys. Rev. C}\ }\textbf {\bibinfo {volume}
  {77}},\ \bibinfo {pages} {024321} (\bibinfo {year} {2008})}\BibitemShut
  {NoStop}%
\bibitem [{\citenamefont {Carbone}\ \emph {et~al.}(2010)\citenamefont
  {Carbone}, \citenamefont {Col\`o}, \citenamefont {Bracco}, \citenamefont
  {Cao}, \citenamefont {Bortignon}, \citenamefont {Camera},\ and\ \citenamefont
  {Wieland}}]{Car10a}%
  \BibitemOpen
  \bibfield  {author} {\bibinfo {author} {\bibfnamefont {A.}~\bibnamefont
  {Carbone}}, \bibinfo {author} {\bibfnamefont {G.}~\bibnamefont {Col\`o}},
  \bibinfo {author} {\bibfnamefont {A.}~\bibnamefont {Bracco}}, \bibinfo
  {author} {\bibfnamefont {L.-G.}\ \bibnamefont {Cao}}, \bibinfo {author}
  {\bibfnamefont {P.~F.}\ \bibnamefont {Bortignon}}, \bibinfo {author}
  {\bibfnamefont {F.}~\bibnamefont {Camera}}, \ and\ \bibinfo {author}
  {\bibfnamefont {O.}~\bibnamefont {Wieland}},\ }\href {\doibase
  10.1103/PhysRevC.81.041301} {\bibfield  {journal} {\bibinfo  {journal} {Phys.
  Rev. C}\ }\textbf {\bibinfo {volume} {81}},\ \bibinfo {pages} {041301}
  (\bibinfo {year} {2010})}\BibitemShut {NoStop}%
\bibitem [{\citenamefont {Piekarewicz}(2011)}]{Pie11a}%
  \BibitemOpen
  \bibfield  {author} {\bibinfo {author} {\bibfnamefont {J.}~\bibnamefont
  {Piekarewicz}},\ }\href {\doibase 10.1103/PhysRevC.83.034319} {\bibfield
  {journal} {\bibinfo  {journal} {Phys. Rev. C}\ }\textbf {\bibinfo {volume}
  {83}},\ \bibinfo {pages} {034319} (\bibinfo {year} {2011})}\BibitemShut
  {NoStop}%
\bibitem [{\citenamefont {Baran}\ \emph {et~al.}(2013)\citenamefont {Baran},
  \citenamefont {Colonna}, \citenamefont {Di~Toro}, \citenamefont {Croitoru},\
  and\ \citenamefont {Dumitru}}]{Bar13a}%
  \BibitemOpen
  \bibfield  {author} {\bibinfo {author} {\bibfnamefont {V.}~\bibnamefont
  {Baran}}, \bibinfo {author} {\bibfnamefont {M.}~\bibnamefont {Colonna}},
  \bibinfo {author} {\bibfnamefont {M.}~\bibnamefont {Di~Toro}}, \bibinfo
  {author} {\bibfnamefont {A.}~\bibnamefont {Croitoru}}, \ and\ \bibinfo
  {author} {\bibfnamefont {D.}~\bibnamefont {Dumitru}},\ }\href {\doibase
  10.1103/PhysRevC.88.044610} {\bibfield  {journal} {\bibinfo  {journal} {Phys.
  Rev. C}\ }\textbf {\bibinfo {volume} {88}},\ \bibinfo {pages} {044610}
  (\bibinfo {year} {2013})}\BibitemShut {NoStop}%
\bibitem [{\citenamefont {Inakura}\ \emph {et~al.}(2011)\citenamefont
  {Inakura}, \citenamefont {Nakatsukasa},\ and\ \citenamefont
  {Yabana}}]{Ina11a}%
  \BibitemOpen
  \bibfield  {author} {\bibinfo {author} {\bibfnamefont {T.}~\bibnamefont
  {Inakura}}, \bibinfo {author} {\bibfnamefont {T.}~\bibnamefont
  {Nakatsukasa}}, \ and\ \bibinfo {author} {\bibfnamefont {K.}~\bibnamefont
  {Yabana}},\ }\href {\doibase 10.1103/PhysRevC.84.021302} {\bibfield
  {journal} {\bibinfo  {journal} {Phys. Rev. C}\ }\textbf {\bibinfo {volume}
  {84}},\ \bibinfo {pages} {021302} (\bibinfo {year} {2011})}\BibitemShut
  {NoStop}%
\bibitem [{\citenamefont {Inakura}\ \emph {et~al.}(2013)\citenamefont
  {Inakura}, \citenamefont {Nakatsukasa},\ and\ \citenamefont
  {Yabana}}]{Ina13a}%
  \BibitemOpen
  \bibfield  {author} {\bibinfo {author} {\bibfnamefont {T.}~\bibnamefont
  {Inakura}}, \bibinfo {author} {\bibfnamefont {T.}~\bibnamefont
  {Nakatsukasa}}, \ and\ \bibinfo {author} {\bibfnamefont {K.}~\bibnamefont
  {Yabana}},\ }\href {\doibase 10.1103/PhysRevC.88.051305} {\bibfield
  {journal} {\bibinfo  {journal} {Phys. Rev. C}\ }\textbf {\bibinfo {volume}
  {88}},\ \bibinfo {pages} {051305} (\bibinfo {year} {2013})}\BibitemShut
  {NoStop}%
\bibitem [{\citenamefont {Reinhard}\ and\ \citenamefont
  {Nazarewicz}(2013)}]{Rei13a}%
  \BibitemOpen
  \bibfield  {author} {\bibinfo {author} {\bibfnamefont {P.-G.}\ \bibnamefont
  {Reinhard}}\ and\ \bibinfo {author} {\bibfnamefont {W.}~\bibnamefont
  {Nazarewicz}},\ }\href {\doibase 10.1103/PhysRevC.87.014324} {\bibfield
  {journal} {\bibinfo  {journal} {Phys. Rev. C}\ }\textbf {\bibinfo {volume}
  {87}},\ \bibinfo {pages} {014324} (\bibinfo {year} {2013})}\BibitemShut
  {NoStop}%
\bibitem [{\citenamefont {Roca-Maza}\ \emph {et~al.}(2015)\citenamefont
  {Roca-Maza}, \citenamefont {Vi\~nas}, \citenamefont {Centelles},
  \citenamefont {Agrawal}, \citenamefont {Col\`o}, \citenamefont {Paar},
  \citenamefont {Piekarewicz},\ and\ \citenamefont {Vretenar}}]{Roc15a}%
  \BibitemOpen
  \bibfield  {author} {\bibinfo {author} {\bibfnamefont {X.}~\bibnamefont
  {Roca-Maza}}, \bibinfo {author} {\bibfnamefont {X.}~\bibnamefont {Vi\~nas}},
  \bibinfo {author} {\bibfnamefont {M.}~\bibnamefont {Centelles}}, \bibinfo
  {author} {\bibfnamefont {B.~K.}\ \bibnamefont {Agrawal}}, \bibinfo {author}
  {\bibfnamefont {G.}~\bibnamefont {Col\`o}}, \bibinfo {author} {\bibfnamefont
  {N.}~\bibnamefont {Paar}}, \bibinfo {author} {\bibfnamefont {J.}~\bibnamefont
  {Piekarewicz}}, \ and\ \bibinfo {author} {\bibfnamefont {D.}~\bibnamefont
  {Vretenar}},\ }\href {\doibase 10.1103/PhysRevC.92.064304} {\bibfield
  {journal} {\bibinfo  {journal} {Phys. Rev. C}\ }\textbf {\bibinfo {volume}
  {92}},\ \bibinfo {pages} {064304} (\bibinfo {year} {2015})}\BibitemShut
  {NoStop}%
\bibitem [{\citenamefont {Endres}\ \emph {et~al.}(2010)\citenamefont {Endres},
  \citenamefont {Litvinova}, \citenamefont {Savran}, \citenamefont {Butler},
  \citenamefont {Harakeh}, \citenamefont {Harissopulos}, \citenamefont
  {Herzberg}, \citenamefont {Kr\"ucken}, \citenamefont {Lagoyannis},
  \citenamefont {Pietralla}, \citenamefont {Ponomarev}, \citenamefont
  {Popescu}, \citenamefont {Ring}, \citenamefont {Scheck}, \citenamefont
  {Sonnabend}, \citenamefont {Stoica}, \citenamefont {W\"ortche},\ and\
  \citenamefont {Zilges}}]{End10a}%
  \BibitemOpen
  \bibfield  {author} {\bibinfo {author} {\bibfnamefont {J.}~\bibnamefont
  {Endres}}, \bibinfo {author} {\bibfnamefont {E.}~\bibnamefont {Litvinova}},
  \bibinfo {author} {\bibfnamefont {D.}~\bibnamefont {Savran}}, \bibinfo
  {author} {\bibfnamefont {P.~A.}\ \bibnamefont {Butler}}, \bibinfo {author}
  {\bibfnamefont {M.~N.}\ \bibnamefont {Harakeh}}, \bibinfo {author}
  {\bibfnamefont {S.}~\bibnamefont {Harissopulos}}, \bibinfo {author}
  {\bibfnamefont {R.-D.}\ \bibnamefont {Herzberg}}, \bibinfo {author}
  {\bibfnamefont {R.}~\bibnamefont {Kr\"ucken}}, \bibinfo {author}
  {\bibfnamefont {A.}~\bibnamefont {Lagoyannis}}, \bibinfo {author}
  {\bibfnamefont {N.}~\bibnamefont {Pietralla}}, \bibinfo {author}
  {\bibfnamefont {V.~Y.}\ \bibnamefont {Ponomarev}}, \bibinfo {author}
  {\bibfnamefont {L.}~\bibnamefont {Popescu}}, \bibinfo {author} {\bibfnamefont
  {P.}~\bibnamefont {Ring}}, \bibinfo {author} {\bibfnamefont {M.}~\bibnamefont
  {Scheck}}, \bibinfo {author} {\bibfnamefont {K.}~\bibnamefont {Sonnabend}},
  \bibinfo {author} {\bibfnamefont {V.~I.}\ \bibnamefont {Stoica}}, \bibinfo
  {author} {\bibfnamefont {H.~J.}\ \bibnamefont {W\"ortche}}, \ and\ \bibinfo
  {author} {\bibfnamefont {A.}~\bibnamefont {Zilges}},\ }\href {\doibase
  10.1103/PhysRevLett.105.212503} {\bibfield  {journal} {\bibinfo  {journal}
  {Phys. Rev. Lett.}\ }\textbf {\bibinfo {volume} {105}},\ \bibinfo {pages}
  {212503} (\bibinfo {year} {2010})}\BibitemShut {NoStop}%
\bibitem [{\citenamefont {Poltoratska}\ \emph {et~al.}(2012)\citenamefont
  {Poltoratska}, \citenamefont {von Neumann-Cosel}, \citenamefont {Tamii},
  \citenamefont {Adachi}, \citenamefont {Bertulani}, \citenamefont {Carter},
  \citenamefont {Dozono}, \citenamefont {Fujita}, \citenamefont {Fujita},
  \citenamefont {Fujita}, \citenamefont {Hatanaka}, \citenamefont {Itoh},
  \citenamefont {Kawabata}, \citenamefont {Kalmykov}, \citenamefont
  {Krumbholz}, \citenamefont {Litvinova}, \citenamefont {Matsubara},
  \citenamefont {Nakanishi}, \citenamefont {Neveling}, \citenamefont {Okamura},
  \citenamefont {Ong}, \citenamefont {\"Ozel-Tashenov}, \citenamefont
  {Ponomarev}, \citenamefont {Richter}, \citenamefont {Rubio}, \citenamefont
  {Sakaguchi}, \citenamefont {Sakemi}, \citenamefont {Sasamoto}, \citenamefont
  {Shimbara}, \citenamefont {Shimizu}, \citenamefont {Smit}, \citenamefont
  {Suzuki}, \citenamefont {Tameshige}, \citenamefont {Wambach}, \citenamefont
  {Yosoi},\ and\ \citenamefont {Zenihiro}}]{Pol12a}%
  \BibitemOpen
  \bibfield  {author} {\bibinfo {author} {\bibfnamefont {I.}~\bibnamefont
  {Poltoratska}}, \bibinfo {author} {\bibfnamefont {P.}~\bibnamefont {von
  Neumann-Cosel}}, \bibinfo {author} {\bibfnamefont {A.}~\bibnamefont {Tamii}},
  \bibinfo {author} {\bibfnamefont {T.}~\bibnamefont {Adachi}}, \bibinfo
  {author} {\bibfnamefont {C.~A.}\ \bibnamefont {Bertulani}}, \bibinfo {author}
  {\bibfnamefont {J.}~\bibnamefont {Carter}}, \bibinfo {author} {\bibfnamefont
  {M.}~\bibnamefont {Dozono}}, \bibinfo {author} {\bibfnamefont
  {H.}~\bibnamefont {Fujita}}, \bibinfo {author} {\bibfnamefont
  {K.}~\bibnamefont {Fujita}}, \bibinfo {author} {\bibfnamefont
  {Y.}~\bibnamefont {Fujita}}, \bibinfo {author} {\bibfnamefont
  {K.}~\bibnamefont {Hatanaka}}, \bibinfo {author} {\bibfnamefont
  {M.}~\bibnamefont {Itoh}}, \bibinfo {author} {\bibfnamefont {T.}~\bibnamefont
  {Kawabata}}, \bibinfo {author} {\bibfnamefont {Y.}~\bibnamefont {Kalmykov}},
  \bibinfo {author} {\bibfnamefont {A.~M.}\ \bibnamefont {Krumbholz}}, \bibinfo
  {author} {\bibfnamefont {E.}~\bibnamefont {Litvinova}}, \bibinfo {author}
  {\bibfnamefont {H.}~\bibnamefont {Matsubara}}, \bibinfo {author}
  {\bibfnamefont {K.}~\bibnamefont {Nakanishi}}, \bibinfo {author}
  {\bibfnamefont {R.}~\bibnamefont {Neveling}}, \bibinfo {author}
  {\bibfnamefont {H.}~\bibnamefont {Okamura}}, \bibinfo {author} {\bibfnamefont
  {H.~J.}\ \bibnamefont {Ong}}, \bibinfo {author} {\bibfnamefont
  {B.}~\bibnamefont {\"Ozel-Tashenov}}, \bibinfo {author} {\bibfnamefont
  {V.~Y.}\ \bibnamefont {Ponomarev}}, \bibinfo {author} {\bibfnamefont
  {A.}~\bibnamefont {Richter}}, \bibinfo {author} {\bibfnamefont
  {B.}~\bibnamefont {Rubio}}, \bibinfo {author} {\bibfnamefont
  {H.}~\bibnamefont {Sakaguchi}}, \bibinfo {author} {\bibfnamefont
  {Y.}~\bibnamefont {Sakemi}}, \bibinfo {author} {\bibfnamefont
  {Y.}~\bibnamefont {Sasamoto}}, \bibinfo {author} {\bibfnamefont
  {Y.}~\bibnamefont {Shimbara}}, \bibinfo {author} {\bibfnamefont
  {Y.}~\bibnamefont {Shimizu}}, \bibinfo {author} {\bibfnamefont {F.~D.}\
  \bibnamefont {Smit}}, \bibinfo {author} {\bibfnamefont {T.}~\bibnamefont
  {Suzuki}}, \bibinfo {author} {\bibfnamefont {Y.}~\bibnamefont {Tameshige}},
  \bibinfo {author} {\bibfnamefont {J.}~\bibnamefont {Wambach}}, \bibinfo
  {author} {\bibfnamefont {M.}~\bibnamefont {Yosoi}}, \ and\ \bibinfo {author}
  {\bibfnamefont {J.}~\bibnamefont {Zenihiro}},\ }\href {\doibase
  10.1103/PhysRevC.85.041304} {\bibfield  {journal} {\bibinfo  {journal} {Phys.
  Rev. C}\ }\textbf {\bibinfo {volume} {85}},\ \bibinfo {pages} {041304}
  (\bibinfo {year} {2012})}\BibitemShut {NoStop}%
\bibitem [{\citenamefont {Savran}\ \emph {et~al.}(2018)\citenamefont {Savran},
  \citenamefont {Derya}, \citenamefont {Bagchi}, \citenamefont {Endres},
  \citenamefont {Harakeh}, \citenamefont {Isaak}, \citenamefont
  {Kalantar-Nayestanaki}, \citenamefont {Lanza}, \citenamefont {Löher},
  \citenamefont {Najafi}, \citenamefont {Pascu}, \citenamefont {Pickstone},
  \citenamefont {Pietralla}, \citenamefont {Ponomarev}, \citenamefont
  {Rigollet}, \citenamefont {Romig}, \citenamefont {Spieker}, \citenamefont
  {Vitturi},\ and\ \citenamefont {Zilges}}]{Sav18a}%
  \BibitemOpen
  \bibfield  {author} {\bibinfo {author} {\bibfnamefont {D.}~\bibnamefont
  {Savran}}, \bibinfo {author} {\bibfnamefont {V.}~\bibnamefont {Derya}},
  \bibinfo {author} {\bibfnamefont {S.}~\bibnamefont {Bagchi}}, \bibinfo
  {author} {\bibfnamefont {J.}~\bibnamefont {Endres}}, \bibinfo {author}
  {\bibfnamefont {M.}~\bibnamefont {Harakeh}}, \bibinfo {author} {\bibfnamefont
  {J.}~\bibnamefont {Isaak}}, \bibinfo {author} {\bibfnamefont
  {N.}~\bibnamefont {Kalantar-Nayestanaki}}, \bibinfo {author} {\bibfnamefont
  {E.}~\bibnamefont {Lanza}}, \bibinfo {author} {\bibfnamefont
  {B.}~\bibnamefont {Löher}}, \bibinfo {author} {\bibfnamefont
  {A.}~\bibnamefont {Najafi}}, \bibinfo {author} {\bibfnamefont
  {S.}~\bibnamefont {Pascu}}, \bibinfo {author} {\bibfnamefont
  {S.}~\bibnamefont {Pickstone}}, \bibinfo {author} {\bibfnamefont
  {N.}~\bibnamefont {Pietralla}}, \bibinfo {author} {\bibfnamefont
  {V.}~\bibnamefont {Ponomarev}}, \bibinfo {author} {\bibfnamefont
  {C.}~\bibnamefont {Rigollet}}, \bibinfo {author} {\bibfnamefont
  {C.}~\bibnamefont {Romig}}, \bibinfo {author} {\bibfnamefont
  {M.}~\bibnamefont {Spieker}}, \bibinfo {author} {\bibfnamefont
  {A.}~\bibnamefont {Vitturi}}, \ and\ \bibinfo {author} {\bibfnamefont
  {A.}~\bibnamefont {Zilges}},\ }\href {\doibase
  https://doi.org/10.1016/j.physletb.2018.09.025} {\bibfield  {journal}
  {\bibinfo  {journal} {Physics Letters B}\ }\textbf {\bibinfo {volume}
  {786}},\ \bibinfo {pages} {16 } (\bibinfo {year} {2018})}\BibitemShut
  {NoStop}%
\bibitem [{\citenamefont {Wieland}\ \emph {et~al.}(2018)\citenamefont
  {Wieland}, \citenamefont {Bracco}, \citenamefont {Camera}, \citenamefont
  {Avigo}, \citenamefont {Baba}, \citenamefont {Nakatsuka}, \citenamefont
  {Aumann}, \citenamefont {Banerjee}, \citenamefont {Benzoni}, \citenamefont
  {Boretzky}, \citenamefont {Caesar}, \citenamefont {Ceruti}, \citenamefont
  {Chen}, \citenamefont {Crespi}, \citenamefont {Derya}, \citenamefont
  {Doornenbal}, \citenamefont {Fukuda}, \citenamefont {Giaz}, \citenamefont
  {Ieki}, \citenamefont {Kobayashi}, \citenamefont {Kondo}, \citenamefont
  {Koyama}, \citenamefont {Kubo}, \citenamefont {Matsushita}, \citenamefont
  {Million}, \citenamefont {Motobayashi}, \citenamefont {Nakamura},
  \citenamefont {Nishimura}, \citenamefont {Otsu}, \citenamefont {Ozaki},
  \citenamefont {Saito}, \citenamefont {Sakurai}, \citenamefont {Scheit},
  \citenamefont {Schindler}, \citenamefont {Schrock}, \citenamefont {Shiga},
  \citenamefont {Shikata}, \citenamefont {Shimoura}, \citenamefont
  {Steppenbeck}, \citenamefont {Sumikama}, \citenamefont {Takeuchi},
  \citenamefont {Taniuchi}, \citenamefont {Togano}, \citenamefont
  {Tscheuschner}, \citenamefont {Tsubota}, \citenamefont {Wang}, \citenamefont
  {Wimmer},\ and\ \citenamefont {Yoneda}}]{Wie18a}%
  \BibitemOpen
  \bibfield  {author} {\bibinfo {author} {\bibfnamefont {O.}~\bibnamefont
  {Wieland}}, \bibinfo {author} {\bibfnamefont {A.}~\bibnamefont {Bracco}},
  \bibinfo {author} {\bibfnamefont {F.}~\bibnamefont {Camera}}, \bibinfo
  {author} {\bibfnamefont {R.}~\bibnamefont {Avigo}}, \bibinfo {author}
  {\bibfnamefont {H.}~\bibnamefont {Baba}}, \bibinfo {author} {\bibfnamefont
  {N.}~\bibnamefont {Nakatsuka}}, \bibinfo {author} {\bibfnamefont
  {T.}~\bibnamefont {Aumann}}, \bibinfo {author} {\bibfnamefont {S.~R.}\
  \bibnamefont {Banerjee}}, \bibinfo {author} {\bibfnamefont {G.}~\bibnamefont
  {Benzoni}}, \bibinfo {author} {\bibfnamefont {K.}~\bibnamefont {Boretzky}},
  \bibinfo {author} {\bibfnamefont {C.}~\bibnamefont {Caesar}}, \bibinfo
  {author} {\bibfnamefont {S.}~\bibnamefont {Ceruti}}, \bibinfo {author}
  {\bibfnamefont {S.}~\bibnamefont {Chen}}, \bibinfo {author} {\bibfnamefont
  {F.~C.~L.}\ \bibnamefont {Crespi}}, \bibinfo {author} {\bibfnamefont
  {V.}~\bibnamefont {Derya}}, \bibinfo {author} {\bibfnamefont
  {P.}~\bibnamefont {Doornenbal}}, \bibinfo {author} {\bibfnamefont
  {N.}~\bibnamefont {Fukuda}}, \bibinfo {author} {\bibfnamefont
  {A.}~\bibnamefont {Giaz}}, \bibinfo {author} {\bibfnamefont {K.}~\bibnamefont
  {Ieki}}, \bibinfo {author} {\bibfnamefont {N.}~\bibnamefont {Kobayashi}},
  \bibinfo {author} {\bibfnamefont {Y.}~\bibnamefont {Kondo}}, \bibinfo
  {author} {\bibfnamefont {S.}~\bibnamefont {Koyama}}, \bibinfo {author}
  {\bibfnamefont {T.}~\bibnamefont {Kubo}}, \bibinfo {author} {\bibfnamefont
  {M.}~\bibnamefont {Matsushita}}, \bibinfo {author} {\bibfnamefont
  {B.}~\bibnamefont {Million}}, \bibinfo {author} {\bibfnamefont
  {T.}~\bibnamefont {Motobayashi}}, \bibinfo {author} {\bibfnamefont
  {T.}~\bibnamefont {Nakamura}}, \bibinfo {author} {\bibfnamefont
  {M.}~\bibnamefont {Nishimura}}, \bibinfo {author} {\bibfnamefont
  {H.}~\bibnamefont {Otsu}}, \bibinfo {author} {\bibfnamefont {T.}~\bibnamefont
  {Ozaki}}, \bibinfo {author} {\bibfnamefont {A.~T.}\ \bibnamefont {Saito}},
  \bibinfo {author} {\bibfnamefont {H.}~\bibnamefont {Sakurai}}, \bibinfo
  {author} {\bibfnamefont {H.}~\bibnamefont {Scheit}}, \bibinfo {author}
  {\bibfnamefont {F.}~\bibnamefont {Schindler}}, \bibinfo {author}
  {\bibfnamefont {P.}~\bibnamefont {Schrock}}, \bibinfo {author} {\bibfnamefont
  {Y.}~\bibnamefont {Shiga}}, \bibinfo {author} {\bibfnamefont
  {M.}~\bibnamefont {Shikata}}, \bibinfo {author} {\bibfnamefont
  {S.}~\bibnamefont {Shimoura}}, \bibinfo {author} {\bibfnamefont
  {D.}~\bibnamefont {Steppenbeck}}, \bibinfo {author} {\bibfnamefont
  {T.}~\bibnamefont {Sumikama}}, \bibinfo {author} {\bibfnamefont
  {S.}~\bibnamefont {Takeuchi}}, \bibinfo {author} {\bibfnamefont
  {R.}~\bibnamefont {Taniuchi}}, \bibinfo {author} {\bibfnamefont
  {Y.}~\bibnamefont {Togano}}, \bibinfo {author} {\bibfnamefont
  {J.}~\bibnamefont {Tscheuschner}}, \bibinfo {author} {\bibfnamefont
  {J.}~\bibnamefont {Tsubota}}, \bibinfo {author} {\bibfnamefont
  {H.}~\bibnamefont {Wang}}, \bibinfo {author} {\bibfnamefont {K.}~\bibnamefont
  {Wimmer}}, \ and\ \bibinfo {author} {\bibfnamefont {K.}~\bibnamefont
  {Yoneda}},\ }\href {\doibase 10.1103/PhysRevC.98.064313} {\bibfield
  {journal} {\bibinfo  {journal} {Phys. Rev. C}\ }\textbf {\bibinfo {volume}
  {98}},\ \bibinfo {pages} {064313} (\bibinfo {year} {2018})}\BibitemShut
  {NoStop}%
\bibitem [{\citenamefont {Goriely}(1998)}]{Gor98a}%
  \BibitemOpen
  \bibfield  {author} {\bibinfo {author} {\bibfnamefont {S.}~\bibnamefont
  {Goriely}},\ }\href {\doibase https://doi.org/10.1016/S0370-2693(98)00907-1}
  {\bibfield  {journal} {\bibinfo  {journal} {Physics Letters B}\ }\textbf
  {\bibinfo {volume} {436}},\ \bibinfo {pages} {10 } (\bibinfo {year}
  {1998})}\BibitemShut {NoStop}%
\bibitem [{\citenamefont {Litvinova}\ \emph
  {et~al.}(2009{\natexlab{a}})\citenamefont {Litvinova}, \citenamefont {Loens},
  \citenamefont {Langanke}, \citenamefont {Martínez-Pinedo}, \citenamefont
  {Rauscher}, \citenamefont {Ring}, \citenamefont {Thielemann},\ and\
  \citenamefont {Tselyaev}}]{Lit09b}%
  \BibitemOpen
  \bibfield  {author} {\bibinfo {author} {\bibfnamefont {E.}~\bibnamefont
  {Litvinova}}, \bibinfo {author} {\bibfnamefont {H.}~\bibnamefont {Loens}},
  \bibinfo {author} {\bibfnamefont {K.}~\bibnamefont {Langanke}}, \bibinfo
  {author} {\bibfnamefont {G.}~\bibnamefont {Martínez-Pinedo}}, \bibinfo
  {author} {\bibfnamefont {T.}~\bibnamefont {Rauscher}}, \bibinfo {author}
  {\bibfnamefont {P.}~\bibnamefont {Ring}}, \bibinfo {author} {\bibfnamefont
  {F.-K.}\ \bibnamefont {Thielemann}}, \ and\ \bibinfo {author} {\bibfnamefont
  {V.}~\bibnamefont {Tselyaev}},\ }\href {\doibase
  https://doi.org/10.1016/j.nuclphysa.2009.03.009} {\bibfield  {journal}
  {\bibinfo  {journal} {Nuclear Physics A}\ }\textbf {\bibinfo {volume}
  {823}},\ \bibinfo {pages} {26 } (\bibinfo {year}
  {2009}{\natexlab{a}})}\BibitemShut {NoStop}%
\bibitem [{\citenamefont {Tsoneva}\ \emph {et~al.}(2015)\citenamefont
  {Tsoneva}, \citenamefont {Goriely}, \citenamefont {Lenske},\ and\
  \citenamefont {Schwengner}}]{Tso15a}%
  \BibitemOpen
  \bibfield  {author} {\bibinfo {author} {\bibfnamefont {N.}~\bibnamefont
  {Tsoneva}}, \bibinfo {author} {\bibfnamefont {S.}~\bibnamefont {Goriely}},
  \bibinfo {author} {\bibfnamefont {H.}~\bibnamefont {Lenske}}, \ and\ \bibinfo
  {author} {\bibfnamefont {R.}~\bibnamefont {Schwengner}},\ }\href {\doibase
  10.1103/PhysRevC.91.044318} {\bibfield  {journal} {\bibinfo  {journal} {Phys.
  Rev. C}\ }\textbf {\bibinfo {volume} {91}},\ \bibinfo {pages} {044318}
  (\bibinfo {year} {2015})}\BibitemShut {NoStop}%
\bibitem [{\citenamefont {Larsen}\ \emph {et~al.}(2019)\citenamefont {Larsen},
  \citenamefont {Spyrou}, \citenamefont {Liddick},\ and\ \citenamefont
  {Guttormsen}}]{Lar19a}%
  \BibitemOpen
  \bibfield  {author} {\bibinfo {author} {\bibfnamefont {A.}~\bibnamefont
  {Larsen}}, \bibinfo {author} {\bibfnamefont {A.}~\bibnamefont {Spyrou}},
  \bibinfo {author} {\bibfnamefont {S.}~\bibnamefont {Liddick}}, \ and\
  \bibinfo {author} {\bibfnamefont {M.}~\bibnamefont {Guttormsen}},\ }\href
  {\doibase https://doi.org/10.1016/j.ppnp.2019.04.002} {\bibfield  {journal}
  {\bibinfo  {journal} {Progress in Particle and Nuclear Physics}\ }\textbf
  {\bibinfo {volume} {107}},\ \bibinfo {pages} {69 } (\bibinfo {year}
  {2019})}\BibitemShut {NoStop}%
\bibitem [{\citenamefont {Angell}\ \emph {et~al.}(2012)\citenamefont {Angell},
  \citenamefont {Hammond}, \citenamefont {Karwowski}, \citenamefont {Kelley},
  \citenamefont {Krti\ifmmode~\check{c}\else \v{c}\fi{}ka}, \citenamefont
  {Kwan}, \citenamefont {Makinaga},\ and\ \citenamefont {Rusev}}]{Ang12a}%
  \BibitemOpen
  \bibfield  {author} {\bibinfo {author} {\bibfnamefont {C.~T.}\ \bibnamefont
  {Angell}}, \bibinfo {author} {\bibfnamefont {S.~L.}\ \bibnamefont {Hammond}},
  \bibinfo {author} {\bibfnamefont {H.~J.}\ \bibnamefont {Karwowski}}, \bibinfo
  {author} {\bibfnamefont {J.~H.}\ \bibnamefont {Kelley}}, \bibinfo {author}
  {\bibfnamefont {M.}~\bibnamefont {Krti\ifmmode~\check{c}\else \v{c}\fi{}ka}},
  \bibinfo {author} {\bibfnamefont {E.}~\bibnamefont {Kwan}}, \bibinfo {author}
  {\bibfnamefont {A.}~\bibnamefont {Makinaga}}, \ and\ \bibinfo {author}
  {\bibfnamefont {G.}~\bibnamefont {Rusev}},\ }\href {\doibase
  10.1103/PhysRevC.86.051302} {\bibfield  {journal} {\bibinfo  {journal} {Phys.
  Rev. C}\ }\textbf {\bibinfo {volume} {86}},\ \bibinfo {pages} {051302}
  (\bibinfo {year} {2012})}\BibitemShut {NoStop}%
\bibitem [{\citenamefont {Bassauer}\ \emph {et~al.}(2016)\citenamefont
  {Bassauer}, \citenamefont {von Neumann-Cosel},\ and\ \citenamefont
  {Tamii}}]{Bas16a}%
  \BibitemOpen
  \bibfield  {author} {\bibinfo {author} {\bibfnamefont {S.}~\bibnamefont
  {Bassauer}}, \bibinfo {author} {\bibfnamefont {P.}~\bibnamefont {von
  Neumann-Cosel}}, \ and\ \bibinfo {author} {\bibfnamefont {A.}~\bibnamefont
  {Tamii}},\ }\href {\doibase 10.1103/PhysRevC.94.054313} {\bibfield  {journal}
  {\bibinfo  {journal} {Phys. Rev. C}\ }\textbf {\bibinfo {volume} {94}},\
  \bibinfo {pages} {054313} (\bibinfo {year} {2016})}\BibitemShut {NoStop}%
\bibitem [{\citenamefont {Guttormsen}\ \emph {et~al.}(2016)\citenamefont
  {Guttormsen}, \citenamefont {Larsen}, \citenamefont {G\"orgen}, \citenamefont
  {Renstr\o{}m}, \citenamefont {Siem}, \citenamefont {Tornyi},\ and\
  \citenamefont {Tveten}}]{Gut16a}%
  \BibitemOpen
  \bibfield  {author} {\bibinfo {author} {\bibfnamefont {M.}~\bibnamefont
  {Guttormsen}}, \bibinfo {author} {\bibfnamefont {A.~C.}\ \bibnamefont
  {Larsen}}, \bibinfo {author} {\bibfnamefont {A.}~\bibnamefont {G\"orgen}},
  \bibinfo {author} {\bibfnamefont {T.}~\bibnamefont {Renstr\o{}m}}, \bibinfo
  {author} {\bibfnamefont {S.}~\bibnamefont {Siem}}, \bibinfo {author}
  {\bibfnamefont {T.~G.}\ \bibnamefont {Tornyi}}, \ and\ \bibinfo {author}
  {\bibfnamefont {G.~M.}\ \bibnamefont {Tveten}},\ }\href {\doibase
  10.1103/PhysRevLett.116.012502} {\bibfield  {journal} {\bibinfo  {journal}
  {Phys. Rev. Lett.}\ }\textbf {\bibinfo {volume} {116}},\ \bibinfo {pages}
  {012502} (\bibinfo {year} {2016})}\BibitemShut {NoStop}%
\bibitem [{\citenamefont {Martin}\ \emph {et~al.}(2017)\citenamefont {Martin},
  \citenamefont {von Neumann-Cosel}, \citenamefont {Tamii}, \citenamefont
  {Aoi}, \citenamefont {Bassauer}, \citenamefont {Bertulani}, \citenamefont
  {Carter}, \citenamefont {Donaldson}, \citenamefont {Fujita}, \citenamefont
  {Fujita}, \citenamefont {Hashimoto}, \citenamefont {Hatanaka}, \citenamefont
  {Ito}, \citenamefont {Krugmann}, \citenamefont {Liu}, \citenamefont {Maeda},
  \citenamefont {Miki}, \citenamefont {Neveling}, \citenamefont {Pietralla},
  \citenamefont {Poltoratska}, \citenamefont {Ponomarev}, \citenamefont
  {Richter}, \citenamefont {Shima}, \citenamefont {Yamamoto},\ and\
  \citenamefont {Zweidinger}}]{Mar17a}%
  \BibitemOpen
  \bibfield  {author} {\bibinfo {author} {\bibfnamefont {D.}~\bibnamefont
  {Martin}}, \bibinfo {author} {\bibfnamefont {P.}~\bibnamefont {von
  Neumann-Cosel}}, \bibinfo {author} {\bibfnamefont {A.}~\bibnamefont {Tamii}},
  \bibinfo {author} {\bibfnamefont {N.}~\bibnamefont {Aoi}}, \bibinfo {author}
  {\bibfnamefont {S.}~\bibnamefont {Bassauer}}, \bibinfo {author}
  {\bibfnamefont {C.~A.}\ \bibnamefont {Bertulani}}, \bibinfo {author}
  {\bibfnamefont {J.}~\bibnamefont {Carter}}, \bibinfo {author} {\bibfnamefont
  {L.}~\bibnamefont {Donaldson}}, \bibinfo {author} {\bibfnamefont
  {H.}~\bibnamefont {Fujita}}, \bibinfo {author} {\bibfnamefont
  {Y.}~\bibnamefont {Fujita}}, \bibinfo {author} {\bibfnamefont
  {T.}~\bibnamefont {Hashimoto}}, \bibinfo {author} {\bibfnamefont
  {K.}~\bibnamefont {Hatanaka}}, \bibinfo {author} {\bibfnamefont
  {T.}~\bibnamefont {Ito}}, \bibinfo {author} {\bibfnamefont {A.}~\bibnamefont
  {Krugmann}}, \bibinfo {author} {\bibfnamefont {B.}~\bibnamefont {Liu}},
  \bibinfo {author} {\bibfnamefont {Y.}~\bibnamefont {Maeda}}, \bibinfo
  {author} {\bibfnamefont {K.}~\bibnamefont {Miki}}, \bibinfo {author}
  {\bibfnamefont {R.}~\bibnamefont {Neveling}}, \bibinfo {author}
  {\bibfnamefont {N.}~\bibnamefont {Pietralla}}, \bibinfo {author}
  {\bibfnamefont {I.}~\bibnamefont {Poltoratska}}, \bibinfo {author}
  {\bibfnamefont {V.~Y.}\ \bibnamefont {Ponomarev}}, \bibinfo {author}
  {\bibfnamefont {A.}~\bibnamefont {Richter}}, \bibinfo {author} {\bibfnamefont
  {T.}~\bibnamefont {Shima}}, \bibinfo {author} {\bibfnamefont
  {T.}~\bibnamefont {Yamamoto}}, \ and\ \bibinfo {author} {\bibfnamefont
  {M.}~\bibnamefont {Zweidinger}},\ }\href {\doibase
  10.1103/PhysRevLett.119.182503} {\bibfield  {journal} {\bibinfo  {journal}
  {Phys. Rev. Lett.}\ }\textbf {\bibinfo {volume} {119}},\ \bibinfo {pages}
  {182503} (\bibinfo {year} {2017})}\BibitemShut {NoStop}%
\bibitem [{\citenamefont {Campo}\ \emph {et~al.}(2018)\citenamefont {Campo},
  \citenamefont {Guttormsen}, \citenamefont {Garrote}, \citenamefont {Eriksen},
  \citenamefont {Giacoppo}, \citenamefont {G\"orgen}, \citenamefont
  {Hadynska-Klek}, \citenamefont {Klintefjord}, \citenamefont {Larsen},
  \citenamefont {Renstr\o{}m}, \citenamefont {Sahin}, \citenamefont {Siem},
  \citenamefont {Springer}, \citenamefont {Tornyi},\ and\ \citenamefont
  {Tveten}}]{Cam18a}%
  \BibitemOpen
  \bibfield  {author} {\bibinfo {author} {\bibfnamefont {L.~C.}\ \bibnamefont
  {Campo}}, \bibinfo {author} {\bibfnamefont {M.}~\bibnamefont {Guttormsen}},
  \bibinfo {author} {\bibfnamefont {F.~L.~B.}\ \bibnamefont {Garrote}},
  \bibinfo {author} {\bibfnamefont {T.~K.}\ \bibnamefont {Eriksen}}, \bibinfo
  {author} {\bibfnamefont {F.}~\bibnamefont {Giacoppo}}, \bibinfo {author}
  {\bibfnamefont {A.}~\bibnamefont {G\"orgen}}, \bibinfo {author}
  {\bibfnamefont {K.}~\bibnamefont {Hadynska-Klek}}, \bibinfo {author}
  {\bibfnamefont {M.}~\bibnamefont {Klintefjord}}, \bibinfo {author}
  {\bibfnamefont {A.~C.}\ \bibnamefont {Larsen}}, \bibinfo {author}
  {\bibfnamefont {T.}~\bibnamefont {Renstr\o{}m}}, \bibinfo {author}
  {\bibfnamefont {E.}~\bibnamefont {Sahin}}, \bibinfo {author} {\bibfnamefont
  {S.}~\bibnamefont {Siem}}, \bibinfo {author} {\bibfnamefont {A.}~\bibnamefont
  {Springer}}, \bibinfo {author} {\bibfnamefont {T.~G.}\ \bibnamefont
  {Tornyi}}, \ and\ \bibinfo {author} {\bibfnamefont {G.~M.}\ \bibnamefont
  {Tveten}},\ }\href {\doibase 10.1103/PhysRevC.98.054303} {\bibfield
  {journal} {\bibinfo  {journal} {Phys. Rev. C}\ }\textbf {\bibinfo {volume}
  {98}},\ \bibinfo {pages} {054303} (\bibinfo {year} {2018})}\BibitemShut
  {NoStop}%
\bibitem [{\citenamefont {Isaak}\ \emph {et~al.}(2019)\citenamefont {Isaak},
  \citenamefont {Savran}, \citenamefont {Löher}, \citenamefont {Beck},
  \citenamefont {Bhike}, \citenamefont {Gayer}, \citenamefont {Krishichayan},
  \citenamefont {Pietralla}, \citenamefont {Scheck}, \citenamefont {Tornow},
  \citenamefont {Werner}, \citenamefont {Zilges},\ and\ \citenamefont
  {Zweidinger}}]{Isa19a}%
  \BibitemOpen
  \bibfield  {author} {\bibinfo {author} {\bibfnamefont {J.}~\bibnamefont
  {Isaak}}, \bibinfo {author} {\bibfnamefont {D.}~\bibnamefont {Savran}},
  \bibinfo {author} {\bibfnamefont {B.}~\bibnamefont {Löher}}, \bibinfo
  {author} {\bibfnamefont {T.}~\bibnamefont {Beck}}, \bibinfo {author}
  {\bibfnamefont {M.}~\bibnamefont {Bhike}}, \bibinfo {author} {\bibfnamefont
  {U.}~\bibnamefont {Gayer}}, \bibinfo {author} {\bibnamefont {Krishichayan}},
  \bibinfo {author} {\bibfnamefont {N.}~\bibnamefont {Pietralla}}, \bibinfo
  {author} {\bibfnamefont {M.}~\bibnamefont {Scheck}}, \bibinfo {author}
  {\bibfnamefont {W.}~\bibnamefont {Tornow}}, \bibinfo {author} {\bibfnamefont
  {V.}~\bibnamefont {Werner}}, \bibinfo {author} {\bibfnamefont
  {A.}~\bibnamefont {Zilges}}, \ and\ \bibinfo {author} {\bibfnamefont
  {M.}~\bibnamefont {Zweidinger}},\ }\href {\doibase
  https://doi.org/10.1016/j.physletb.2018.11.038} {\bibfield  {journal}
  {\bibinfo  {journal} {Physics Letters B}\ }\textbf {\bibinfo {volume}
  {788}},\ \bibinfo {pages} {225} (\bibinfo {year} {2019})}\BibitemShut
  {NoStop}%
\bibitem [{\citenamefont {Simbirtseva}\ \emph {et~al.}(2020)\citenamefont
  {Simbirtseva}, \citenamefont {Krti\ifmmode~\check{c}\else \v{c}\fi{}ka},
  \citenamefont {Casten}, \citenamefont {Couture}, \citenamefont {Furman},
  \citenamefont {Knapov\'a}, \citenamefont {O'Donnell}, \citenamefont {Rusev},
  \citenamefont {Ullmann},\ and\ \citenamefont {Valenta}}]{Sim20a}%
  \BibitemOpen
  \bibfield  {author} {\bibinfo {author} {\bibfnamefont {N.}~\bibnamefont
  {Simbirtseva}}, \bibinfo {author} {\bibfnamefont {M.}~\bibnamefont
  {Krti\ifmmode~\check{c}\else \v{c}\fi{}ka}}, \bibinfo {author} {\bibfnamefont
  {R.~F.}\ \bibnamefont {Casten}}, \bibinfo {author} {\bibfnamefont
  {A.}~\bibnamefont {Couture}}, \bibinfo {author} {\bibfnamefont {W.~I.}\
  \bibnamefont {Furman}}, \bibinfo {author} {\bibfnamefont {I.}~\bibnamefont
  {Knapov\'a}}, \bibinfo {author} {\bibfnamefont {J.~M.}\ \bibnamefont
  {O'Donnell}}, \bibinfo {author} {\bibfnamefont {G.}~\bibnamefont {Rusev}},
  \bibinfo {author} {\bibfnamefont {J.~L.}\ \bibnamefont {Ullmann}}, \ and\
  \bibinfo {author} {\bibfnamefont {S.}~\bibnamefont {Valenta}},\ }\href
  {\doibase 10.1103/PhysRevC.101.024302} {\bibfield  {journal} {\bibinfo
  {journal} {Phys. Rev. C}\ }\textbf {\bibinfo {volume} {101}},\ \bibinfo
  {pages} {024302} (\bibinfo {year} {2020})}\BibitemShut {NoStop}%
\bibitem [{\citenamefont {Scholz}\ \emph {et~al.}(2020)\citenamefont {Scholz},
  \citenamefont {Guttormsen}, \citenamefont {Heim}, \citenamefont {Larsen},
  \citenamefont {Mayer}, \citenamefont {Savran}, \citenamefont {Spieker},
  \citenamefont {Tveten}, \citenamefont {Voinov}, \citenamefont {Wilhelmy},
  \citenamefont {Zeiser},\ and\ \citenamefont {Zilges}}]{Sch20a}%
  \BibitemOpen
  \bibfield  {author} {\bibinfo {author} {\bibfnamefont {P.}~\bibnamefont
  {Scholz}}, \bibinfo {author} {\bibfnamefont {M.}~\bibnamefont {Guttormsen}},
  \bibinfo {author} {\bibfnamefont {F.}~\bibnamefont {Heim}}, \bibinfo {author}
  {\bibfnamefont {A.~C.}\ \bibnamefont {Larsen}}, \bibinfo {author}
  {\bibfnamefont {J.}~\bibnamefont {Mayer}}, \bibinfo {author} {\bibfnamefont
  {D.}~\bibnamefont {Savran}}, \bibinfo {author} {\bibfnamefont
  {M.}~\bibnamefont {Spieker}}, \bibinfo {author} {\bibfnamefont {G.~M.}\
  \bibnamefont {Tveten}}, \bibinfo {author} {\bibfnamefont {A.~V.}\
  \bibnamefont {Voinov}}, \bibinfo {author} {\bibfnamefont {J.}~\bibnamefont
  {Wilhelmy}}, \bibinfo {author} {\bibfnamefont {F.}~\bibnamefont {Zeiser}}, \
  and\ \bibinfo {author} {\bibfnamefont {A.}~\bibnamefont {Zilges}},\ }\href
  {\doibase 10.1103/PhysRevC.101.045806} {\bibfield  {journal} {\bibinfo
  {journal} {Phys. Rev. C}\ }\textbf {\bibinfo {volume} {101}},\ \bibinfo
  {pages} {045806} (\bibinfo {year} {2020})}\BibitemShut {NoStop}%
\bibitem [{\citenamefont {Savran}\ \emph {et~al.}(2006)\citenamefont {Savran},
  \citenamefont {Babilon}, \citenamefont {van~den Berg}, \citenamefont
  {Harakeh}, \citenamefont {Hasper}, \citenamefont {Matic}, \citenamefont
  {W\"ortche},\ and\ \citenamefont {Zilges}}]{Sav06a}%
  \BibitemOpen
  \bibfield  {author} {\bibinfo {author} {\bibfnamefont {D.}~\bibnamefont
  {Savran}}, \bibinfo {author} {\bibfnamefont {M.}~\bibnamefont {Babilon}},
  \bibinfo {author} {\bibfnamefont {A.~M.}\ \bibnamefont {van~den Berg}},
  \bibinfo {author} {\bibfnamefont {M.~N.}\ \bibnamefont {Harakeh}}, \bibinfo
  {author} {\bibfnamefont {J.}~\bibnamefont {Hasper}}, \bibinfo {author}
  {\bibfnamefont {A.}~\bibnamefont {Matic}}, \bibinfo {author} {\bibfnamefont
  {H.~J.}\ \bibnamefont {W\"ortche}}, \ and\ \bibinfo {author} {\bibfnamefont
  {A.}~\bibnamefont {Zilges}},\ }\href {\doibase 10.1103/PhysRevLett.97.172502}
  {\bibfield  {journal} {\bibinfo  {journal} {Phys. Rev. Lett.}\ }\textbf
  {\bibinfo {volume} {97}},\ \bibinfo {pages} {172502} (\bibinfo {year}
  {2006})}\BibitemShut {NoStop}%
\bibitem [{\citenamefont {Endres}\ \emph {et~al.}(2009)\citenamefont {Endres},
  \citenamefont {Savran}, \citenamefont {Berg}, \citenamefont {Dendooven},
  \citenamefont {Fritzsche}, \citenamefont {Harakeh}, \citenamefont {Hasper},
  \citenamefont {W\"ortche},\ and\ \citenamefont {Zilges}}]{End09a}%
  \BibitemOpen
  \bibfield  {author} {\bibinfo {author} {\bibfnamefont {J.}~\bibnamefont
  {Endres}}, \bibinfo {author} {\bibfnamefont {D.}~\bibnamefont {Savran}},
  \bibinfo {author} {\bibfnamefont {A.~M. v.~d.}\ \bibnamefont {Berg}},
  \bibinfo {author} {\bibfnamefont {P.}~\bibnamefont {Dendooven}}, \bibinfo
  {author} {\bibfnamefont {M.}~\bibnamefont {Fritzsche}}, \bibinfo {author}
  {\bibfnamefont {M.~N.}\ \bibnamefont {Harakeh}}, \bibinfo {author}
  {\bibfnamefont {J.}~\bibnamefont {Hasper}}, \bibinfo {author} {\bibfnamefont
  {H.~J.}\ \bibnamefont {W\"ortche}}, \ and\ \bibinfo {author} {\bibfnamefont
  {A.}~\bibnamefont {Zilges}},\ }\href {\doibase 10.1103/PhysRevC.80.034302}
  {\bibfield  {journal} {\bibinfo  {journal} {Phys. Rev. C}\ }\textbf {\bibinfo
  {volume} {80}},\ \bibinfo {pages} {034302} (\bibinfo {year}
  {2009})}\BibitemShut {NoStop}%
\bibitem [{\citenamefont {Derya}\ \emph {et~al.}(2013)\citenamefont {Derya},
  \citenamefont {Endres}, \citenamefont {Elvers}, \citenamefont {Harakeh},
  \citenamefont {Pietralla}, \citenamefont {Romig}, \citenamefont {Savran},
  \citenamefont {Scheck}, \citenamefont {Siebenhühner}, \citenamefont
  {Stoica}, \citenamefont {Wörtche},\ and\ \citenamefont {Zilges}}]{Der13a}%
  \BibitemOpen
  \bibfield  {author} {\bibinfo {author} {\bibfnamefont {V.}~\bibnamefont
  {Derya}}, \bibinfo {author} {\bibfnamefont {J.}~\bibnamefont {Endres}},
  \bibinfo {author} {\bibfnamefont {M.}~\bibnamefont {Elvers}}, \bibinfo
  {author} {\bibfnamefont {M.}~\bibnamefont {Harakeh}}, \bibinfo {author}
  {\bibfnamefont {N.}~\bibnamefont {Pietralla}}, \bibinfo {author}
  {\bibfnamefont {C.}~\bibnamefont {Romig}}, \bibinfo {author} {\bibfnamefont
  {D.}~\bibnamefont {Savran}}, \bibinfo {author} {\bibfnamefont
  {M.}~\bibnamefont {Scheck}}, \bibinfo {author} {\bibfnamefont
  {F.}~\bibnamefont {Siebenhühner}}, \bibinfo {author} {\bibfnamefont
  {V.}~\bibnamefont {Stoica}}, \bibinfo {author} {\bibfnamefont
  {H.}~\bibnamefont {Wörtche}}, \ and\ \bibinfo {author} {\bibfnamefont
  {A.}~\bibnamefont {Zilges}},\ }\href {\doibase
  https://doi.org/10.1016/j.nuclphysa.2013.02.018} {\bibfield  {journal}
  {\bibinfo  {journal} {Nuclear Physics A}\ }\textbf {\bibinfo {volume}
  {906}},\ \bibinfo {pages} {94 } (\bibinfo {year} {2013})}\BibitemShut
  {NoStop}%
\bibitem [{\citenamefont {Derya}\ \emph {et~al.}(2014)\citenamefont {Derya},
  \citenamefont {Savran}, \citenamefont {Endres}, \citenamefont {Harakeh},
  \citenamefont {Hergert}, \citenamefont {Kelley}, \citenamefont
  {Papakonstantinou}, \citenamefont {Pietralla}, \citenamefont {Ponomarev},
  \citenamefont {Roth}, \citenamefont {Rusev}, \citenamefont {Tonchev},
  \citenamefont {Tornow}, \citenamefont {Wörtche},\ and\ \citenamefont
  {Zilges}}]{Der14a}%
  \BibitemOpen
  \bibfield  {author} {\bibinfo {author} {\bibfnamefont {V.}~\bibnamefont
  {Derya}}, \bibinfo {author} {\bibfnamefont {D.}~\bibnamefont {Savran}},
  \bibinfo {author} {\bibfnamefont {J.}~\bibnamefont {Endres}}, \bibinfo
  {author} {\bibfnamefont {M.}~\bibnamefont {Harakeh}}, \bibinfo {author}
  {\bibfnamefont {H.}~\bibnamefont {Hergert}}, \bibinfo {author} {\bibfnamefont
  {J.}~\bibnamefont {Kelley}}, \bibinfo {author} {\bibfnamefont
  {P.}~\bibnamefont {Papakonstantinou}}, \bibinfo {author} {\bibfnamefont
  {N.}~\bibnamefont {Pietralla}}, \bibinfo {author} {\bibfnamefont
  {V.}~\bibnamefont {Ponomarev}}, \bibinfo {author} {\bibfnamefont
  {R.}~\bibnamefont {Roth}}, \bibinfo {author} {\bibfnamefont {G.}~\bibnamefont
  {Rusev}}, \bibinfo {author} {\bibfnamefont {A.}~\bibnamefont {Tonchev}},
  \bibinfo {author} {\bibfnamefont {W.}~\bibnamefont {Tornow}}, \bibinfo
  {author} {\bibfnamefont {H.}~\bibnamefont {Wörtche}}, \ and\ \bibinfo
  {author} {\bibfnamefont {A.}~\bibnamefont {Zilges}},\ }\href {\doibase
  https://doi.org/10.1016/j.physletb.2014.01.050} {\bibfield  {journal}
  {\bibinfo  {journal} {Physics Letters B}\ }\textbf {\bibinfo {volume}
  {730}},\ \bibinfo {pages} {288 } (\bibinfo {year} {2014})}\BibitemShut
  {NoStop}%
\bibitem [{\citenamefont {Crespi}\ \emph {et~al.}(2014)\citenamefont {Crespi},
  \citenamefont {Bracco}, \citenamefont {Nicolini}, \citenamefont {Mengoni},
  \citenamefont {Pellegri}, \citenamefont {Lanza}, \citenamefont {Leoni},
  \citenamefont {Maj}, \citenamefont {Kmiecik}, \citenamefont {Avigo},
  \citenamefont {Benzoni}, \citenamefont {Blasi}, \citenamefont {Boiano},
  \citenamefont {Bottoni}, \citenamefont {Brambilla}, \citenamefont {Camera},
  \citenamefont {Ceruti}, \citenamefont {Giaz}, \citenamefont {Million},
  \citenamefont {Morales}, \citenamefont {Vandone}, \citenamefont {Wieland},
  \citenamefont {Bednarczyk}, \citenamefont {Ciema\l{}a}, \citenamefont
  {Grebosz}, \citenamefont {Krzysiek}, \citenamefont {Mazurek}, \citenamefont
  {Zieblinski}, \citenamefont {Bazzacco}, \citenamefont {Bellato},
  \citenamefont {Birkenbach}, \citenamefont {Bortolato}, \citenamefont
  {Calore}, \citenamefont {Cederwall}, \citenamefont {Charles}, \citenamefont
  {de~Angelis}, \citenamefont {D\'esesquelles}, \citenamefont {Eberth},
  \citenamefont {Farnea}, \citenamefont {Gadea}, \citenamefont {G\"orgen},
  \citenamefont {Gottardo}, \citenamefont {Isocrate}, \citenamefont {Jolie},
  \citenamefont {Jungclaus}, \citenamefont {Karkour}, \citenamefont {Korten},
  \citenamefont {Menegazzo}, \citenamefont {Michelagnoli}, \citenamefont
  {Molini}, \citenamefont {Napoli}, \citenamefont {Pullia}, \citenamefont
  {Recchia}, \citenamefont {Reiter}, \citenamefont {Rosso}, \citenamefont
  {Sahin}, \citenamefont {Salsac}, \citenamefont {Siebeck}, \citenamefont
  {Siem}, \citenamefont {Simpson}, \citenamefont {S\"oderstr\"om},
  \citenamefont {Stezowski}, \citenamefont {Theisen}, \citenamefont {Ur},\ and\
  \citenamefont {Valiente-Dob\'on}}]{Cre14a}%
  \BibitemOpen
  \bibfield  {author} {\bibinfo {author} {\bibfnamefont {F.~C.~L.}\
  \bibnamefont {Crespi}}, \bibinfo {author} {\bibfnamefont {A.}~\bibnamefont
  {Bracco}}, \bibinfo {author} {\bibfnamefont {R.}~\bibnamefont {Nicolini}},
  \bibinfo {author} {\bibfnamefont {D.}~\bibnamefont {Mengoni}}, \bibinfo
  {author} {\bibfnamefont {L.}~\bibnamefont {Pellegri}}, \bibinfo {author}
  {\bibfnamefont {E.~G.}\ \bibnamefont {Lanza}}, \bibinfo {author}
  {\bibfnamefont {S.}~\bibnamefont {Leoni}}, \bibinfo {author} {\bibfnamefont
  {A.}~\bibnamefont {Maj}}, \bibinfo {author} {\bibfnamefont {M.}~\bibnamefont
  {Kmiecik}}, \bibinfo {author} {\bibfnamefont {R.}~\bibnamefont {Avigo}},
  \bibinfo {author} {\bibfnamefont {G.}~\bibnamefont {Benzoni}}, \bibinfo
  {author} {\bibfnamefont {N.}~\bibnamefont {Blasi}}, \bibinfo {author}
  {\bibfnamefont {C.}~\bibnamefont {Boiano}}, \bibinfo {author} {\bibfnamefont
  {S.}~\bibnamefont {Bottoni}}, \bibinfo {author} {\bibfnamefont
  {S.}~\bibnamefont {Brambilla}}, \bibinfo {author} {\bibfnamefont
  {F.}~\bibnamefont {Camera}}, \bibinfo {author} {\bibfnamefont
  {S.}~\bibnamefont {Ceruti}}, \bibinfo {author} {\bibfnamefont
  {A.}~\bibnamefont {Giaz}}, \bibinfo {author} {\bibfnamefont {B.}~\bibnamefont
  {Million}}, \bibinfo {author} {\bibfnamefont {A.~I.}\ \bibnamefont
  {Morales}}, \bibinfo {author} {\bibfnamefont {V.}~\bibnamefont {Vandone}},
  \bibinfo {author} {\bibfnamefont {O.}~\bibnamefont {Wieland}}, \bibinfo
  {author} {\bibfnamefont {P.}~\bibnamefont {Bednarczyk}}, \bibinfo {author}
  {\bibfnamefont {M.}~\bibnamefont {Ciema\l{}a}}, \bibinfo {author}
  {\bibfnamefont {J.}~\bibnamefont {Grebosz}}, \bibinfo {author} {\bibfnamefont
  {M.}~\bibnamefont {Krzysiek}}, \bibinfo {author} {\bibfnamefont
  {K.}~\bibnamefont {Mazurek}}, \bibinfo {author} {\bibfnamefont
  {M.}~\bibnamefont {Zieblinski}}, \bibinfo {author} {\bibfnamefont
  {D.}~\bibnamefont {Bazzacco}}, \bibinfo {author} {\bibfnamefont
  {M.}~\bibnamefont {Bellato}}, \bibinfo {author} {\bibfnamefont
  {B.}~\bibnamefont {Birkenbach}}, \bibinfo {author} {\bibfnamefont
  {D.}~\bibnamefont {Bortolato}}, \bibinfo {author} {\bibfnamefont
  {E.}~\bibnamefont {Calore}}, \bibinfo {author} {\bibfnamefont
  {B.}~\bibnamefont {Cederwall}}, \bibinfo {author} {\bibfnamefont
  {L.}~\bibnamefont {Charles}}, \bibinfo {author} {\bibfnamefont
  {G.}~\bibnamefont {de~Angelis}}, \bibinfo {author} {\bibfnamefont
  {P.}~\bibnamefont {D\'esesquelles}}, \bibinfo {author} {\bibfnamefont
  {J.}~\bibnamefont {Eberth}}, \bibinfo {author} {\bibfnamefont
  {E.}~\bibnamefont {Farnea}}, \bibinfo {author} {\bibfnamefont
  {A.}~\bibnamefont {Gadea}}, \bibinfo {author} {\bibfnamefont
  {A.}~\bibnamefont {G\"orgen}}, \bibinfo {author} {\bibfnamefont
  {A.}~\bibnamefont {Gottardo}}, \bibinfo {author} {\bibfnamefont
  {R.}~\bibnamefont {Isocrate}}, \bibinfo {author} {\bibfnamefont
  {J.}~\bibnamefont {Jolie}}, \bibinfo {author} {\bibfnamefont
  {A.}~\bibnamefont {Jungclaus}}, \bibinfo {author} {\bibfnamefont
  {N.}~\bibnamefont {Karkour}}, \bibinfo {author} {\bibfnamefont
  {W.}~\bibnamefont {Korten}}, \bibinfo {author} {\bibfnamefont
  {R.}~\bibnamefont {Menegazzo}}, \bibinfo {author} {\bibfnamefont
  {C.}~\bibnamefont {Michelagnoli}}, \bibinfo {author} {\bibfnamefont
  {P.}~\bibnamefont {Molini}}, \bibinfo {author} {\bibfnamefont {D.~R.}\
  \bibnamefont {Napoli}}, \bibinfo {author} {\bibfnamefont {A.}~\bibnamefont
  {Pullia}}, \bibinfo {author} {\bibfnamefont {F.}~\bibnamefont {Recchia}},
  \bibinfo {author} {\bibfnamefont {P.}~\bibnamefont {Reiter}}, \bibinfo
  {author} {\bibfnamefont {D.}~\bibnamefont {Rosso}}, \bibinfo {author}
  {\bibfnamefont {E.}~\bibnamefont {Sahin}}, \bibinfo {author} {\bibfnamefont
  {M.~D.}\ \bibnamefont {Salsac}}, \bibinfo {author} {\bibfnamefont
  {B.}~\bibnamefont {Siebeck}}, \bibinfo {author} {\bibfnamefont
  {S.}~\bibnamefont {Siem}}, \bibinfo {author} {\bibfnamefont {J.}~\bibnamefont
  {Simpson}}, \bibinfo {author} {\bibfnamefont {P.-A.}\ \bibnamefont
  {S\"oderstr\"om}}, \bibinfo {author} {\bibfnamefont {O.}~\bibnamefont
  {Stezowski}}, \bibinfo {author} {\bibfnamefont {C.}~\bibnamefont {Theisen}},
  \bibinfo {author} {\bibfnamefont {C.}~\bibnamefont {Ur}}, \ and\ \bibinfo
  {author} {\bibfnamefont {J.~J.}\ \bibnamefont {Valiente-Dob\'on}},\ }\href
  {\doibase 10.1103/PhysRevLett.113.012501} {\bibfield  {journal} {\bibinfo
  {journal} {Phys. Rev. Lett.}\ }\textbf {\bibinfo {volume} {113}},\ \bibinfo
  {pages} {012501} (\bibinfo {year} {2014})}\BibitemShut {NoStop}%
\bibitem [{\citenamefont {Pellegri}\ \emph {et~al.}(2014)\citenamefont
  {Pellegri}, \citenamefont {Bracco}, \citenamefont {Crespi}, \citenamefont
  {Leoni}, \citenamefont {Camera}, \citenamefont {Lanza}, \citenamefont
  {Kmiecik}, \citenamefont {Maj}, \citenamefont {Avigo}, \citenamefont
  {Benzoni}, \citenamefont {Blasi}, \citenamefont {Boiano}, \citenamefont
  {Bottoni}, \citenamefont {Brambilla}, \citenamefont {Ceruti}, \citenamefont
  {Giaz}, \citenamefont {Million}, \citenamefont {Morales}, \citenamefont
  {Nicolini}, \citenamefont {Vandone}, \citenamefont {Wieland}, \citenamefont
  {Bazzacco}, \citenamefont {Bednarczyk}, \citenamefont {Bellato},
  \citenamefont {Birkenbach}, \citenamefont {Bortolato}, \citenamefont
  {Cederwall}, \citenamefont {Charles}, \citenamefont {Ciemala}, \citenamefont
  {Angelis]}, \citenamefont {Désesquelles}, \citenamefont {Eberth},
  \citenamefont {Farnea}, \citenamefont {Gadea}, \citenamefont {Gernhäuser},
  \citenamefont {Görgen}, \citenamefont {Gottardo}, \citenamefont {Grebosz},
  \citenamefont {Hess}, \citenamefont {Isocrate}, \citenamefont {Jolie},
  \citenamefont {Judson}, \citenamefont {Jungclaus}, \citenamefont {Karkour},
  \citenamefont {Krzysiek}, \citenamefont {Litvinova}, \citenamefont {Lunardi},
  \citenamefont {Mazurek}, \citenamefont {Mengoni}, \citenamefont
  {Michelagnoli}, \citenamefont {Menegazzo}, \citenamefont {Molini},
  \citenamefont {Napoli}, \citenamefont {Pullia}, \citenamefont {Quintana},
  \citenamefont {Recchia}, \citenamefont {Reiter}, \citenamefont {Salsac},
  \citenamefont {Siebeck}, \citenamefont {Siem}, \citenamefont {Simpson},
  \citenamefont {Söderström}, \citenamefont {Stezowski}, \citenamefont
  {Theisen}, \citenamefont {Ur}, \citenamefont {Dobon]},\ and\ \citenamefont
  {Zieblinski}}]{Pel14a}%
  \BibitemOpen
  \bibfield  {author} {\bibinfo {author} {\bibfnamefont {L.}~\bibnamefont
  {Pellegri}}, \bibinfo {author} {\bibfnamefont {A.}~\bibnamefont {Bracco}},
  \bibinfo {author} {\bibfnamefont {F.}~\bibnamefont {Crespi}}, \bibinfo
  {author} {\bibfnamefont {S.}~\bibnamefont {Leoni}}, \bibinfo {author}
  {\bibfnamefont {F.}~\bibnamefont {Camera}}, \bibinfo {author} {\bibfnamefont
  {E.}~\bibnamefont {Lanza}}, \bibinfo {author} {\bibfnamefont
  {M.}~\bibnamefont {Kmiecik}}, \bibinfo {author} {\bibfnamefont
  {A.}~\bibnamefont {Maj}}, \bibinfo {author} {\bibfnamefont {R.}~\bibnamefont
  {Avigo}}, \bibinfo {author} {\bibfnamefont {G.}~\bibnamefont {Benzoni}},
  \bibinfo {author} {\bibfnamefont {N.}~\bibnamefont {Blasi}}, \bibinfo
  {author} {\bibfnamefont {C.}~\bibnamefont {Boiano}}, \bibinfo {author}
  {\bibfnamefont {S.}~\bibnamefont {Bottoni}}, \bibinfo {author} {\bibfnamefont
  {S.}~\bibnamefont {Brambilla}}, \bibinfo {author} {\bibfnamefont
  {S.}~\bibnamefont {Ceruti}}, \bibinfo {author} {\bibfnamefont
  {A.}~\bibnamefont {Giaz}}, \bibinfo {author} {\bibfnamefont {B.}~\bibnamefont
  {Million}}, \bibinfo {author} {\bibfnamefont {A.}~\bibnamefont {Morales}},
  \bibinfo {author} {\bibfnamefont {R.}~\bibnamefont {Nicolini}}, \bibinfo
  {author} {\bibfnamefont {V.}~\bibnamefont {Vandone}}, \bibinfo {author}
  {\bibfnamefont {O.}~\bibnamefont {Wieland}}, \bibinfo {author} {\bibfnamefont
  {D.}~\bibnamefont {Bazzacco}}, \bibinfo {author} {\bibfnamefont
  {P.}~\bibnamefont {Bednarczyk}}, \bibinfo {author} {\bibfnamefont
  {M.}~\bibnamefont {Bellato}}, \bibinfo {author} {\bibfnamefont
  {B.}~\bibnamefont {Birkenbach}}, \bibinfo {author} {\bibfnamefont
  {D.}~\bibnamefont {Bortolato}}, \bibinfo {author} {\bibfnamefont
  {B.}~\bibnamefont {Cederwall}}, \bibinfo {author} {\bibfnamefont
  {L.}~\bibnamefont {Charles}}, \bibinfo {author} {\bibfnamefont
  {M.}~\bibnamefont {Ciemala}}, \bibinfo {author} {\bibfnamefont {G.~D.}\
  \bibnamefont {Angelis]}}, \bibinfo {author} {\bibfnamefont {P.}~\bibnamefont
  {Désesquelles}}, \bibinfo {author} {\bibfnamefont {J.}~\bibnamefont
  {Eberth}}, \bibinfo {author} {\bibfnamefont {E.}~\bibnamefont {Farnea}},
  \bibinfo {author} {\bibfnamefont {A.}~\bibnamefont {Gadea}}, \bibinfo
  {author} {\bibfnamefont {R.}~\bibnamefont {Gernhäuser}}, \bibinfo {author}
  {\bibfnamefont {A.}~\bibnamefont {Görgen}}, \bibinfo {author} {\bibfnamefont
  {A.}~\bibnamefont {Gottardo}}, \bibinfo {author} {\bibfnamefont
  {J.}~\bibnamefont {Grebosz}}, \bibinfo {author} {\bibfnamefont
  {H.}~\bibnamefont {Hess}}, \bibinfo {author} {\bibfnamefont {R.}~\bibnamefont
  {Isocrate}}, \bibinfo {author} {\bibfnamefont {J.}~\bibnamefont {Jolie}},
  \bibinfo {author} {\bibfnamefont {D.}~\bibnamefont {Judson}}, \bibinfo
  {author} {\bibfnamefont {A.}~\bibnamefont {Jungclaus}}, \bibinfo {author}
  {\bibfnamefont {N.}~\bibnamefont {Karkour}}, \bibinfo {author} {\bibfnamefont
  {M.}~\bibnamefont {Krzysiek}}, \bibinfo {author} {\bibfnamefont
  {E.}~\bibnamefont {Litvinova}}, \bibinfo {author} {\bibfnamefont
  {S.}~\bibnamefont {Lunardi}}, \bibinfo {author} {\bibfnamefont
  {K.}~\bibnamefont {Mazurek}}, \bibinfo {author} {\bibfnamefont
  {D.}~\bibnamefont {Mengoni}}, \bibinfo {author} {\bibfnamefont
  {C.}~\bibnamefont {Michelagnoli}}, \bibinfo {author} {\bibfnamefont
  {R.}~\bibnamefont {Menegazzo}}, \bibinfo {author} {\bibfnamefont
  {P.}~\bibnamefont {Molini}}, \bibinfo {author} {\bibfnamefont
  {D.}~\bibnamefont {Napoli}}, \bibinfo {author} {\bibfnamefont
  {A.}~\bibnamefont {Pullia}}, \bibinfo {author} {\bibfnamefont
  {B.}~\bibnamefont {Quintana}}, \bibinfo {author} {\bibfnamefont
  {F.}~\bibnamefont {Recchia}}, \bibinfo {author} {\bibfnamefont
  {P.}~\bibnamefont {Reiter}}, \bibinfo {author} {\bibfnamefont
  {M.}~\bibnamefont {Salsac}}, \bibinfo {author} {\bibfnamefont
  {B.}~\bibnamefont {Siebeck}}, \bibinfo {author} {\bibfnamefont
  {S.}~\bibnamefont {Siem}}, \bibinfo {author} {\bibfnamefont {J.}~\bibnamefont
  {Simpson}}, \bibinfo {author} {\bibfnamefont {P.-A.}\ \bibnamefont
  {Söderström}}, \bibinfo {author} {\bibfnamefont {O.}~\bibnamefont
  {Stezowski}}, \bibinfo {author} {\bibfnamefont {C.}~\bibnamefont {Theisen}},
  \bibinfo {author} {\bibfnamefont {C.}~\bibnamefont {Ur}}, \bibinfo {author}
  {\bibfnamefont {J.~V.}\ \bibnamefont {Dobon]}}, \ and\ \bibinfo {author}
  {\bibfnamefont {M.}~\bibnamefont {Zieblinski}},\ }\href {\doibase
  https://doi.org/10.1016/j.physletb.2014.08.029} {\bibfield  {journal}
  {\bibinfo  {journal} {Physics Letters B}\ }\textbf {\bibinfo {volume}
  {738}},\ \bibinfo {pages} {519 } (\bibinfo {year} {2014})}\BibitemShut
  {NoStop}%
\bibitem [{\citenamefont {Lanza}\ \emph {et~al.}(2014)\citenamefont {Lanza},
  \citenamefont {Vitturi}, \citenamefont {Litvinova},\ and\ \citenamefont
  {Savran}}]{Lan14a}%
  \BibitemOpen
  \bibfield  {author} {\bibinfo {author} {\bibfnamefont {E.~G.}\ \bibnamefont
  {Lanza}}, \bibinfo {author} {\bibfnamefont {A.}~\bibnamefont {Vitturi}},
  \bibinfo {author} {\bibfnamefont {E.}~\bibnamefont {Litvinova}}, \ and\
  \bibinfo {author} {\bibfnamefont {D.}~\bibnamefont {Savran}},\ }\href
  {\doibase 10.1103/PhysRevC.89.041601} {\bibfield  {journal} {\bibinfo
  {journal} {Phys. Rev. C}\ }\textbf {\bibinfo {volume} {89}},\ \bibinfo
  {pages} {041601} (\bibinfo {year} {2014})}\BibitemShut {NoStop}%
\bibitem [{\citenamefont {Crespi}\ \emph {et~al.}(2015)\citenamefont {Crespi},
  \citenamefont {Bracco}, \citenamefont {Nicolini}, \citenamefont {Lanza},
  \citenamefont {Vitturi}, \citenamefont {Mengoni}, \citenamefont {Leoni},
  \citenamefont {Benzoni}, \citenamefont {Blasi}, \citenamefont {Boiano},
  \citenamefont {Bottoni}, \citenamefont {Brambilla}, \citenamefont {Camera},
  \citenamefont {Corsi}, \citenamefont {Giaz}, \citenamefont {Million},
  \citenamefont {Pellegri}, \citenamefont {Vandone}, \citenamefont {Wieland},
  \citenamefont {Bednarczyk}, \citenamefont {Ciema\l{}a}, \citenamefont
  {Kmiecik}, \citenamefont {Krzysiek}, \citenamefont {Maj}, \citenamefont
  {Bazzacco}, \citenamefont {Bellato}, \citenamefont {Birkenbach},
  \citenamefont {Bortolato}, \citenamefont {Calore}, \citenamefont {Cederwall},
  \citenamefont {de~Angelis}, \citenamefont {D\'esesquelles}, \citenamefont
  {Eberth}, \citenamefont {Farnea}, \citenamefont {Gadea}, \citenamefont
  {G\"orgen}, \citenamefont {Gottardo}, \citenamefont {Hess}, \citenamefont
  {Isocrate}, \citenamefont {Jolie}, \citenamefont {Jungclaus}, \citenamefont
  {Kempley}, \citenamefont {Labiche}, \citenamefont {Menegazzo}, \citenamefont
  {Michelagnoli}, \citenamefont {Molini}, \citenamefont {Napoli}, \citenamefont
  {Pullia}, \citenamefont {Quintana}, \citenamefont {Recchia}, \citenamefont
  {Reiter}, \citenamefont {Sahin}, \citenamefont {Siem}, \citenamefont
  {S\"oderstr\"om}, \citenamefont {Stezowski}, \citenamefont {Theisen},
  \citenamefont {Ur},\ and\ \citenamefont {Valiente-Dob\'on}}]{Cre15a}%
  \BibitemOpen
  \bibfield  {author} {\bibinfo {author} {\bibfnamefont {F.~C.~L.}\
  \bibnamefont {Crespi}}, \bibinfo {author} {\bibfnamefont {A.}~\bibnamefont
  {Bracco}}, \bibinfo {author} {\bibfnamefont {R.}~\bibnamefont {Nicolini}},
  \bibinfo {author} {\bibfnamefont {E.~G.}\ \bibnamefont {Lanza}}, \bibinfo
  {author} {\bibfnamefont {A.}~\bibnamefont {Vitturi}}, \bibinfo {author}
  {\bibfnamefont {D.}~\bibnamefont {Mengoni}}, \bibinfo {author} {\bibfnamefont
  {S.}~\bibnamefont {Leoni}}, \bibinfo {author} {\bibfnamefont
  {G.}~\bibnamefont {Benzoni}}, \bibinfo {author} {\bibfnamefont
  {N.}~\bibnamefont {Blasi}}, \bibinfo {author} {\bibfnamefont
  {C.}~\bibnamefont {Boiano}}, \bibinfo {author} {\bibfnamefont
  {S.}~\bibnamefont {Bottoni}}, \bibinfo {author} {\bibfnamefont
  {S.}~\bibnamefont {Brambilla}}, \bibinfo {author} {\bibfnamefont
  {F.}~\bibnamefont {Camera}}, \bibinfo {author} {\bibfnamefont
  {A.}~\bibnamefont {Corsi}}, \bibinfo {author} {\bibfnamefont
  {A.}~\bibnamefont {Giaz}}, \bibinfo {author} {\bibfnamefont {B.}~\bibnamefont
  {Million}}, \bibinfo {author} {\bibfnamefont {L.}~\bibnamefont {Pellegri}},
  \bibinfo {author} {\bibfnamefont {V.}~\bibnamefont {Vandone}}, \bibinfo
  {author} {\bibfnamefont {O.}~\bibnamefont {Wieland}}, \bibinfo {author}
  {\bibfnamefont {P.}~\bibnamefont {Bednarczyk}}, \bibinfo {author}
  {\bibfnamefont {M.}~\bibnamefont {Ciema\l{}a}}, \bibinfo {author}
  {\bibfnamefont {M.}~\bibnamefont {Kmiecik}}, \bibinfo {author} {\bibfnamefont
  {M.}~\bibnamefont {Krzysiek}}, \bibinfo {author} {\bibfnamefont
  {A.}~\bibnamefont {Maj}}, \bibinfo {author} {\bibfnamefont {D.}~\bibnamefont
  {Bazzacco}}, \bibinfo {author} {\bibfnamefont {M.}~\bibnamefont {Bellato}},
  \bibinfo {author} {\bibfnamefont {B.}~\bibnamefont {Birkenbach}}, \bibinfo
  {author} {\bibfnamefont {D.}~\bibnamefont {Bortolato}}, \bibinfo {author}
  {\bibfnamefont {E.}~\bibnamefont {Calore}}, \bibinfo {author} {\bibfnamefont
  {B.}~\bibnamefont {Cederwall}}, \bibinfo {author} {\bibfnamefont
  {G.}~\bibnamefont {de~Angelis}}, \bibinfo {author} {\bibfnamefont
  {P.}~\bibnamefont {D\'esesquelles}}, \bibinfo {author} {\bibfnamefont
  {J.}~\bibnamefont {Eberth}}, \bibinfo {author} {\bibfnamefont
  {E.}~\bibnamefont {Farnea}}, \bibinfo {author} {\bibfnamefont
  {A.}~\bibnamefont {Gadea}}, \bibinfo {author} {\bibfnamefont
  {A.}~\bibnamefont {G\"orgen}}, \bibinfo {author} {\bibfnamefont
  {A.}~\bibnamefont {Gottardo}}, \bibinfo {author} {\bibfnamefont
  {H.}~\bibnamefont {Hess}}, \bibinfo {author} {\bibfnamefont {R.}~\bibnamefont
  {Isocrate}}, \bibinfo {author} {\bibfnamefont {J.}~\bibnamefont {Jolie}},
  \bibinfo {author} {\bibfnamefont {A.}~\bibnamefont {Jungclaus}}, \bibinfo
  {author} {\bibfnamefont {R.~S.}\ \bibnamefont {Kempley}}, \bibinfo {author}
  {\bibfnamefont {M.}~\bibnamefont {Labiche}}, \bibinfo {author} {\bibfnamefont
  {R.}~\bibnamefont {Menegazzo}}, \bibinfo {author} {\bibfnamefont
  {C.}~\bibnamefont {Michelagnoli}}, \bibinfo {author} {\bibfnamefont
  {P.}~\bibnamefont {Molini}}, \bibinfo {author} {\bibfnamefont {D.~R.}\
  \bibnamefont {Napoli}}, \bibinfo {author} {\bibfnamefont {A.}~\bibnamefont
  {Pullia}}, \bibinfo {author} {\bibfnamefont {B.}~\bibnamefont {Quintana}},
  \bibinfo {author} {\bibfnamefont {F.}~\bibnamefont {Recchia}}, \bibinfo
  {author} {\bibfnamefont {P.}~\bibnamefont {Reiter}}, \bibinfo {author}
  {\bibfnamefont {E.}~\bibnamefont {Sahin}}, \bibinfo {author} {\bibfnamefont
  {S.}~\bibnamefont {Siem}}, \bibinfo {author} {\bibfnamefont {P.-A.}\
  \bibnamefont {S\"oderstr\"om}}, \bibinfo {author} {\bibfnamefont
  {O.}~\bibnamefont {Stezowski}}, \bibinfo {author} {\bibfnamefont
  {C.}~\bibnamefont {Theisen}}, \bibinfo {author} {\bibfnamefont
  {C.}~\bibnamefont {Ur}}, \ and\ \bibinfo {author} {\bibfnamefont {J.~J.}\
  \bibnamefont {Valiente-Dob\'on}},\ }\href {\doibase
  10.1103/PhysRevC.91.024323} {\bibfield  {journal} {\bibinfo  {journal} {Phys.
  Rev. C}\ }\textbf {\bibinfo {volume} {91}},\ \bibinfo {pages} {024323}
  (\bibinfo {year} {2015})}\BibitemShut {NoStop}%
\bibitem [{\citenamefont {Krzysiek}\ \emph {et~al.}(2016)\citenamefont
  {Krzysiek}, \citenamefont {Kmiecik}, \citenamefont {Maj}, \citenamefont
  {Bednarczyk}, \citenamefont {Bracco}, \citenamefont {Crespi}, \citenamefont
  {Lanza}, \citenamefont {Litvinova}, \citenamefont {Paar}, \citenamefont
  {Avigo}, \citenamefont {Bazzacco}, \citenamefont {Benzoni}, \citenamefont
  {Birkenbach}, \citenamefont {Blasi}, \citenamefont {Bottoni}, \citenamefont
  {Brambilla}, \citenamefont {Camera}, \citenamefont {Ceruti}, \citenamefont
  {Ciema\l{}a}, \citenamefont {de~Angelis}, \citenamefont {D\'esesquelles},
  \citenamefont {Eberth}, \citenamefont {Farnea}, \citenamefont {Gadea},
  \citenamefont {Giaz}, \citenamefont {G\"orgen}, \citenamefont {Gottardo},
  \citenamefont {Grebosz}, \citenamefont {Hess}, \citenamefont {Isocarte},
  \citenamefont {Jungclaus}, \citenamefont {Leoni}, \citenamefont {Ljungvall},
  \citenamefont {Lunardi}, \citenamefont {Mazurek}, \citenamefont {Menegazzo},
  \citenamefont {Mengoni}, \citenamefont {Michelagnoli}, \citenamefont
  {Milion}, \citenamefont {Morales}, \citenamefont {Napoli}, \citenamefont
  {Nicolini}, \citenamefont {Pellegri}, \citenamefont {Pullia}, \citenamefont
  {Quintana}, \citenamefont {Recchia}, \citenamefont {Reiter}, \citenamefont
  {Rosso}, \citenamefont {Salsac}, \citenamefont {Siebeck}, \citenamefont
  {Siem}, \citenamefont {S\"oderstr\"om}, \citenamefont {Ur}, \citenamefont
  {Valiente-Dobon}, \citenamefont {Wieland},\ and\ \citenamefont
  {Zieblinski}}]{Krz16a}%
  \BibitemOpen
  \bibfield  {author} {\bibinfo {author} {\bibfnamefont {M.}~\bibnamefont
  {Krzysiek}}, \bibinfo {author} {\bibfnamefont {M.}~\bibnamefont {Kmiecik}},
  \bibinfo {author} {\bibfnamefont {A.}~\bibnamefont {Maj}}, \bibinfo {author}
  {\bibfnamefont {P.}~\bibnamefont {Bednarczyk}}, \bibinfo {author}
  {\bibfnamefont {A.}~\bibnamefont {Bracco}}, \bibinfo {author} {\bibfnamefont
  {F.~C.~L.}\ \bibnamefont {Crespi}}, \bibinfo {author} {\bibfnamefont {E.~G.}\
  \bibnamefont {Lanza}}, \bibinfo {author} {\bibfnamefont {E.}~\bibnamefont
  {Litvinova}}, \bibinfo {author} {\bibfnamefont {N.}~\bibnamefont {Paar}},
  \bibinfo {author} {\bibfnamefont {R.}~\bibnamefont {Avigo}}, \bibinfo
  {author} {\bibfnamefont {D.}~\bibnamefont {Bazzacco}}, \bibinfo {author}
  {\bibfnamefont {G.}~\bibnamefont {Benzoni}}, \bibinfo {author} {\bibfnamefont
  {B.}~\bibnamefont {Birkenbach}}, \bibinfo {author} {\bibfnamefont
  {N.}~\bibnamefont {Blasi}}, \bibinfo {author} {\bibfnamefont
  {S.}~\bibnamefont {Bottoni}}, \bibinfo {author} {\bibfnamefont
  {S.}~\bibnamefont {Brambilla}}, \bibinfo {author} {\bibfnamefont
  {F.}~\bibnamefont {Camera}}, \bibinfo {author} {\bibfnamefont
  {S.}~\bibnamefont {Ceruti}}, \bibinfo {author} {\bibfnamefont
  {M.}~\bibnamefont {Ciema\l{}a}}, \bibinfo {author} {\bibfnamefont
  {G.}~\bibnamefont {de~Angelis}}, \bibinfo {author} {\bibfnamefont
  {P.}~\bibnamefont {D\'esesquelles}}, \bibinfo {author} {\bibfnamefont
  {J.}~\bibnamefont {Eberth}}, \bibinfo {author} {\bibfnamefont
  {E.}~\bibnamefont {Farnea}}, \bibinfo {author} {\bibfnamefont
  {A.}~\bibnamefont {Gadea}}, \bibinfo {author} {\bibfnamefont
  {A.}~\bibnamefont {Giaz}}, \bibinfo {author} {\bibfnamefont {A.}~\bibnamefont
  {G\"orgen}}, \bibinfo {author} {\bibfnamefont {A.}~\bibnamefont {Gottardo}},
  \bibinfo {author} {\bibfnamefont {J.}~\bibnamefont {Grebosz}}, \bibinfo
  {author} {\bibfnamefont {H.}~\bibnamefont {Hess}}, \bibinfo {author}
  {\bibfnamefont {R.}~\bibnamefont {Isocarte}}, \bibinfo {author}
  {\bibfnamefont {A.}~\bibnamefont {Jungclaus}}, \bibinfo {author}
  {\bibfnamefont {S.}~\bibnamefont {Leoni}}, \bibinfo {author} {\bibfnamefont
  {J.}~\bibnamefont {Ljungvall}}, \bibinfo {author} {\bibfnamefont
  {S.}~\bibnamefont {Lunardi}}, \bibinfo {author} {\bibfnamefont
  {K.}~\bibnamefont {Mazurek}}, \bibinfo {author} {\bibfnamefont
  {R.}~\bibnamefont {Menegazzo}}, \bibinfo {author} {\bibfnamefont
  {D.}~\bibnamefont {Mengoni}}, \bibinfo {author} {\bibfnamefont
  {C.}~\bibnamefont {Michelagnoli}}, \bibinfo {author} {\bibfnamefont
  {B.}~\bibnamefont {Milion}}, \bibinfo {author} {\bibfnamefont {A.~I.}\
  \bibnamefont {Morales}}, \bibinfo {author} {\bibfnamefont {D.~R.}\
  \bibnamefont {Napoli}}, \bibinfo {author} {\bibfnamefont {R.}~\bibnamefont
  {Nicolini}}, \bibinfo {author} {\bibfnamefont {L.}~\bibnamefont {Pellegri}},
  \bibinfo {author} {\bibfnamefont {A.}~\bibnamefont {Pullia}}, \bibinfo
  {author} {\bibfnamefont {B.}~\bibnamefont {Quintana}}, \bibinfo {author}
  {\bibfnamefont {F.}~\bibnamefont {Recchia}}, \bibinfo {author} {\bibfnamefont
  {P.}~\bibnamefont {Reiter}}, \bibinfo {author} {\bibfnamefont
  {D.}~\bibnamefont {Rosso}}, \bibinfo {author} {\bibfnamefont {M.~D.}\
  \bibnamefont {Salsac}}, \bibinfo {author} {\bibfnamefont {B.}~\bibnamefont
  {Siebeck}}, \bibinfo {author} {\bibfnamefont {S.}~\bibnamefont {Siem}},
  \bibinfo {author} {\bibfnamefont {P.-A.}\ \bibnamefont {S\"oderstr\"om}},
  \bibinfo {author} {\bibfnamefont {C.}~\bibnamefont {Ur}}, \bibinfo {author}
  {\bibfnamefont {J.~J.}\ \bibnamefont {Valiente-Dobon}}, \bibinfo {author}
  {\bibfnamefont {O.}~\bibnamefont {Wieland}}, \ and\ \bibinfo {author}
  {\bibfnamefont {M.}~\bibnamefont {Zieblinski}},\ }\href {\doibase
  10.1103/PhysRevC.93.044330} {\bibfield  {journal} {\bibinfo  {journal} {Phys.
  Rev. C}\ }\textbf {\bibinfo {volume} {93}},\ \bibinfo {pages} {044330}
  (\bibinfo {year} {2016})}\BibitemShut {NoStop}%
\bibitem [{\citenamefont {Nakatsuka}\ \emph {et~al.}(2017)\citenamefont
  {Nakatsuka}, \citenamefont {Baba}, \citenamefont {Aumann}, \citenamefont
  {Avigo}, \citenamefont {Banerjee}, \citenamefont {Bracco}, \citenamefont
  {Caesar}, \citenamefont {Camera}, \citenamefont {Ceruti}, \citenamefont
  {Chen}, \citenamefont {Derya}, \citenamefont {Doornenbal}, \citenamefont
  {Giaz}, \citenamefont {Horvat}, \citenamefont {Ieki}, \citenamefont
  {Inakura}, \citenamefont {Imai}, \citenamefont {Kawabata}, \citenamefont
  {Kobayashi}, \citenamefont {Kondo}, \citenamefont {Koyama}, \citenamefont
  {Kurata-Nishimura}, \citenamefont {Masuoka}, \citenamefont {Matsushita},
  \citenamefont {Michimasa}, \citenamefont {Million}, \citenamefont
  {Motobayashi}, \citenamefont {Murakami}, \citenamefont {Nakamura},
  \citenamefont {Ohnishi}, \citenamefont {Ong}, \citenamefont {Ota},
  \citenamefont {Otsu}, \citenamefont {Ozaki}, \citenamefont {Saito},
  \citenamefont {Sakurai}, \citenamefont {Scheit}, \citenamefont {Schindler},
  \citenamefont {Schrock}, \citenamefont {Shiga}, \citenamefont {Shikata},
  \citenamefont {Shimoura}, \citenamefont {Steppenbeck}, \citenamefont
  {Sumikama}, \citenamefont {Syndikus}, \citenamefont {Takeda}, \citenamefont
  {Takeuchi}, \citenamefont {Tamii}, \citenamefont {Taniuchi}, \citenamefont
  {Togano}, \citenamefont {Tscheuschner}, \citenamefont {Tsubota},
  \citenamefont {Wang}, \citenamefont {Wieland}, \citenamefont {Wimmer},
  \citenamefont {Yamaguchi}, \citenamefont {Yoneda},\ and\ \citenamefont
  {Zenihiro}}]{Nak17a}%
  \BibitemOpen
  \bibfield  {author} {\bibinfo {author} {\bibfnamefont {N.}~\bibnamefont
  {Nakatsuka}}, \bibinfo {author} {\bibfnamefont {H.}~\bibnamefont {Baba}},
  \bibinfo {author} {\bibfnamefont {T.}~\bibnamefont {Aumann}}, \bibinfo
  {author} {\bibfnamefont {R.}~\bibnamefont {Avigo}}, \bibinfo {author}
  {\bibfnamefont {S.}~\bibnamefont {Banerjee}}, \bibinfo {author}
  {\bibfnamefont {A.}~\bibnamefont {Bracco}}, \bibinfo {author} {\bibfnamefont
  {C.}~\bibnamefont {Caesar}}, \bibinfo {author} {\bibfnamefont
  {F.}~\bibnamefont {Camera}}, \bibinfo {author} {\bibfnamefont
  {S.}~\bibnamefont {Ceruti}}, \bibinfo {author} {\bibfnamefont
  {S.}~\bibnamefont {Chen}}, \bibinfo {author} {\bibfnamefont {V.}~\bibnamefont
  {Derya}}, \bibinfo {author} {\bibfnamefont {P.}~\bibnamefont {Doornenbal}},
  \bibinfo {author} {\bibfnamefont {A.}~\bibnamefont {Giaz}}, \bibinfo {author}
  {\bibfnamefont {A.}~\bibnamefont {Horvat}}, \bibinfo {author} {\bibfnamefont
  {K.}~\bibnamefont {Ieki}}, \bibinfo {author} {\bibfnamefont {T.}~\bibnamefont
  {Inakura}}, \bibinfo {author} {\bibfnamefont {N.}~\bibnamefont {Imai}},
  \bibinfo {author} {\bibfnamefont {T.}~\bibnamefont {Kawabata}}, \bibinfo
  {author} {\bibfnamefont {N.}~\bibnamefont {Kobayashi}}, \bibinfo {author}
  {\bibfnamefont {Y.}~\bibnamefont {Kondo}}, \bibinfo {author} {\bibfnamefont
  {S.}~\bibnamefont {Koyama}}, \bibinfo {author} {\bibfnamefont
  {M.}~\bibnamefont {Kurata-Nishimura}}, \bibinfo {author} {\bibfnamefont
  {S.}~\bibnamefont {Masuoka}}, \bibinfo {author} {\bibfnamefont
  {M.}~\bibnamefont {Matsushita}}, \bibinfo {author} {\bibfnamefont
  {S.}~\bibnamefont {Michimasa}}, \bibinfo {author} {\bibfnamefont
  {B.}~\bibnamefont {Million}}, \bibinfo {author} {\bibfnamefont
  {T.}~\bibnamefont {Motobayashi}}, \bibinfo {author} {\bibfnamefont
  {T.}~\bibnamefont {Murakami}}, \bibinfo {author} {\bibfnamefont
  {T.}~\bibnamefont {Nakamura}}, \bibinfo {author} {\bibfnamefont
  {T.}~\bibnamefont {Ohnishi}}, \bibinfo {author} {\bibfnamefont
  {H.}~\bibnamefont {Ong}}, \bibinfo {author} {\bibfnamefont {S.}~\bibnamefont
  {Ota}}, \bibinfo {author} {\bibfnamefont {H.}~\bibnamefont {Otsu}}, \bibinfo
  {author} {\bibfnamefont {T.}~\bibnamefont {Ozaki}}, \bibinfo {author}
  {\bibfnamefont {A.}~\bibnamefont {Saito}}, \bibinfo {author} {\bibfnamefont
  {H.}~\bibnamefont {Sakurai}}, \bibinfo {author} {\bibfnamefont
  {H.}~\bibnamefont {Scheit}}, \bibinfo {author} {\bibfnamefont
  {F.}~\bibnamefont {Schindler}}, \bibinfo {author} {\bibfnamefont
  {P.}~\bibnamefont {Schrock}}, \bibinfo {author} {\bibfnamefont
  {Y.}~\bibnamefont {Shiga}}, \bibinfo {author} {\bibfnamefont
  {M.}~\bibnamefont {Shikata}}, \bibinfo {author} {\bibfnamefont
  {S.}~\bibnamefont {Shimoura}}, \bibinfo {author} {\bibfnamefont
  {D.}~\bibnamefont {Steppenbeck}}, \bibinfo {author} {\bibfnamefont
  {T.}~\bibnamefont {Sumikama}}, \bibinfo {author} {\bibfnamefont
  {I.}~\bibnamefont {Syndikus}}, \bibinfo {author} {\bibfnamefont
  {H.}~\bibnamefont {Takeda}}, \bibinfo {author} {\bibfnamefont
  {S.}~\bibnamefont {Takeuchi}}, \bibinfo {author} {\bibfnamefont
  {A.}~\bibnamefont {Tamii}}, \bibinfo {author} {\bibfnamefont
  {R.}~\bibnamefont {Taniuchi}}, \bibinfo {author} {\bibfnamefont
  {Y.}~\bibnamefont {Togano}}, \bibinfo {author} {\bibfnamefont
  {J.}~\bibnamefont {Tscheuschner}}, \bibinfo {author} {\bibfnamefont
  {J.}~\bibnamefont {Tsubota}}, \bibinfo {author} {\bibfnamefont
  {H.}~\bibnamefont {Wang}}, \bibinfo {author} {\bibfnamefont {O.}~\bibnamefont
  {Wieland}}, \bibinfo {author} {\bibfnamefont {K.}~\bibnamefont {Wimmer}},
  \bibinfo {author} {\bibfnamefont {Y.}~\bibnamefont {Yamaguchi}}, \bibinfo
  {author} {\bibfnamefont {K.}~\bibnamefont {Yoneda}}, \ and\ \bibinfo {author}
  {\bibfnamefont {J.}~\bibnamefont {Zenihiro}},\ }\href {\doibase
  https://doi.org/10.1016/j.physletb.2017.03.017} {\bibfield  {journal}
  {\bibinfo  {journal} {Physics Letters B}\ }\textbf {\bibinfo {volume}
  {768}},\ \bibinfo {pages} {387 } (\bibinfo {year} {2017})}\BibitemShut
  {NoStop}%
\bibitem [{\citenamefont {Crespi}\ \emph {et~al.}(2018)\citenamefont {Crespi},
  \citenamefont {Bracco}, \citenamefont {Tamii}, \citenamefont {Blasi},
  \citenamefont {Camera}, \citenamefont {Wieland}, \citenamefont {Aoi},
  \citenamefont {Balabanski}, \citenamefont {Bassauer}, \citenamefont {Brown},
  \citenamefont {Carpenter}, \citenamefont {Carroll}, \citenamefont {Ciemala},
  \citenamefont {Czeszumska}, \citenamefont {Davies}, \citenamefont
  {Donaldson}, \citenamefont {Fang}, \citenamefont {Fujita}, \citenamefont
  {Gey}, \citenamefont {Hoang}, \citenamefont {Ichige}, \citenamefont
  {Ideguchi}, \citenamefont {Inoue}, \citenamefont {Isaak}, \citenamefont
  {Iwamoto}, \citenamefont {Jenkins}, \citenamefont {Jin}, \citenamefont
  {Klaus}, \citenamefont {Kobayashi}, \citenamefont {Koike}, \citenamefont
  {Krzysiek}, \citenamefont {Raju}, \citenamefont {Liu}, \citenamefont {Maj},
  \citenamefont {Montanari}, \citenamefont {Morris}, \citenamefont {Noji},
  \citenamefont {Pickstone}, \citenamefont {Savran}, \citenamefont {Spieker},
  \citenamefont {Steinhilber}, \citenamefont {Sullivan}, \citenamefont
  {Wasilewska}, \citenamefont {Werner}, \citenamefont {Yamamoto}, \citenamefont
  {Yamamoto}, \citenamefont {Zhou},\ and\ \citenamefont {Zhu}}]{Cre18a}%
  \BibitemOpen
  \bibfield  {author} {\bibinfo {author} {\bibfnamefont {F.}~\bibnamefont
  {Crespi}}, \bibinfo {author} {\bibfnamefont {A.}~\bibnamefont {Bracco}},
  \bibinfo {author} {\bibfnamefont {A.}~\bibnamefont {Tamii}}, \bibinfo
  {author} {\bibfnamefont {N.}~\bibnamefont {Blasi}}, \bibinfo {author}
  {\bibfnamefont {F.}~\bibnamefont {Camera}}, \bibinfo {author} {\bibfnamefont
  {O.}~\bibnamefont {Wieland}}, \bibinfo {author} {\bibfnamefont
  {N.}~\bibnamefont {Aoi}}, \bibinfo {author} {\bibfnamefont {D.}~\bibnamefont
  {Balabanski}}, \bibinfo {author} {\bibfnamefont {S.}~\bibnamefont
  {Bassauer}}, \bibinfo {author} {\bibfnamefont {A.~S.}\ \bibnamefont {Brown}},
  \bibinfo {author} {\bibfnamefont {M.~P.}\ \bibnamefont {Carpenter}}, \bibinfo
  {author} {\bibfnamefont {J.~J.}\ \bibnamefont {Carroll}}, \bibinfo {author}
  {\bibfnamefont {M.}~\bibnamefont {Ciemala}}, \bibinfo {author} {\bibfnamefont
  {A.}~\bibnamefont {Czeszumska}}, \bibinfo {author} {\bibfnamefont {P.~J.}\
  \bibnamefont {Davies}}, \bibinfo {author} {\bibfnamefont {L.}~\bibnamefont
  {Donaldson}}, \bibinfo {author} {\bibfnamefont {Y.}~\bibnamefont {Fang}},
  \bibinfo {author} {\bibfnamefont {H.}~\bibnamefont {Fujita}}, \bibinfo
  {author} {\bibfnamefont {G.}~\bibnamefont {Gey}}, \bibinfo {author}
  {\bibfnamefont {T.~H.}\ \bibnamefont {Hoang}}, \bibinfo {author}
  {\bibfnamefont {N.}~\bibnamefont {Ichige}}, \bibinfo {author} {\bibfnamefont
  {E.}~\bibnamefont {Ideguchi}}, \bibinfo {author} {\bibfnamefont
  {A.}~\bibnamefont {Inoue}}, \bibinfo {author} {\bibfnamefont
  {J.}~\bibnamefont {Isaak}}, \bibinfo {author} {\bibfnamefont
  {C.}~\bibnamefont {Iwamoto}}, \bibinfo {author} {\bibfnamefont {D.~G.}\
  \bibnamefont {Jenkins}}, \bibinfo {author} {\bibfnamefont {O.~H.}\
  \bibnamefont {Jin}}, \bibinfo {author} {\bibfnamefont {T.}~\bibnamefont
  {Klaus}}, \bibinfo {author} {\bibfnamefont {N.}~\bibnamefont {Kobayashi}},
  \bibinfo {author} {\bibfnamefont {T.}~\bibnamefont {Koike}}, \bibinfo
  {author} {\bibfnamefont {M.}~\bibnamefont {Krzysiek}}, \bibinfo {author}
  {\bibfnamefont {M.~K.}\ \bibnamefont {Raju}}, \bibinfo {author}
  {\bibfnamefont {M.}~\bibnamefont {Liu}}, \bibinfo {author} {\bibfnamefont
  {A.}~\bibnamefont {Maj}}, \bibinfo {author} {\bibfnamefont {D.}~\bibnamefont
  {Montanari}}, \bibinfo {author} {\bibfnamefont {L.}~\bibnamefont {Morris}},
  \bibinfo {author} {\bibfnamefont {S.}~\bibnamefont {Noji}}, \bibinfo {author}
  {\bibfnamefont {S.~G.}\ \bibnamefont {Pickstone}}, \bibinfo {author}
  {\bibfnamefont {D.}~\bibnamefont {Savran}}, \bibinfo {author} {\bibfnamefont
  {M.}~\bibnamefont {Spieker}}, \bibinfo {author} {\bibfnamefont
  {G.}~\bibnamefont {Steinhilber}}, \bibinfo {author} {\bibfnamefont
  {C.}~\bibnamefont {Sullivan}}, \bibinfo {author} {\bibfnamefont
  {B.}~\bibnamefont {Wasilewska}}, \bibinfo {author} {\bibfnamefont
  {V.}~\bibnamefont {Werner}}, \bibinfo {author} {\bibfnamefont
  {T.}~\bibnamefont {Yamamoto}}, \bibinfo {author} {\bibfnamefont
  {Y.}~\bibnamefont {Yamamoto}}, \bibinfo {author} {\bibfnamefont
  {X.}~\bibnamefont {Zhou}}, \ and\ \bibinfo {author} {\bibfnamefont
  {S.}~\bibnamefont {Zhu}},\ }\href {\doibase 10.1088/1742-6596/1014/1/012002}
  {\bibfield  {journal} {\bibinfo  {journal} {Journal of Physics: Conference
  Series}\ }\textbf {\bibinfo {volume} {1014}},\ \bibinfo {pages} {012002}
  (\bibinfo {year} {2018})}\BibitemShut {NoStop}%
\bibitem [{\citenamefont {Vretenar}\ \emph {et~al.}(2001)\citenamefont
  {Vretenar}, \citenamefont {Paar}, \citenamefont {Ring},\ and\ \citenamefont
  {Lalazissis}}]{Vre01a}%
  \BibitemOpen
  \bibfield  {author} {\bibinfo {author} {\bibfnamefont {D.}~\bibnamefont
  {Vretenar}}, \bibinfo {author} {\bibfnamefont {N.}~\bibnamefont {Paar}},
  \bibinfo {author} {\bibfnamefont {P.}~\bibnamefont {Ring}}, \ and\ \bibinfo
  {author} {\bibfnamefont {G.~A.}\ \bibnamefont {Lalazissis}},\ }\href
  {\doibase 10.1103/PhysRevC.63.047301} {\bibfield  {journal} {\bibinfo
  {journal} {Phys. Rev. C}\ }\textbf {\bibinfo {volume} {63}},\ \bibinfo
  {pages} {047301} (\bibinfo {year} {2001})}\BibitemShut {NoStop}%
\bibitem [{\citenamefont {Ryezayeva}\ \emph {et~al.}(2002)\citenamefont
  {Ryezayeva}, \citenamefont {Hartmann}, \citenamefont {Kalmykov},
  \citenamefont {Lenske}, \citenamefont {von Neumann-Cosel}, \citenamefont
  {Ponomarev}, \citenamefont {Richter}, \citenamefont {Shevchenko},
  \citenamefont {Volz},\ and\ \citenamefont {Wambach}}]{Rye02a}%
  \BibitemOpen
  \bibfield  {author} {\bibinfo {author} {\bibfnamefont {N.}~\bibnamefont
  {Ryezayeva}}, \bibinfo {author} {\bibfnamefont {T.}~\bibnamefont {Hartmann}},
  \bibinfo {author} {\bibfnamefont {Y.}~\bibnamefont {Kalmykov}}, \bibinfo
  {author} {\bibfnamefont {H.}~\bibnamefont {Lenske}}, \bibinfo {author}
  {\bibfnamefont {P.}~\bibnamefont {von Neumann-Cosel}}, \bibinfo {author}
  {\bibfnamefont {V.~Y.}\ \bibnamefont {Ponomarev}}, \bibinfo {author}
  {\bibfnamefont {A.}~\bibnamefont {Richter}}, \bibinfo {author} {\bibfnamefont
  {A.}~\bibnamefont {Shevchenko}}, \bibinfo {author} {\bibfnamefont
  {S.}~\bibnamefont {Volz}}, \ and\ \bibinfo {author} {\bibfnamefont
  {J.}~\bibnamefont {Wambach}},\ }\href {\doibase
  10.1103/PhysRevLett.89.272502} {\bibfield  {journal} {\bibinfo  {journal}
  {Phys. Rev. Lett.}\ }\textbf {\bibinfo {volume} {89}},\ \bibinfo {pages}
  {272502} (\bibinfo {year} {2002})}\BibitemShut {NoStop}%
\bibitem [{\citenamefont {Litvinova}\ \emph
  {et~al.}(2009{\natexlab{b}})\citenamefont {Litvinova}, \citenamefont {Ring},
  \citenamefont {Tselyaev},\ and\ \citenamefont {Langanke}}]{Lit09a}%
  \BibitemOpen
  \bibfield  {author} {\bibinfo {author} {\bibfnamefont {E.}~\bibnamefont
  {Litvinova}}, \bibinfo {author} {\bibfnamefont {P.}~\bibnamefont {Ring}},
  \bibinfo {author} {\bibfnamefont {V.}~\bibnamefont {Tselyaev}}, \ and\
  \bibinfo {author} {\bibfnamefont {K.}~\bibnamefont {Langanke}},\ }\href
  {\doibase 10.1103/PhysRevC.79.054312} {\bibfield  {journal} {\bibinfo
  {journal} {Phys. Rev. C}\ }\textbf {\bibinfo {volume} {79}},\ \bibinfo
  {pages} {054312} (\bibinfo {year} {2009}{\natexlab{b}})}\BibitemShut
  {NoStop}%
\bibitem [{\citenamefont {Lanza}\ \emph {et~al.}(2009)\citenamefont {Lanza},
  \citenamefont {Catara}, \citenamefont {Gambacurta}, \citenamefont
  {Andr\'es},\ and\ \citenamefont {Chomaz}}]{Lan09a}%
  \BibitemOpen
  \bibfield  {author} {\bibinfo {author} {\bibfnamefont {E.~G.}\ \bibnamefont
  {Lanza}}, \bibinfo {author} {\bibfnamefont {F.}~\bibnamefont {Catara}},
  \bibinfo {author} {\bibfnamefont {D.}~\bibnamefont {Gambacurta}}, \bibinfo
  {author} {\bibfnamefont {M.~V.}\ \bibnamefont {Andr\'es}}, \ and\ \bibinfo
  {author} {\bibfnamefont {P.}~\bibnamefont {Chomaz}},\ }\href {\doibase
  10.1103/PhysRevC.79.054615} {\bibfield  {journal} {\bibinfo  {journal} {Phys.
  Rev. C}\ }\textbf {\bibinfo {volume} {79}},\ \bibinfo {pages} {054615}
  (\bibinfo {year} {2009})}\BibitemShut {NoStop}%
\bibitem [{\citenamefont {Roca-Maza}\ \emph {et~al.}(2012)\citenamefont
  {Roca-Maza}, \citenamefont {Pozzi}, \citenamefont {Brenna}, \citenamefont
  {Mizuyama},\ and\ \citenamefont {Col\`o}}]{Roc12a}%
  \BibitemOpen
  \bibfield  {author} {\bibinfo {author} {\bibfnamefont {X.}~\bibnamefont
  {Roca-Maza}}, \bibinfo {author} {\bibfnamefont {G.}~\bibnamefont {Pozzi}},
  \bibinfo {author} {\bibfnamefont {M.}~\bibnamefont {Brenna}}, \bibinfo
  {author} {\bibfnamefont {K.}~\bibnamefont {Mizuyama}}, \ and\ \bibinfo
  {author} {\bibfnamefont {G.}~\bibnamefont {Col\`o}},\ }\href {\doibase
  10.1103/PhysRevC.85.024601} {\bibfield  {journal} {\bibinfo  {journal} {Phys.
  Rev. C}\ }\textbf {\bibinfo {volume} {85}},\ \bibinfo {pages} {024601}
  (\bibinfo {year} {2012})}\BibitemShut {NoStop}%
\bibitem [{\citenamefont {Vretenar}\ \emph {et~al.}(2012)\citenamefont
  {Vretenar}, \citenamefont {Niu}, \citenamefont {Paar},\ and\ \citenamefont
  {Meng}}]{Vre12a}%
  \BibitemOpen
  \bibfield  {author} {\bibinfo {author} {\bibfnamefont {D.}~\bibnamefont
  {Vretenar}}, \bibinfo {author} {\bibfnamefont {Y.~F.}\ \bibnamefont {Niu}},
  \bibinfo {author} {\bibfnamefont {N.}~\bibnamefont {Paar}}, \ and\ \bibinfo
  {author} {\bibfnamefont {J.}~\bibnamefont {Meng}},\ }\href {\doibase
  10.1103/PhysRevC.85.044317} {\bibfield  {journal} {\bibinfo  {journal} {Phys.
  Rev. C}\ }\textbf {\bibinfo {volume} {85}},\ \bibinfo {pages} {044317}
  (\bibinfo {year} {2012})}\BibitemShut {NoStop}%
\bibitem [{\citenamefont {Bianco}\ \emph {et~al.}(2012)\citenamefont {Bianco},
  \citenamefont {Knapp}, \citenamefont {Lo~Iudice}, \citenamefont {Andreozzi},
  \citenamefont {Porrino},\ and\ \citenamefont {Vesely}}]{Bia12a}%
  \BibitemOpen
  \bibfield  {author} {\bibinfo {author} {\bibfnamefont {D.}~\bibnamefont
  {Bianco}}, \bibinfo {author} {\bibfnamefont {F.}~\bibnamefont {Knapp}},
  \bibinfo {author} {\bibfnamefont {N.}~\bibnamefont {Lo~Iudice}}, \bibinfo
  {author} {\bibfnamefont {F.}~\bibnamefont {Andreozzi}}, \bibinfo {author}
  {\bibfnamefont {A.}~\bibnamefont {Porrino}}, \ and\ \bibinfo {author}
  {\bibfnamefont {P.}~\bibnamefont {Vesely}},\ }\href {\doibase
  10.1103/PhysRevC.86.044327} {\bibfield  {journal} {\bibinfo  {journal} {Phys.
  Rev. C}\ }\textbf {\bibinfo {volume} {86}},\ \bibinfo {pages} {044327}
  (\bibinfo {year} {2012})}\BibitemShut {NoStop}%
\bibitem [{\citenamefont {Baran}\ \emph {et~al.}(2015)\citenamefont {Baran},
  \citenamefont {Palade}, \citenamefont {Colonna}, \citenamefont {Di~Toro},
  \citenamefont {Croitoru},\ and\ \citenamefont {Nicolin}}]{Bar15a}%
  \BibitemOpen
  \bibfield  {author} {\bibinfo {author} {\bibfnamefont {V.}~\bibnamefont
  {Baran}}, \bibinfo {author} {\bibfnamefont {D.~I.}\ \bibnamefont {Palade}},
  \bibinfo {author} {\bibfnamefont {M.}~\bibnamefont {Colonna}}, \bibinfo
  {author} {\bibfnamefont {M.}~\bibnamefont {Di~Toro}}, \bibinfo {author}
  {\bibfnamefont {A.}~\bibnamefont {Croitoru}}, \ and\ \bibinfo {author}
  {\bibfnamefont {A.~I.}\ \bibnamefont {Nicolin}},\ }\href {\doibase
  10.1103/PhysRevC.91.054303} {\bibfield  {journal} {\bibinfo  {journal} {Phys.
  Rev. C}\ }\textbf {\bibinfo {volume} {91}},\ \bibinfo {pages} {054303}
  (\bibinfo {year} {2015})}\BibitemShut {NoStop}%
\bibitem [{\citenamefont {Papakonstantinou}\ \emph {et~al.}(2015)\citenamefont
  {Papakonstantinou}, \citenamefont {Hergert},\ and\ \citenamefont
  {Roth}}]{Pap15a}%
  \BibitemOpen
  \bibfield  {author} {\bibinfo {author} {\bibfnamefont {P.}~\bibnamefont
  {Papakonstantinou}}, \bibinfo {author} {\bibfnamefont {H.}~\bibnamefont
  {Hergert}}, \ and\ \bibinfo {author} {\bibfnamefont {R.}~\bibnamefont
  {Roth}},\ }\href {\doibase 10.1103/PhysRevC.92.034311} {\bibfield  {journal}
  {\bibinfo  {journal} {Phys. Rev. C}\ }\textbf {\bibinfo {volume} {92}},\
  \bibinfo {pages} {034311} (\bibinfo {year} {2015})}\BibitemShut {NoStop}%
\bibitem [{\citenamefont {Ries}\ \emph {et~al.}(2019)\citenamefont {Ries},
  \citenamefont {Pai}, \citenamefont {Beck}, \citenamefont {Beller},
  \citenamefont {Bhike}, \citenamefont {Derya}, \citenamefont {Gayer},
  \citenamefont {Isaak}, \citenamefont {L\"oher}, \citenamefont {Krishichayan},
  \citenamefont {Mertes}, \citenamefont {Pietralla}, \citenamefont {Romig},
  \citenamefont {Savran}, \citenamefont {Schilling}, \citenamefont {Tornow},
  \citenamefont {Typel}, \citenamefont {Werner}, \citenamefont {Wilhelmy},
  \citenamefont {Zilges},\ and\ \citenamefont {Zweidinger}}]{Rie19a}%
  \BibitemOpen
  \bibfield  {author} {\bibinfo {author} {\bibfnamefont {P.~C.}\ \bibnamefont
  {Ries}}, \bibinfo {author} {\bibfnamefont {H.}~\bibnamefont {Pai}}, \bibinfo
  {author} {\bibfnamefont {T.}~\bibnamefont {Beck}}, \bibinfo {author}
  {\bibfnamefont {J.}~\bibnamefont {Beller}}, \bibinfo {author} {\bibfnamefont
  {M.}~\bibnamefont {Bhike}}, \bibinfo {author} {\bibfnamefont
  {V.}~\bibnamefont {Derya}}, \bibinfo {author} {\bibfnamefont
  {U.}~\bibnamefont {Gayer}}, \bibinfo {author} {\bibfnamefont
  {J.}~\bibnamefont {Isaak}}, \bibinfo {author} {\bibfnamefont
  {B.}~\bibnamefont {L\"oher}}, \bibinfo {author} {\bibnamefont
  {Krishichayan}}, \bibinfo {author} {\bibfnamefont {L.}~\bibnamefont
  {Mertes}}, \bibinfo {author} {\bibfnamefont {N.}~\bibnamefont {Pietralla}},
  \bibinfo {author} {\bibfnamefont {C.}~\bibnamefont {Romig}}, \bibinfo
  {author} {\bibfnamefont {D.}~\bibnamefont {Savran}}, \bibinfo {author}
  {\bibfnamefont {M.}~\bibnamefont {Schilling}}, \bibinfo {author}
  {\bibfnamefont {W.}~\bibnamefont {Tornow}}, \bibinfo {author} {\bibfnamefont
  {S.}~\bibnamefont {Typel}}, \bibinfo {author} {\bibfnamefont
  {V.}~\bibnamefont {Werner}}, \bibinfo {author} {\bibfnamefont
  {J.}~\bibnamefont {Wilhelmy}}, \bibinfo {author} {\bibfnamefont
  {A.}~\bibnamefont {Zilges}}, \ and\ \bibinfo {author} {\bibfnamefont
  {M.}~\bibnamefont {Zweidinger}},\ }\href {\doibase
  10.1103/PhysRevC.100.021301} {\bibfield  {journal} {\bibinfo  {journal}
  {Phys. Rev. C}\ }\textbf {\bibinfo {volume} {100}},\ \bibinfo {pages}
  {021301} (\bibinfo {year} {2019})}\BibitemShut {NoStop}%
\bibitem [{\citenamefont {Heusler}\ \emph {et~al.}(2007)\citenamefont
  {Heusler}, \citenamefont {Graw}, \citenamefont {Hertenberger}, \citenamefont
  {Riess}, \citenamefont {Wirth}, \citenamefont {Kr\"ucken},\ and\
  \citenamefont {Brentano}}]{Heu07a}%
  \BibitemOpen
  \bibfield  {author} {\bibinfo {author} {\bibfnamefont {A.}~\bibnamefont
  {Heusler}}, \bibinfo {author} {\bibfnamefont {G.}~\bibnamefont {Graw}},
  \bibinfo {author} {\bibfnamefont {R.}~\bibnamefont {Hertenberger}}, \bibinfo
  {author} {\bibfnamefont {F.}~\bibnamefont {Riess}}, \bibinfo {author}
  {\bibfnamefont {H.-F.}\ \bibnamefont {Wirth}}, \bibinfo {author}
  {\bibfnamefont {R.}~\bibnamefont {Kr\"ucken}}, \ and\ \bibinfo {author}
  {\bibfnamefont {P.~v.}\ \bibnamefont {Brentano}},\ }\href {\doibase
  10.1103/PhysRevC.75.024312} {\bibfield  {journal} {\bibinfo  {journal} {Phys.
  Rev. C}\ }\textbf {\bibinfo {volume} {75}},\ \bibinfo {pages} {024312}
  (\bibinfo {year} {2007})}\BibitemShut {NoStop}%
\bibitem [{\citenamefont {Heusler}\ \emph {et~al.}(2010)\citenamefont
  {Heusler}, \citenamefont {Faestermann}, \citenamefont {Hertenberger},
  \citenamefont {Kr{\"u}cken}, \citenamefont {Wirth},\ and\ \citenamefont {von
  Brentano}}]{Heu10a}%
  \BibitemOpen
  \bibfield  {author} {\bibinfo {author} {\bibfnamefont {A.}~\bibnamefont
  {Heusler}}, \bibinfo {author} {\bibfnamefont {T.}~\bibnamefont
  {Faestermann}}, \bibinfo {author} {\bibfnamefont {R.}~\bibnamefont
  {Hertenberger}}, \bibinfo {author} {\bibfnamefont {R.}~\bibnamefont
  {Kr{\"u}cken}}, \bibinfo {author} {\bibfnamefont {H.~F.}\ \bibnamefont
  {Wirth}}, \ and\ \bibinfo {author} {\bibfnamefont {P.}~\bibnamefont {von
  Brentano}},\ }\href {\doibase 10.1140/epja/i2010-11019-8} {\bibfield
  {journal} {\bibinfo  {journal} {Eur. Phys. J. A}\ }\textbf {\bibinfo {volume}
  {46}},\ \bibinfo {pages} {17} (\bibinfo {year} {2010})}\BibitemShut {NoStop}%
\bibitem [{\citenamefont {Heusler}\ \emph
  {et~al.}(2014{\natexlab{a}})\citenamefont {Heusler}, \citenamefont
  {Faestermann}, \citenamefont {Hertenberger}, \citenamefont {Wirth},\ and\
  \citenamefont {von Brentano}}]{Heu14a}%
  \BibitemOpen
  \bibfield  {author} {\bibinfo {author} {\bibfnamefont {A.}~\bibnamefont
  {Heusler}}, \bibinfo {author} {\bibfnamefont {T.}~\bibnamefont
  {Faestermann}}, \bibinfo {author} {\bibfnamefont {R.}~\bibnamefont
  {Hertenberger}}, \bibinfo {author} {\bibfnamefont {H.-F.}\ \bibnamefont
  {Wirth}}, \ and\ \bibinfo {author} {\bibfnamefont {P.}~\bibnamefont {von
  Brentano}},\ }\href {\doibase 10.1103/PhysRevC.89.024322} {\bibfield
  {journal} {\bibinfo  {journal} {Phys. Rev. C}\ }\textbf {\bibinfo {volume}
  {89}},\ \bibinfo {pages} {024322} (\bibinfo {year}
  {2014}{\natexlab{a}})}\BibitemShut {NoStop}%
\bibitem [{\citenamefont {Heusler}\ \emph
  {et~al.}(2014{\natexlab{b}})\citenamefont {Heusler}, \citenamefont
  {Gl{\"o}ckner}, \citenamefont {Grosse}, \citenamefont {Moore}, \citenamefont
  {Solf},\ and\ \citenamefont {von Brentano}}]{Heu14b}%
  \BibitemOpen
  \bibfield  {author} {\bibinfo {author} {\bibfnamefont {A.}~\bibnamefont
  {Heusler}}, \bibinfo {author} {\bibfnamefont {H.-J.}\ \bibnamefont
  {Gl{\"o}ckner}}, \bibinfo {author} {\bibfnamefont {E.}~\bibnamefont
  {Grosse}}, \bibinfo {author} {\bibfnamefont {C.}~\bibnamefont {Moore}},
  \bibinfo {author} {\bibfnamefont {J.}~\bibnamefont {Solf}}, \ and\ \bibinfo
  {author} {\bibfnamefont {P.}~\bibnamefont {von Brentano}},\ }\href {\doibase
  10.1140/epja/i2014-14092-y} {\bibfield  {journal} {\bibinfo  {journal} {Eur.
  Phys. J. A}\ }\textbf {\bibinfo {volume} {50}},\ \bibinfo {pages} {92}
  (\bibinfo {year} {2014}{\natexlab{b}})}\BibitemShut {NoStop}%
\bibitem [{\citenamefont {Heusler}(2020)}]{Heu20a}%
  \BibitemOpen
  \bibfield  {author} {\bibinfo {author} {\bibfnamefont {A.}~\bibnamefont
  {Heusler}},\ }\href@noop {} {}\bibinfo {howpublished} {Eur. Phys. J. A,
  accepted for publication} (\bibinfo {year} {2020})\BibitemShut {NoStop}%
\bibitem [{\citenamefont {Heusler}()}]{Heu20b}%
  \BibitemOpen
  \bibfield  {author} {\bibinfo {author} {\bibfnamefont {A.}~\bibnamefont
  {Heusler}},\ }\href@noop {} {}\bibinfo {howpublished} {to be
  published}\BibitemShut {NoStop}%
\bibitem [{\citenamefont {Schwengner}\ \emph {et~al.}(2010)\citenamefont
  {Schwengner}, \citenamefont {Massarczyk}, \citenamefont {Brown},
  \citenamefont {Beyer}, \citenamefont {D\"onau}, \citenamefont {Erhard},
  \citenamefont {Grosse}, \citenamefont {Junghans}, \citenamefont {Kosev},
  \citenamefont {Nair}, \citenamefont {Rusev}, \citenamefont {Schilling},\ and\
  \citenamefont {Wagner}}]{Sch10a}%
  \BibitemOpen
  \bibfield  {author} {\bibinfo {author} {\bibfnamefont {R.}~\bibnamefont
  {Schwengner}}, \bibinfo {author} {\bibfnamefont {R.}~\bibnamefont
  {Massarczyk}}, \bibinfo {author} {\bibfnamefont {B.~A.}\ \bibnamefont
  {Brown}}, \bibinfo {author} {\bibfnamefont {R.}~\bibnamefont {Beyer}},
  \bibinfo {author} {\bibfnamefont {F.}~\bibnamefont {D\"onau}}, \bibinfo
  {author} {\bibfnamefont {M.}~\bibnamefont {Erhard}}, \bibinfo {author}
  {\bibfnamefont {E.}~\bibnamefont {Grosse}}, \bibinfo {author} {\bibfnamefont
  {A.~R.}\ \bibnamefont {Junghans}}, \bibinfo {author} {\bibfnamefont
  {K.}~\bibnamefont {Kosev}}, \bibinfo {author} {\bibfnamefont
  {C.}~\bibnamefont {Nair}}, \bibinfo {author} {\bibfnamefont {G.}~\bibnamefont
  {Rusev}}, \bibinfo {author} {\bibfnamefont {K.~D.}\ \bibnamefont
  {Schilling}}, \ and\ \bibinfo {author} {\bibfnamefont {A.}~\bibnamefont
  {Wagner}},\ }\href {\doibase 10.1103/PhysRevC.81.054315} {\bibfield
  {journal} {\bibinfo  {journal} {Phys. Rev. C}\ }\textbf {\bibinfo {volume}
  {81}},\ \bibinfo {pages} {054315} (\bibinfo {year} {2010})}\BibitemShut
  {NoStop}%
\bibitem [{\citenamefont {Brown}(2000)}]{Bro00a}%
  \BibitemOpen
  \bibfield  {author} {\bibinfo {author} {\bibfnamefont {B.~A.}\ \bibnamefont
  {Brown}},\ }\href {\doibase 10.1103/PhysRevLett.85.5300} {\bibfield
  {journal} {\bibinfo  {journal} {Phys. Rev. Lett.}\ }\textbf {\bibinfo
  {volume} {85}},\ \bibinfo {pages} {5300} (\bibinfo {year}
  {2000})}\BibitemShut {NoStop}%
\bibitem [{\citenamefont {Heusler}(2017)}]{Heu17a}%
  \BibitemOpen
  \bibfield  {author} {\bibinfo {author} {\bibfnamefont {A.}~\bibnamefont
  {Heusler}},\ }\href {\doibase 10.1140/epja/i2017-12416-1} {\bibfield
  {journal} {\bibinfo  {journal} {Eur. Phys. J. A}\ }\textbf {\bibinfo {volume}
  {53}},\ \bibinfo {pages} {215} (\bibinfo {year} {2017})}\BibitemShut
  {NoStop}%
\bibitem [{\citenamefont {{Maier-Leibnitz-Laboratorium
  M{\"u}nchen}}(2020)}]{MLL}%
  \BibitemOpen
  \bibfield  {author} {\bibinfo {author} {\bibnamefont
  {{Maier-Leibnitz-Laboratorium M{\"u}nchen}}},\ }\href@noop {} {}\bibinfo
  {howpublished} {\url{https://www.bl.physik.uni-muenchen.de/}} (\bibinfo
  {year} {2020})\BibitemShut {NoStop}%
\bibitem [{\citenamefont {Dollinger}\ and\ \citenamefont
  {Faestermann}(2018)}]{MLL2}%
  \BibitemOpen
  \bibfield  {author} {\bibinfo {author} {\bibfnamefont {G.}~\bibnamefont
  {Dollinger}}\ and\ \bibinfo {author} {\bibfnamefont {T.}~\bibnamefont
  {Faestermann}},\ }\href@noop {} {\bibfield  {journal} {\bibinfo  {journal}
  {Nucl. Phys. News}\ }\textbf {\bibinfo {volume} {28}},\ \bibinfo {pages} {5}
  (\bibinfo {year} {2018})}\BibitemShut {NoStop}%
\bibitem [{\citenamefont {Wirth}\ \emph {et~al.}()\citenamefont {Wirth} \emph
  {et~al.}}]{Wir00}%
  \BibitemOpen
  \bibfield  {author} {\bibinfo {author} {\bibfnamefont {H.-F.}\ \bibnamefont
  {Wirth}} \emph {et~al.},\ }\href@noop {} {}\bibinfo {howpublished} {Annual
  Report 2000, Beschleunigerlaboratorium M\"unchen, 71 (2000),
  {\url{https://www-old.mll-muenchen.de/bl_rep/jb2000/index.html}}}\BibitemShut
  {NoStop}%
\bibitem [{\citenamefont {Wirth}(2001)}]{Wir01}%
  \BibitemOpen
  \bibfield  {author} {\bibinfo {author} {\bibfnamefont {H.-F.}\ \bibnamefont
  {Wirth}},\ }\emph {\bibinfo {title} {Bau eines hochaufl{\"o}senden
  Fokalebenendetektors fur den M{\"u}nchener Q3D-Magnetspektrographen und
  Untersuchung zur Kernstruktur von $^{129}$Te}},\ \href
  {http://mediatum.ub.tum.de/?id=602907} {Ph.D. thesis},\ \bibinfo  {school}
  {Technical University Munich (Germany),
  {\url{http://mediatum.ub.tum.de/?id=602907}}} (\bibinfo {year}
  {2001})\BibitemShut {NoStop}%
\bibitem [{\citenamefont {Enge}(1979)}]{Eng79a}%
  \BibitemOpen
  \bibfield  {author} {\bibinfo {author} {\bibfnamefont {H.~A.}\ \bibnamefont
  {Enge}},\ }\href {\doibase https://doi.org/10.1016/0029-554X(79)90711-0}
  {\bibfield  {journal} {\bibinfo  {journal} {Nuclear Instruments and Methods}\
  }\textbf {\bibinfo {volume} {162}},\ \bibinfo {pages} {161} (\bibinfo {year}
  {1979})}\BibitemShut {NoStop}%
\bibitem [{\citenamefont {Martin}(2007)}]{Mar07a}%
  \BibitemOpen
  \bibfield  {author} {\bibinfo {author} {\bibfnamefont {M.}~\bibnamefont
  {Martin}},\ }\href {\doibase https://doi.org/10.1016/j.nds.2007.07.001}
  {\bibfield  {journal} {\bibinfo  {journal} {Nuclear Data Sheets}\ }\textbf
  {\bibinfo {volume} {108}},\ \bibinfo {pages} {1583 } (\bibinfo {year}
  {2007})}\BibitemShut {NoStop}%
\bibitem [{ENSDF()}]{ENSDF}%
  \BibitemOpen
  ENSDF,\ \href@noop {} {}\bibinfo {howpublished} {NNDC Online Data Service,
  ENSDF database, \newline \url{http://www.nndc.bnl.gov/ensdf/}} (\bibinfo
  {year} {2020})\BibitemShut {NoStop}%
\bibitem [{\citenamefont {Heusler}\ \emph {et~al.}(2016)\citenamefont
  {Heusler}, \citenamefont {Jolos}, \citenamefont {Faestermann}, \citenamefont
  {Hertenberger}, \citenamefont {Wirth},\ and\ \citenamefont {von
  Brentano}}]{Heu16a}%
  \BibitemOpen
  \bibfield  {author} {\bibinfo {author} {\bibfnamefont {A.}~\bibnamefont
  {Heusler}}, \bibinfo {author} {\bibfnamefont {R.~V.}\ \bibnamefont {Jolos}},
  \bibinfo {author} {\bibfnamefont {T.}~\bibnamefont {Faestermann}}, \bibinfo
  {author} {\bibfnamefont {R.}~\bibnamefont {Hertenberger}}, \bibinfo {author}
  {\bibfnamefont {H.-F.}\ \bibnamefont {Wirth}}, \ and\ \bibinfo {author}
  {\bibfnamefont {P.}~\bibnamefont {von Brentano}},\ }\href {\doibase
  10.1103/PhysRevC.93.054321} {\bibfield  {journal} {\bibinfo  {journal} {Phys.
  Rev. C}\ }\textbf {\bibinfo {volume} {93}},\ \bibinfo {pages} {054321}
  (\bibinfo {year} {2016})}\BibitemShut {NoStop}%
\bibitem [{\citenamefont {Vold}\ \emph {et~al.}(1973)\citenamefont {Vold},
  \citenamefont {Andreassen}, \citenamefont {Lien}, \citenamefont {Graue},
  \citenamefont {Cosman}, \citenamefont {Dünnweber}, \citenamefont {Schmitt},\
  and\ \citenamefont {Nüsslin}}]{Vol73a}%
  \BibitemOpen
  \bibfield  {author} {\bibinfo {author} {\bibfnamefont {P.}~\bibnamefont
  {Vold}}, \bibinfo {author} {\bibfnamefont {J.}~\bibnamefont {Andreassen}},
  \bibinfo {author} {\bibfnamefont {J.}~\bibnamefont {Lien}}, \bibinfo {author}
  {\bibfnamefont {A.}~\bibnamefont {Graue}}, \bibinfo {author} {\bibfnamefont
  {E.}~\bibnamefont {Cosman}}, \bibinfo {author} {\bibfnamefont
  {W.}~\bibnamefont {Dünnweber}}, \bibinfo {author} {\bibfnamefont
  {D.}~\bibnamefont {Schmitt}}, \ and\ \bibinfo {author} {\bibfnamefont
  {F.}~\bibnamefont {Nüsslin}},\ }\href {\doibase
  https://doi.org/10.1016/0375-9474(73)90101-2} {\bibfield  {journal} {\bibinfo
   {journal} {Nuclear Physics A}\ }\textbf {\bibinfo {volume} {215}},\ \bibinfo
  {pages} {61 } (\bibinfo {year} {1973})}\BibitemShut {NoStop}%
\bibitem [{\citenamefont {Schramm}\ \emph {et~al.}(1997)\citenamefont
  {Schramm}, \citenamefont {Maier}, \citenamefont {Rejmund}, \citenamefont
  {Wood}, \citenamefont {Roy}, \citenamefont {Kuhnert}, \citenamefont
  {Aprahamian}, \citenamefont {Becker}, \citenamefont {Brinkman}, \citenamefont
  {Decman}, \citenamefont {Henry}, \citenamefont {Hoff}, \citenamefont
  {Manatt}, \citenamefont {Mann}, \citenamefont {Meyer}, \citenamefont
  {Stoeffl}, \citenamefont {Struble},\ and\ \citenamefont {Wang}}]{Sch97a}%
  \BibitemOpen
  \bibfield  {author} {\bibinfo {author} {\bibfnamefont {M.}~\bibnamefont
  {Schramm}}, \bibinfo {author} {\bibfnamefont {K.~H.}\ \bibnamefont {Maier}},
  \bibinfo {author} {\bibfnamefont {M.}~\bibnamefont {Rejmund}}, \bibinfo
  {author} {\bibfnamefont {L.~D.}\ \bibnamefont {Wood}}, \bibinfo {author}
  {\bibfnamefont {N.}~\bibnamefont {Roy}}, \bibinfo {author} {\bibfnamefont
  {A.}~\bibnamefont {Kuhnert}}, \bibinfo {author} {\bibfnamefont
  {A.}~\bibnamefont {Aprahamian}}, \bibinfo {author} {\bibfnamefont
  {J.}~\bibnamefont {Becker}}, \bibinfo {author} {\bibfnamefont
  {M.}~\bibnamefont {Brinkman}}, \bibinfo {author} {\bibfnamefont {D.~J.}\
  \bibnamefont {Decman}}, \bibinfo {author} {\bibfnamefont {E.~A.}\
  \bibnamefont {Henry}}, \bibinfo {author} {\bibfnamefont {R.}~\bibnamefont
  {Hoff}}, \bibinfo {author} {\bibfnamefont {D.}~\bibnamefont {Manatt}},
  \bibinfo {author} {\bibfnamefont {L.~G.}\ \bibnamefont {Mann}}, \bibinfo
  {author} {\bibfnamefont {R.~A.}\ \bibnamefont {Meyer}}, \bibinfo {author}
  {\bibfnamefont {W.}~\bibnamefont {Stoeffl}}, \bibinfo {author} {\bibfnamefont
  {G.~L.}\ \bibnamefont {Struble}}, \ and\ \bibinfo {author} {\bibfnamefont
  {T.-F.}\ \bibnamefont {Wang}},\ }\href {\doibase 10.1103/PhysRevC.56.1320}
  {\bibfield  {journal} {\bibinfo  {journal} {Phys. Rev. C}\ }\textbf {\bibinfo
  {volume} {56}},\ \bibinfo {pages} {1320} (\bibinfo {year}
  {1997})}\BibitemShut {NoStop}%
\bibitem [{\citenamefont {Valnion}\ \emph {et~al.}(2001)\citenamefont
  {Valnion}, \citenamefont {Ponomarev}, \citenamefont {Eisermann},
  \citenamefont {Gollwitzer}, \citenamefont {Hertenberger}, \citenamefont
  {Metz}, \citenamefont {Schiemenz},\ and\ \citenamefont {Graw}}]{Val01a}%
  \BibitemOpen
  \bibfield  {author} {\bibinfo {author} {\bibfnamefont {B.~D.}\ \bibnamefont
  {Valnion}}, \bibinfo {author} {\bibfnamefont {V.~Y.}\ \bibnamefont
  {Ponomarev}}, \bibinfo {author} {\bibfnamefont {Y.}~\bibnamefont
  {Eisermann}}, \bibinfo {author} {\bibfnamefont {A.}~\bibnamefont
  {Gollwitzer}}, \bibinfo {author} {\bibfnamefont {R.}~\bibnamefont
  {Hertenberger}}, \bibinfo {author} {\bibfnamefont {A.}~\bibnamefont {Metz}},
  \bibinfo {author} {\bibfnamefont {P.}~\bibnamefont {Schiemenz}}, \ and\
  \bibinfo {author} {\bibfnamefont {G.}~\bibnamefont {Graw}},\ }\href {\doibase
  10.1103/PhysRevC.63.024318} {\bibfield  {journal} {\bibinfo  {journal} {Phys.
  Rev. C}\ }\textbf {\bibinfo {volume} {63}},\ \bibinfo {pages} {024318}
  (\bibinfo {year} {2001})}\BibitemShut {NoStop}%
\bibitem [{\citenamefont {Kunz}\ and\ \citenamefont {Comfort}()}]{chuck}%
  \BibitemOpen
  \bibfield  {author} {\bibinfo {author} {\bibfnamefont {P.~D.}\ \bibnamefont
  {Kunz}}\ and\ \bibinfo {author} {\bibfnamefont {J.~R.}\ \bibnamefont
  {Comfort}},\ }\href@noop {} {\bibinfo  {journal} {Program CHUCK}\ ,\ \bibinfo
  {pages} {unpublished}}\BibitemShut {NoStop}%
\bibitem [{\citenamefont {Becchetti}\ and\ \citenamefont
  {Greenlees}(1969)}]{bec69a}%
  \BibitemOpen
\bibfield  {journal} {  }\bibfield  {author} {\bibinfo {author} {\bibfnamefont
  {F.~D.}\ \bibnamefont {Becchetti}}\ and\ \bibinfo {author} {\bibfnamefont
  {G.~W.}\ \bibnamefont {Greenlees}},\ }\href@noop {} {\bibfield  {journal}
  {\bibinfo  {journal} {Phys. Rev.}\ }\textbf {\bibinfo {volume} {182}},\
  \bibinfo {pages} {1190} (\bibinfo {year} {1969})}\BibitemShut {NoStop}%
\bibitem [{\citenamefont {Daehnick}\ \emph {et~al.}(1980)\citenamefont
  {Daehnick}, \citenamefont {Childs},\ and\ \citenamefont {Vrcelj}}]{Dae80a}%
  \BibitemOpen
  \bibfield  {author} {\bibinfo {author} {\bibfnamefont {W.~W.}\ \bibnamefont
  {Daehnick}}, \bibinfo {author} {\bibfnamefont {J.~D.}\ \bibnamefont
  {Childs}}, \ and\ \bibinfo {author} {\bibfnamefont {Z.}~\bibnamefont
  {Vrcelj}},\ }\href {\doibase 10.1103/PhysRevC.21.2253} {\bibfield  {journal}
  {\bibinfo  {journal} {Phys. Rev. C}\ }\textbf {\bibinfo {volume} {21}},\
  \bibinfo {pages} {2253} (\bibinfo {year} {1980})}\BibitemShut {NoStop}%
\bibitem [{\citenamefont {Debenham}\ \emph {et~al.}(1970)\citenamefont
  {Debenham}, \citenamefont {Griffith}, \citenamefont {Irshad},\ and\
  \citenamefont {Roman}}]{Deb70a}%
  \BibitemOpen
  \bibfield  {author} {\bibinfo {author} {\bibfnamefont {A.}~\bibnamefont
  {Debenham}}, \bibinfo {author} {\bibfnamefont {J.}~\bibnamefont {Griffith}},
  \bibinfo {author} {\bibfnamefont {M.}~\bibnamefont {Irshad}}, \ and\ \bibinfo
  {author} {\bibfnamefont {S.}~\bibnamefont {Roman}},\ }\href {\doibase
  https://doi.org/10.1016/0375-9474(70)90969-3} {\bibfield  {journal} {\bibinfo
   {journal} {Nuclear Physics A}\ }\textbf {\bibinfo {volume} {151}},\ \bibinfo
  {pages} {81 } (\bibinfo {year} {1970})}\BibitemShut {NoStop}%
\bibitem [{\citenamefont {Valnion}(1998)}]{Val98a}%
  \BibitemOpen
  \bibfield  {author} {\bibinfo {author} {\bibfnamefont {B.~D.}\ \bibnamefont
  {Valnion}},\ }\emph {\bibinfo {title} {Leichtionen-induzierte Anregungen in
  $^{178}$Hf und $^{208}$Pb}},\ \href@noop {} {Ph.D. thesis},\ \bibinfo
  {school} {Ludwig-Maximilians-Universit{\"a}t M{\"u}nchen (Germany)} (\bibinfo
  {year} {1998})\BibitemShut {NoStop}%
\bibitem [{\citenamefont {Kovar}\ \emph {et~al.}(1974)\citenamefont {Kovar},
  \citenamefont {Stein},\ and\ \citenamefont {Bockelman}}]{Kov74a}%
  \BibitemOpen
  \bibfield  {author} {\bibinfo {author} {\bibfnamefont {D.~G.}\ \bibnamefont
  {Kovar}}, \bibinfo {author} {\bibfnamefont {N.}~\bibnamefont {Stein}}, \ and\
  \bibinfo {author} {\bibfnamefont {C.~K.}\ \bibnamefont {Bockelman}},\ }\href
  {\doibase https://doi.org/10.1016/0375-9474(74)90522-3} {\bibfield  {journal}
  {\bibinfo  {journal} {Nuclear Physics A}\ }\textbf {\bibinfo {volume}
  {231}},\ \bibinfo {pages} {266} (\bibinfo {year} {1974})}\BibitemShut
  {NoStop}%
\bibitem [{\citenamefont {Tsoneva}\ and\ \citenamefont
  {Lenske}(2016)}]{Tso16a}%
  \BibitemOpen
  \bibfield  {author} {\bibinfo {author} {\bibfnamefont {N.}~\bibnamefont
  {Tsoneva}}\ and\ \bibinfo {author} {\bibfnamefont {H.}~\bibnamefont
  {Lenske}},\ }\href {\doibase 10.1134/S1063778816060247} {\bibfield  {journal}
  {\bibinfo  {journal} {Phys. Atom. Nuclei}\ }\textbf {\bibinfo {volume}
  {79}},\ \bibinfo {pages} {885} (\bibinfo {year} {2016})}\BibitemShut
  {NoStop}%
\bibitem [{sup()}]{suppl}%
  \BibitemOpen
  \href@noop {} {}\bibinfo {howpublished} {See Supplemental Material at
  http://link.aps.org/supplemental/10.1103/yyy.zz.xxxxxx for additional
  discussion of the theoretical $B(E1)$ strengths and additional information on
  the LSSM, EDF+QRPA and EDF+QPM wave functions.}\BibitemShut {Stop}%
\bibitem [{\citenamefont {Ratkiewicz}\ \emph {et~al.}(2019)\citenamefont
  {Ratkiewicz}, \citenamefont {Cizewski}, \citenamefont {Escher}, \citenamefont
  {Potel}, \citenamefont {Burke}, \citenamefont {Casperson}, \citenamefont
  {McCleskey}, \citenamefont {Austin}, \citenamefont {Burcher}, \citenamefont
  {Hughes}, \citenamefont {Manning}, \citenamefont {Pain}, \citenamefont
  {Peters}, \citenamefont {Rice}, \citenamefont {Ross}, \citenamefont
  {Scielzo}, \citenamefont {Shand},\ and\ \citenamefont {Smith}}]{Rat19a}%
  \BibitemOpen
  \bibfield  {author} {\bibinfo {author} {\bibfnamefont {A.}~\bibnamefont
  {Ratkiewicz}}, \bibinfo {author} {\bibfnamefont {J.~A.}\ \bibnamefont
  {Cizewski}}, \bibinfo {author} {\bibfnamefont {J.~E.}\ \bibnamefont
  {Escher}}, \bibinfo {author} {\bibfnamefont {G.}~\bibnamefont {Potel}},
  \bibinfo {author} {\bibfnamefont {J.~T.}\ \bibnamefont {Burke}}, \bibinfo
  {author} {\bibfnamefont {R.~J.}\ \bibnamefont {Casperson}}, \bibinfo {author}
  {\bibfnamefont {M.}~\bibnamefont {McCleskey}}, \bibinfo {author}
  {\bibfnamefont {R.~A.~E.}\ \bibnamefont {Austin}}, \bibinfo {author}
  {\bibfnamefont {S.}~\bibnamefont {Burcher}}, \bibinfo {author} {\bibfnamefont
  {R.~O.}\ \bibnamefont {Hughes}}, \bibinfo {author} {\bibfnamefont
  {B.}~\bibnamefont {Manning}}, \bibinfo {author} {\bibfnamefont {S.~D.}\
  \bibnamefont {Pain}}, \bibinfo {author} {\bibfnamefont {W.~A.}\ \bibnamefont
  {Peters}}, \bibinfo {author} {\bibfnamefont {S.}~\bibnamefont {Rice}},
  \bibinfo {author} {\bibfnamefont {T.~J.}\ \bibnamefont {Ross}}, \bibinfo
  {author} {\bibfnamefont {N.~D.}\ \bibnamefont {Scielzo}}, \bibinfo {author}
  {\bibfnamefont {C.}~\bibnamefont {Shand}}, \ and\ \bibinfo {author}
  {\bibfnamefont {K.}~\bibnamefont {Smith}},\ }\href {\doibase
  10.1103/PhysRevLett.122.052502} {\bibfield  {journal} {\bibinfo  {journal}
  {Phys. Rev. Lett.}\ }\textbf {\bibinfo {volume} {122}},\ \bibinfo {pages}
  {052502} (\bibinfo {year} {2019})}\BibitemShut {NoStop}%
\bibitem [{\citenamefont {Escher}\ \emph {et~al.}(2012)\citenamefont {Escher},
  \citenamefont {Burke}, \citenamefont {Dietrich}, \citenamefont {Scielzo},
  \citenamefont {Thompson},\ and\ \citenamefont {Younes}}]{Esch12a}%
  \BibitemOpen
  \bibfield  {author} {\bibinfo {author} {\bibfnamefont {J.~E.}\ \bibnamefont
  {Escher}}, \bibinfo {author} {\bibfnamefont {J.~T.}\ \bibnamefont {Burke}},
  \bibinfo {author} {\bibfnamefont {F.~S.}\ \bibnamefont {Dietrich}}, \bibinfo
  {author} {\bibfnamefont {N.~D.}\ \bibnamefont {Scielzo}}, \bibinfo {author}
  {\bibfnamefont {I.~J.}\ \bibnamefont {Thompson}}, \ and\ \bibinfo {author}
  {\bibfnamefont {W.}~\bibnamefont {Younes}},\ }\href {\doibase
  10.1103/RevModPhys.84.353} {\bibfield  {journal} {\bibinfo  {journal} {Rev.
  Mod. Phys.}\ }\textbf {\bibinfo {volume} {84}},\ \bibinfo {pages} {353}
  (\bibinfo {year} {2012})}\BibitemShut {NoStop}%
\bibitem [{hel(2020)}]{hel20a}%
  \BibitemOpen
  \href@noop {} {}\bibinfo {howpublished} {Helical Orbit Spectrometer (HELIOS),
  \newline \url{https://www.phy.anl.gov/atlas/helios/}} (\bibinfo {year}
  {2020})\BibitemShut {NoStop}%
\bibitem [{sol(2020)}]{sol20a}%
  \BibitemOpen
  \href@noop {} {}\bibinfo {howpublished} {SOLARIS, \newline
  \url{https://www.anl.gov/phy/solaris}} (\bibinfo {year} {2020})\BibitemShut
  {NoStop}%
\bibitem [{sol(2018)}]{sol18a}%
  \BibitemOpen
  \href@noop {} {}\bibinfo {howpublished} {SOLARIS White Paper, \newline
  \url{https://www.anl.gov/sites/www/files/2018-11/solaris_white_paper_final.pdf}}
  (\bibinfo {year} {2018})\BibitemShut {NoStop}%
\bibitem [{iss(2020)}]{iss20a}%
  \BibitemOpen
  \href@noop {} {}\bibinfo {howpublished} {ISOLDE Solenoidal Spectrometer
  (ISS), \newline
  \url{https://isolde.web.cern.ch/experiments/isolde-solenoidal-spectrometer-iss}}
  (\bibinfo {year} {2020})\BibitemShut {NoStop}%
\bibitem [{\citenamefont {Tang}\ \emph {et~al.}(2020)\citenamefont {Tang},
  \citenamefont {Kay}, \citenamefont {Hoffman}, \citenamefont {Schiffer},
  \citenamefont {Sharp}, \citenamefont {Gaffney}, \citenamefont {Freeman},
  \citenamefont {Mumpower}, \citenamefont {Arokiaraj}, \citenamefont {Baader},
  \citenamefont {Butler}, \citenamefont {Catford}, \citenamefont {de~Angelis},
  \citenamefont {Flavigny}, \citenamefont {Gott}, \citenamefont {Gregor},
  \citenamefont {Konki}, \citenamefont {Labiche}, \citenamefont {Lazarus},
  \citenamefont {MacGregor}, \citenamefont {Martel}, \citenamefont {Page},
  \citenamefont {Podoly\'ak}, \citenamefont {Poleshchuk}, \citenamefont
  {Raabe}, \citenamefont {Recchia}, \citenamefont {Smith}, \citenamefont
  {Szwec},\ and\ \citenamefont {Yang}}]{Tan20a}%
  \BibitemOpen
  \bibfield  {author} {\bibinfo {author} {\bibfnamefont {T.~L.}\ \bibnamefont
  {Tang}}, \bibinfo {author} {\bibfnamefont {B.~P.}\ \bibnamefont {Kay}},
  \bibinfo {author} {\bibfnamefont {C.~R.}\ \bibnamefont {Hoffman}}, \bibinfo
  {author} {\bibfnamefont {J.~P.}\ \bibnamefont {Schiffer}}, \bibinfo {author}
  {\bibfnamefont {D.~K.}\ \bibnamefont {Sharp}}, \bibinfo {author}
  {\bibfnamefont {L.~P.}\ \bibnamefont {Gaffney}}, \bibinfo {author}
  {\bibfnamefont {S.~J.}\ \bibnamefont {Freeman}}, \bibinfo {author}
  {\bibfnamefont {M.~R.}\ \bibnamefont {Mumpower}}, \bibinfo {author}
  {\bibfnamefont {A.}~\bibnamefont {Arokiaraj}}, \bibinfo {author}
  {\bibfnamefont {E.~F.}\ \bibnamefont {Baader}}, \bibinfo {author}
  {\bibfnamefont {P.~A.}\ \bibnamefont {Butler}}, \bibinfo {author}
  {\bibfnamefont {W.~N.}\ \bibnamefont {Catford}}, \bibinfo {author}
  {\bibfnamefont {G.}~\bibnamefont {de~Angelis}}, \bibinfo {author}
  {\bibfnamefont {F.}~\bibnamefont {Flavigny}}, \bibinfo {author}
  {\bibfnamefont {M.~D.}\ \bibnamefont {Gott}}, \bibinfo {author}
  {\bibfnamefont {E.~T.}\ \bibnamefont {Gregor}}, \bibinfo {author}
  {\bibfnamefont {J.}~\bibnamefont {Konki}}, \bibinfo {author} {\bibfnamefont
  {M.}~\bibnamefont {Labiche}}, \bibinfo {author} {\bibfnamefont {I.~H.}\
  \bibnamefont {Lazarus}}, \bibinfo {author} {\bibfnamefont {P.~T.}\
  \bibnamefont {MacGregor}}, \bibinfo {author} {\bibfnamefont {I.}~\bibnamefont
  {Martel}}, \bibinfo {author} {\bibfnamefont {R.~D.}\ \bibnamefont {Page}},
  \bibinfo {author} {\bibfnamefont {Z.}~\bibnamefont {Podoly\'ak}}, \bibinfo
  {author} {\bibfnamefont {O.}~\bibnamefont {Poleshchuk}}, \bibinfo {author}
  {\bibfnamefont {R.}~\bibnamefont {Raabe}}, \bibinfo {author} {\bibfnamefont
  {F.}~\bibnamefont {Recchia}}, \bibinfo {author} {\bibfnamefont {J.~F.}\
  \bibnamefont {Smith}}, \bibinfo {author} {\bibfnamefont {S.~V.}\ \bibnamefont
  {Szwec}}, \ and\ \bibinfo {author} {\bibfnamefont {J.}~\bibnamefont {Yang}},\
  }\href {\doibase 10.1103/PhysRevLett.124.062502} {\bibfield  {journal}
  {\bibinfo  {journal} {Phys. Rev. Lett.}\ }\textbf {\bibinfo {volume} {124}},\
  \bibinfo {pages} {062502} (\bibinfo {year} {2020})}\BibitemShut {NoStop}%
\end{thebibliography}%

\end{document}